\documentclass{aa}  
\usepackage{todonotes}
\usepackage{booktabs}
\usepackage{txfonts}
\usepackage{graphicx,amsmath,amssymb,multirow}
\usepackage{xspace}
\usepackage{color}
\usepackage{natbib}
\usepackage{nicefrac}
\usepackage{ulem}

\newcommand{\Vband}{{\it V}\xspace}
\newcommand{\Bband}{{\it B}\xspace}

\newcommand{\Jband}{{\it J}\xspace}
\newcommand{\Hband}{{\it H}\xspace}

\newcommand{\Ksband}{{$K_s$}\xspace}

\newcommand{\eg}{e.g.\xspace}
\newcommand{\sn}{SN\xspace}
\newcommand{\sne}{SNe\xspace}
\newcommand{\snia}{SN~Ia\xspace}
\newcommand{\sneia}{SNe~Ia\xspace}

\def\msun{M_\odot}

\def\Ncluster{2}

\newcommand{\myemail}{tpetr@fysik.su.se}
\defcitealias{Second}{G09}

\begin{document}

 \title{High-redshift supernova rates measured with the gravitational telescope A1689}
 
 \subtitle{}

 \author{T.~Petrushevska\inst{1}\inst{,2}\fnmsep\thanks\myemail \and 
         R.~Amanullah\inst{1}\inst{,2} \and 
         A.~Goobar\inst{1}\inst{,2} \and
         S.~Fabbro\inst{3} \and
         J.~Johansson\inst{4} \and 
         T.~Kjellsson\inst{2}\and
         C.~Lidman\inst{5}\and
         K.~Paech\inst{6}\inst{,7} \and
         J.~Richard\inst{8}\and
         H.~Dahle\inst{9} \and
         R.~Ferretti\inst{1}\inst{,2} \and 
         J.~P.~Kneib\inst{10}\and
         M.~Limousin\inst{11} \and
         J.~Nordin\inst{12} \and
         V.~Stanishev\inst{13}
}
\institute{Oskar Klein Centre, Physics Department, Stockholm University, SE 106 91 Stockholm, Sweden 
\and Physics Department, Stockholm University, SE 106 91 Stockholm, Sweden
\and NRC Herzberg Institute for Astrophysics, 5071 West Saanich Road, Victoria V9E 2E7, British Columbia, Canada
\and Department of Particle Physics and Astrophysics, Weizmann Institute of Science, Rehovot 7610001, Israel
\and Australian Astronomical Observatory, PO Box 915, North Ryde NSW 1670, Australia.
\and Universit\"ats-Sternwarte, Fakult\"at f\"ur Physik, Ludwig-Maximilians Universitaet M\"unchen, Scheinerstr. 1, D-81679 Muenchen, Germany
\and Excellence Cluster Universe, Boltzmannstr. 2, D-85748 Garching, Germany
\and Univ Lyon, Univ Lyon1, Ens de Lyon, CNRS, Centre de Recherche Astrophysique de Lyon UMR5574, F-69230, Saint-Genis-Laval, France
\and Institute of Theoretical Astrophysics, University of Oslo, P.O. Box 1029, Blindern, N-0315 Oslo, Norway
\and Laboratoire d'Astrophysique, Ecole Polytechnique F\'ed\'erale de Lausanne (EPFL), Observatoire de Sauverny, CH-1290 Versoix, Switzerland
\and Laboratoire d'Astrophysique de Marseille, UMR\,6610, CNRS-Universit\'e de Provence,
38 rue Fr\'ed\'eric Joliot-Curie, 13\,388 Marseille Cedex 13, France
\and Institut f\"ur Physik, Humboldt-Universitat zu Berlin, Newtonstr. 15, 12489 Berlin, Germany 
\and Department of Physics, Chemistry and Biology, IFM, Link\"oping University, 581 83 Link\"oping, Sweden
\fnmsep\thanks{Based on observations made with  European Southern
	Observatory (ESO) telescopes at the Paranal Observatory under programme ID 082.A-0431; 0.83.A-0398, 091.A-0108 and ID 093.A-0278  PI: A. Goobar.}}
 
 \date{Received May, 13, 2016; accepted July, 5, 2016}

 \abstract
{}
{We present a ground-based, near-infrared search for lensed supernovae behind the massive 
cluster Abell 1689 at $z=0.18$, which is one of the most powerful gravitational telescopes that nature provides.}
{Our survey was based on multi-epoch $J$-band observations with the HAWK-I instrument on VLT, with supporting optical data from the 
Nordic Optical Telescope.}
{Our search resulted in the discovery of five photometrically classified, core-collapse supernovae with high redshifts of $0.671<z<1.703$ and magnifications in the range $\Delta m$ =  $-0.31$ to $-1.58$ mag, as calculated from lensing models in the literature.
Owing to the power of the lensing cluster, the survey had the sensitivity
to detect supernovae up to very high redshifts, $z$$\sim$$3$, albeit for a
limited region of space.  
We present a study of the core-collapse supernova rates for $0.4\leq z
< 2.9$, and find good agreement with previous estimates and predictions from star formation history.
During our survey, we also discovered two Type Ia supernovae in A1689 cluster members, which allowed us to determine the cluster Ia rate to be $0.14^{+0.19}_{-0.09}\pm0.01$ $\rm{SNuB}$$\,h^2$ (SNuB$
\equiv 10^{-12} \,\rm{SNe} \, L^{-1}_{\odot,B} yr^{-1}$), where the
error bars indicate $1\sigma$ confidence intervals, statistical and
systematic, respectively. The cluster rate normalized by the stellar
mass is $0.10^{+0.13}_{-0.06}\pm0.02$ in $\rm SNuM$$\,h^2$ (SNuM$
\equiv 10^{-12} \,\rm{SNe} \, M^{-1}_{\odot} yr^{-1}$).
Furthermore, we explore the optimal future survey for improving the core-collapse supernova rate measurements 
at $z\gtrsim2$ using gravitational telescopes, and for detections with multiply lensed images, and we
find that the planned WFIRST space mission has excellent prospects.} 
{Massive clusters can be used  as gravitational telescopes to significantly expand the survey range of supernova searches,
with important implications for the study of the high-$z$ transient universe. }


\maketitle

\section{Introduction}
Supernovae (\sne) have proved to be exceptionally useful for different astrophysical and cosmological applications. For example, Type Ia supernovae (\sneia) have been used as distance indicators  to show that the expansion rate of the Universe is accelerating \citep[see \eg][for a review]{2011ARNPS..61..251G},  while the rate of core-collapse \sne (CC~\sn)  can be used to trace the star formation history (SFH; see \eg \citealt[]{2008ApJ...681..462D, 2012ApJ...757...70D}).  Further, \sne contribute to the heavy elements in the Universe, so understanding the redshift dependence of \sn rates informs us about the chemical enrichment of galaxies over cosmic time. 

Measurements of SN rates at very high redshifts, $z \gtrsim 1$, are particularly difficult because of the limited light-gathering power of existing telescopes. This has been especially problematic for the study of CC~\sn rates, since 
they are on average intrinsically fainter than \sneia and often embedded in dusty environments 
\citep[see \eg][]{Mattila12}. Using ground-based facilities, only a few surveys have been
 able to measure CC~\sn rates beyond $z\gtrsim0.4$
 \citep{Graur11,Melinder}. An advance in redshift was provided by
 the Hubble Space Telescope ({\it HST}) by extending the CC~SN rates measurements up to
 $z\approx2.5$ \citep{2004ApJ...613..189D,2012ApJ...757...70D,Strolger15} .  This progress, however, was only made possible by the advent
 of very large {\it HST} multi-cycle treasury programmes: the Great Observatories Origins Deep Survey (GOODS), the Cosmic
 Assembly Near-Infrared Deep Extragalactic Legacy Survey (CANDELS) and the Cluster Lensing And Supernova survey with Hubble
 (CLASH).

The magnification power of galaxy clusters can be used as gravitational
 telescopes to enhance survey depth
 \citep{1988ApJ...332...75N,1988ApJ...335L...9K,1998MNRAS.296..763K,
   2000ApJ...532..679P,2003A&A...405..859G}. Galaxy clusters are the
 most massive gravitationally bound objects in the Universe,
 distorting and magnifying objects behind them. Gravitational lensing magnifies both the
 area and the flux of background objects, thereby increasing the depth of the survey; also the
 ability to find very high-$z$ \sne is enhanced. However, conservation of flux ensures that the source-plane area behind the cluster is shrunk owing to lensing, implying that the expected number of \sne does not necessarily increase in the
 field of view  \citep[see][for a
 description of possible optimizations]{2003A&A...405..859G}.

Even though \citet{1937ApJ....86..217Z} suggested
the use of gravitational telescopes nearly 80 years ago,  
it is only recently that systematic
\sn searches have been performed in background galaxies behind clusters. These investigations were initiated by feasibility studies by  \citet{2000MNRAS.319..549S} and \citet{2003A&A...405..859G}.

\citet{2002MNRAS.332...37G} searched {\it HST} archival images of
galaxy cluster fields for lensed \sne, finding one \sn candidate
at $z=0.985$. With its 524-orbit survey aimed at 25 galaxy
clusters, one of the main objectives of the  CLASH survey was to
find lensed \sne behind the clusters
\citep{2012ApJS..199...25P}. Three transients, out of which two were classified as secure \sneia,  were detected and used as direct tests of independently derived lensing magnification maps \citep{2014MNRAS.440.2742N,2014ApJ...786....9P}. The Frontier Fields
survey\footnote{www.stsci.edu/hst/campaigns/frontier-fields/} is an ongoing
effort devoting almost a thousand {\it HST} orbits and
targeting six lensing galaxy clusters. One of its most remarkable discoveries was a
\snia at $z=1.346$ behind Abell 2744 \citep{2015arXiv150506211R}.

Strong lensing also gives multiple images of the galaxies behind the cluster that host \sne. Even though the probability of observing such events is very low, multiple images of a strongly lensed \sn
from  Grism Lens-Amplified Survey from Space (GLASS; \citealt{2015ApJ...812..114T})
were detected. GLASS is a complementary {\it HST} spectroscopic survey targeting ten clusters, including those covered by Frontier Fields. The \sn was at $z=1.489$ (dubbed
'SN Refsdal') behind MACS J1149.6+2223, which re-appeared almost a
year later \citep{2015ApJ...811...70R,2015Sci...347.1123K,
  2015arXiv151104093G, 2015arXiv151204654K}.
 
The ground-based near-infrared (NIR) search for lensed \sne behind
galaxy clusters was pioneered by 
\citet{First} and \citet{Second} (hereafter G09). A search using the ISAAC instrument at the Very Large Telescope 
(VLT) was carried out targeting
Abell~1689, Abell~1835, and AC114, with observations separated by one
month. 
For Abell~1689, the gravitational telescope used in this work,
a total of six hours of observations in $SZ$ band (similar to $Y$ band)
were used, along with archival data used as a reference for image
subtractions. This survey resulted in the discovery of one reddened, highly magnified SN~IIP at $z=0.6$ with a high lensing magnification from archival data taken in 2003.

In addition to searching for transients in lensed galaxies, monitoring galaxy
clusters offers the opportunity to detect \sne that originate from 
cluster members. These are mostly thermonuclear \sne, since
clusters are dominated by early-type galaxies. Cluster \snia rates
have been proposed to help disentangle the proposed scenarios for
\snia progenitors  and
are essential in understanding several astrophysical processes such as
the iron abundance in intracluster medium \citep{2007ApJ...660.1165S,2008AJ....135.1343G,2008MNRAS.383.1121M, 2010ApJ...722.1879M,  2010ApJ...713.1026D,2012ApJ...745...32B}.

Here, we present the continuation of the effort of \citet{First} and \citet{Second} with HAWK-I on the VLT, which has greater sensitivity and a wider field of view than ISAAC.  This paper is organized as follows. In Sect.\,\ref{sec:surveys} our surveys are presented and in Sect.\,3 the transient search strategy and \sn candidates are described. In Sect.\,4 the volumetric \sn rates are calculated;  Sect.\,5  regards the connection between CC~\sn rates and SFH. Sect.\,6 concerns the rates in the galaxies with multiple images, while in Sect.\,7 the cluster \snia rates are calculated. In Sect.\,8 the possibilities of future surveys detecting \sne in the strongly lensed galaxies behind galaxy clusters are discussed. Also, given the lack of \sne at $z\gtrsim 2$, the requirements of finding very high-$z$ \sne are investigated. In Sect.\,9, summary and conclusions are drawn. Throughout the paper we assume the cosmology
$\Omega_{\Lambda}=0.7$, $\Omega_{\rm M}=0.3$ and $h=0.7$, unless stated otherwise. All the magnitudes are given in the Vega system.

\section{A1689 lensed supernova surveys}\label{sec:surveys}
Most of the \sn luminosity is emitted at optical and
UV rest-frame wavelengths, which is why, historically, optical bands have been
used for finding low $z$.  At high redshifts, \sne are more difficult to observe since they are fainter and
the light is shifted to longer wavelengths. This makes optical observations inefficient
in discovering \sneia at $z\gtrsim2$, even from space
\cite[\eg][]{2011ApJ...742L...7A}. From the ground, \sn surveys are difficult to carry out
in the NIR because of the bright and variable atmospheric
foreground. There are still somewhat stable parts of the NIR transmissive
windows of the atmosphere where the commonly used filters \Jband, \Hband and \Ksband are
centred. In order to maximize our sensitivity to high-$z$ \sne, we
conducted the survey described here with the 8~m Very Large Telescope (VLT) using one of the most efficient NIR cameras available, accompanied by a supporting optical programme at the 2.56~m Nordic Optical Telescope.

\subsection{Lens model and galaxy catalogues}
\label{sec:catalog}
\begin{figure*}[htbp]
	\begin{center}
		\includegraphics[width=6in]{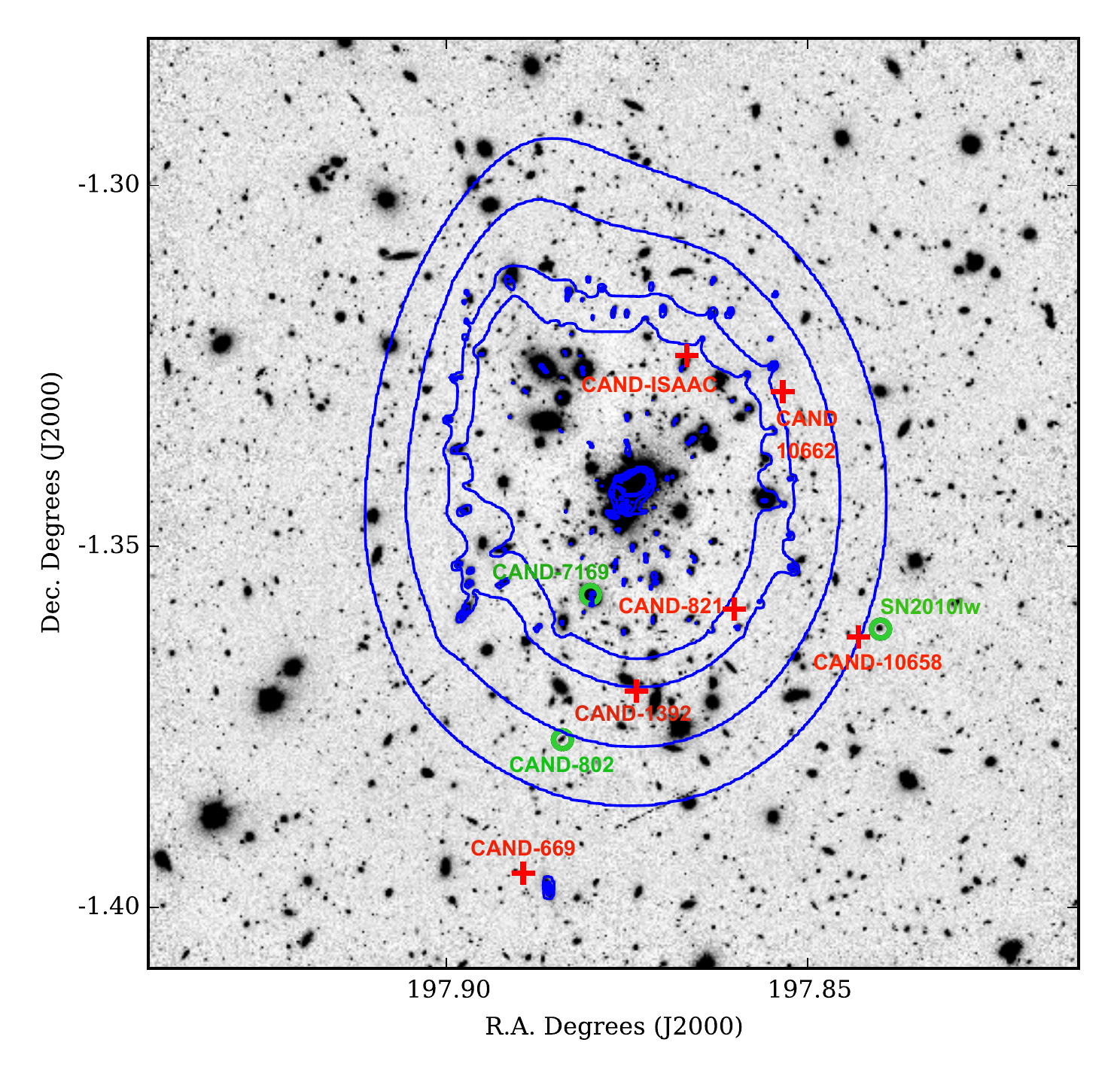}
		\caption{Near-infrared VLT/HAWK-I image of A1689 with critical lines for a source at $z_s$=2.0 plotted as contours for magnifications (from inside to outside) 2.0, 1.5, 1.0, and 0.75 magnitudes. The red crosses indicate the position of the core-collapse supernovae behind A1689 detected by our surveys, while the green circles denote those classified as Type Ia supernovae. }
		\label{lensing}
	\end{center}
\end{figure*}

We used the massive galaxy cluster Abell~1689 ($z=0.18$ or 670 Mpc) as the gravitational telescope. It is one of the best studied gravitational lenses. It has an extended Einstein radius of $\theta_E=47\arcsec.0\pm1.2$ for a source at $z_s = 2$,  which makes it very well suited for the task of finding high $z$ \sne \citep{2005ApJ...621...53B, 2007ApJ...668..643L,2010ApJ...723.1678C,2015ApJ...806..207U}.
The mass distribution profile was modeled using  strong lensing features detected with deep {\it HST} observations and ground-based spectroscopy.  
We used magnification maps (see Figure~\ref{lensing}) produced with the public software LENSTOOL\footnote{LENSTOOL is developed at the Laboratoire
	d'Astrophysique de Marseille, projets.lam.fr/projects/lenstool}
\citep{2007NJPh....9..447J} and the mass profile described in \citet{2007ApJ...668..643L}.  The mass distribution model predicts magnifications in the field of view  to an accuracy of $~10-15\%$. The magnification depends on
the distance to the lens (the galaxy cluster), $d_{LS}$, and the distance to the background
source, $d_{S}$. The regions of high magnification scale approximately as the Einsten radius $\theta _{E} \propto \sqrt{d_{LS}/d_S} $.

A  photometric and spectroscopic catalogue of the sources in the field of view of A1689 was previously compiled and presented in G09. 
The catalogue contains photometric redshifts calculated from archival optical and NIR {\it HST} data. In addition to this, we added spectroscopic redshifts derived from data taken with the VLT/FORS spectrograph (ESO programme IDs 65.O-0566 and 67.A-0095). An active galactic nuclei (AGN) catalogue is also included from \citet{2007ApJ...664..761M}, which is based on a spectroscopic survey of Chandra X-ray point sources.

\subsection{Near-infrared VLT/HAWK-I observations}
The NIR data were obtained with the High Acuity Wide field K-band imager  (HAWK-I; \citealt{2004SPIE.5492.1763P,2006SPIE.6269E..0WC}) mounted on the VLT (Programmes ID 082.A-0431, 0.83.A-0398, 090.A-0492, 091.A-0108, P.I. Goobar). The HAWK-I has an array of four $2048\times2048$ HgCdTe detectors covering a total area of $7.5' \times 7.5'$ with a sampling of  $0.106''/\rm {pix}$ per pixel. The chips are separated by a $15''$ gap. 

The search was carried out in the \Jband band, $1.17 - 1.34$  $\mu m$, with a pointing such that the galaxy cluster was placed in the centre of the field of view.   The observations were executed in blocks of 29~exposures made of six 20 second integrations.  Between each exposure, the telescope was offset in a semi-random manner. This allowed us to perform accurate sky subtraction and to fill in the gaps between the four arrays. The strategy was to execute two such blocks for each epoch, but when the observing constraints of $\leq0.8\arcsec$ and photometric conditions could not be obtained for the full two~hours during the same night, the observations for one epoch were spread over multiple nights.  Nightly standard stars were also observed as part of the  ESO standard calibration programme.

\begin{figure*}[htbp]
 	\begin{center}
 		\includegraphics[width=\hsize]{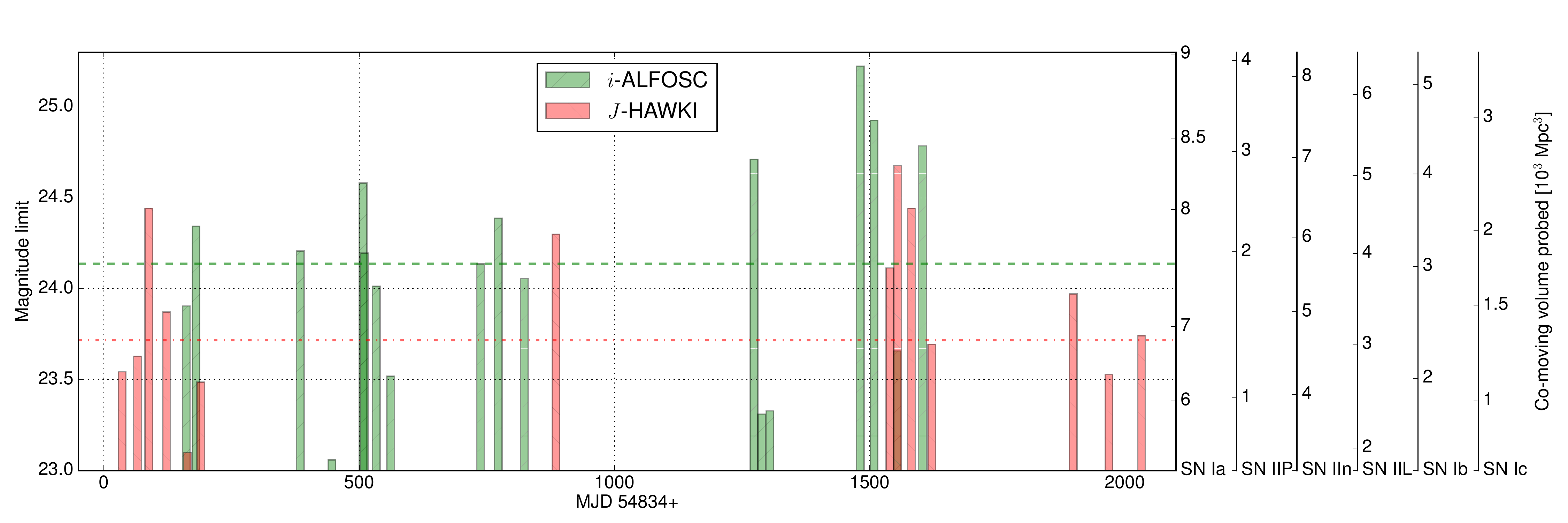}
 		\caption{Magnitude limit for the HAWK-I and ALFOSC observations as presented in Tables~\ref{HAWKI_data}~and~\ref{ALFOSC_data}. The red (green) dotted line indicates the average magnitude limit for the HAWK-I (ALFOSC) survey. On the right y-axis the probed co-moving volume in the \Jband band is shown for each supernova type.  }
 		\label{magnitude_limit}
 	\end{center}
 \end{figure*}

The \Jband band data of A1689 were taken over 29 separate nights, starting on December 2008 and ending on July 2014, with an average seeing of $\sim$$0.6\arcsec$ (see Table~\ref{HAWKI_data} and Figure~\ref{magnitude_limit}). The raw images were retrieved, immediately reduced and searched for transients as explained below.  
Additionally, a few epochs in \Ksband band and the narrow {\it NB1060} band were obtained by other programmes (ID: 085.A-0909, PI: D. Watson and ID: 181.A-0485, PI: J. G. Cuby) that observed the cluster during the same period.

\subsection{Near-IR data reduction}\label{sec:reduction}
The reduction of the VLT/HAWK-I data was carried out  using a
pipeline that includes procedures both from ESO and written by our group. The individual frames of
each epoch were dark-subtracted and flat-fielded with the procedures
from ESO Common Pipeline Library. For each frame a bad pixel mask consisting of saturated and cosmic ray affected pixels was produced.

To account for the contamination from the bright and variable
atmosphere, standard NIR sky-subtraction was performed.  For each
exposure, the sky level in each pixel was estimated as the running
median using the seven preceding and seven succeeding exposures after
rejecting the three highest and lowest pixel values in the stack.  In
addition to these, we used an object mask to exclude any pixel
that contained light from stars and galaxies.  Contrary to
the standard two-step NIR image reduction, the object mask was not
created using the images obtained during the same night.  Instead, we
used the full set of images from the programme to create one deep
image stack and built the object mask using SExtractor
\citep{1996A&AS..117..393B}.  This deep stack was continuously updated
during the programme as new data were obtained.

We performed aperture photometry of isolated stars in the field for
each night that also had standard star observations.  From this we
created a catalogue of the average stellar magnitudes. After excluding
outliers, it was used as a tertiary standard star catalogue, as
described below.

The individual frames were corrected for geometrical distortions
by applying the distortion map provided by ESO. 
The distortion-corrected and
sky-subtracted frames were then geometrically aligned and combined using SExtractor, SCAMP \citep{2006ASPC..351..112B}, and Swarp \citep{2002ASPC..281..228B}.
The frames from the four arrays were combined into a mosaic image (dubbed \textit{new} image). For the transient search purpose (presented in Sect.\,\ref{sec:transient}, we split the list of frames into two parts and combined them for two more search images (\textit{new1} and \textit{new2}). The mosaic images allowed us to search for transients between the chips, since large enough dithering steps were used to fill the gaps. However, the combined images were slightly shallower in these areas. Given that the core of the cluster was placed in the centre, many of the strongly lensed galaxies were located in these regions. It was a compromise to have the same pointing as archival data of the field. 

To perform image differencing we used the software described in \citet{Sebastian,2006A&A...447...31A, 2008A&A...486..375A}, which is based on the image subtraction algorithm from \citet{Alard1} and \citet{Alardi2}.
 As a reference image (\textit{ref}) in the subtraction process, we used either a single epoch or a deeper stack of several epochs that were significantly separated from the search epoch. The reference image was subtracted from \textit{new}, \textit{new1}, and \textit{new2} to create \textit{sub}, \textit{sub1}, and \textit{sub2}, respectively.

\begin{table}
	\begin{center}
		\caption{A1689 observations with VLT/HAWK-I in the \Jband band \label{HAWKI_data}}
		\begin{tabular}{lcccc}
				\hline
				\hline
			Date & Exp. time & Seeing   & Det. eff. & 
			$m_{lim}$  \\
			&  &  &  90\% & average \\
			& (sec) & (arcsec) &  (mag) & (mag) \\
			\midrule
			2008-12-30 & 3420&	0.69 &  \\
			2008-12-31 & 2400&  0.57 &  &  \\			
			2009-01-02 & 2400& 0.65 &  23.05 & 22.97 \\
			\midrule
			2009-01-30 & 7200& 0.61 & 23.63 & 23.54\\
						\midrule
			2009-03-01 & 4800& 0.53 & 23.85 & 23.67\\
						\midrule
			2009-03-23 & 4800& 0.62 & 24.15 & 24.44\\
			2009-03-25 & 4800& 0.38 & \\
						\midrule
			2009-04-27 & 7200& 0.56 & 23.87 & 23.87\\
						\midrule
			2009-06-07 & 2400& 0.66 & 23.28 & 23.10\\
										\midrule
			2009-07-03 & 2400& 0.61 & \\
			2009-07-05 & 2400& 0.63 & 23.53 & 23.49 \\
			2009-07-07 & 2400& 0.73 & \\
						\midrule
			2011-05-29 & 2400& 0.87 & Used as & 24.30\\
			2011-05-30 & 7200& 0.44 &  a reference \\
										\midrule

			2013-03-14 & 6960& 0.61 & 23.92 & 24.11\\
						\midrule
			2013-03-29 & 3480& 0.44 & \\
			2013-03-31 & 3480& 0.57 & 24.26 & 24.44 \\
			2013-04-02 & 3480& 0.36 & \\
					\midrule
			2013-04-25 & 3480& 0.61 & \\
			2013-05-06 & 3480& 0.44 & 24.19 & 24.69 \\
			2013-05-07 & 3480& 0.44 & \\
								\midrule
			2013-06-04 & 3480& 0.64 & 23.48 & 23.97\\
			2013-06-05 & 3480& 0.61& \\
								\midrule

			2014-03-08 & 3480& 0.48& 23.95 & 23.62\\
			2014-03-15 & 2160& 0.59& \\
						\midrule
			2014-05-17 & 3480& 0.82	& 23.52 & 23.28 \\
			2014-05-19 & 3480& 	0.91& \\
					\midrule
			2014-06-19 & 3480&	0.78	& 22.55 & 22.77\\
					\midrule
			2014-07-20 & 3480&	0.37	& 23.94 & 23.74\\
			
		\hline
		\multicolumn{5}{c}{{\bf Notes.} The line indicates that a stack of those observations} \\ \multicolumn{1}{c}{was made. }\\
		\end{tabular}

	\end{center}
\end{table}

\subsection{Optical NOT/ALFOSC survey}

Adjacent to the NIR survey, we performed a complementary optical
survey at the Nordic Optical Telescope (NOT; programmes P39-011,
P46-008, P47-014, PI: A. Goobar).  The NOT is a 2.56~m telescope at the
Observatorio del Roque de los Muchachos at La Palma in the Canary
Islands.  We used the Andalucia Faint Object Spectrograph and Camera
(ALFOSC), a $2048 \times 2048$ pixel CCD with a field of view of $6.4'
\times 6.4'$ and pixel scale of $0.19''/\rm{pix}$. We monitored A1689
in the $i$ band and searched for transients. We obtained
additional data in the $g$ and $r$ bands. The images were centred on the galaxy cluster and a 9 point dithering pattern with the integration times listed in Table~\ref{ALFOSC_data} was used.

The reduction of the optical images was carried out following standard
IRAF\footnote{IRAF: The Image Reduction and Analysis Facility is distributed by the National Optical Astronomy Observatory, which is operated by the Association of Universities for Research in Astronomy (AURA) under cooperative agreement with the National Science Foundation (NSF).} procedures. The transient search was carried out using the method described above
for the NIR data. The catalogue of tertiary standards with NIR
photometry mentioned above was extended with the corresponding $gri$ magnitudes measured by the Sloan Digital Sky Survey.

\begin{table}
	\begin{center}
		\caption{A1689 data from NOT/ALFOSC in the $i$ band searched for transients}
		\label{ALFOSC_data}
		\begin{tabular}{lcccc}
			\hline
			\hline
			Date  &  Exp. time & Seeing & Det. eff. & $m_{lim}$  \\
			  &    &   &  90\% &  \\
			    &   (sec) &  (arcsec) &  (mag) & (mag) \\
			\hline


			2009-06-05&	3600 & 0.56 & 23.41 & 23.87	\\
			2009-06-24&	3600 & 0.69 & 23.64 & 24.3 \\
			2010-01-15&	3600 & 0.94 & 23.22 & 24.07 \\
			2010-03-17&	3600 & 1.04 & 22.38 & 23.02 \\
			2010-05-17&	3600 & 0.69 & 23.86 & 24.54	\\
			2010-05-21&	3600 & 0.88 & 23.16 & 24.16 \\
			2010-06-12&	3600 & 1.05 & 22.95 & 23.97 \\
			2010-07-11&	3600 & 1.01 & 22.71 & 23.48 \\
			2011-01-02&	3600 & 0.72 & 23.61 & 24.10 \\
			2011-02-06&	3600 & 0.68 & Ref  & \\
			2011-03-29&	3600 & 1.07 & 23.59 & 24.01 \\

			2012-05-14&	3600 & 1.37 & 22.34 & 22.80	\\
			2012-06-21&	3600 & 0.81 & 24.14 & 24.67 \\ 
			2012-07-06&	3600 & 0.79 & 22.41 & 23.27 \\
			2012-07-22&	3600 & 0.83 & 22.64 & 23.29 \\
			2013-01-15& 3600 & 0.72 & 23.76 & 25.18 \\
			2013-02-11& 3600 & 0.90 & 23.71 & 24.89 \\
			2013-03-28&	3600 & 0.99 & 22.14 & 23.62 \\
			2013-05-17&	3600 & 0.80 & 23.46 & 24.75\\
			\hline
		\end{tabular}

	\end{center}
\end{table}

\section{Transient search}
\label{sec:transient}
After the image differencing, we ran an automated \sn candidate detection algorithm on the subtracted images. Our search criteria were a S/N $\ge5$ detection in the \textit{sub} image and for the candidate to
be present in both \textit{sub1} and \textit{sub2} with S/N $\ge4$. 
As a first step, all candidates were methodically scanned by eye and ranked.  The information available for each candidate includes S/N (for stacks and sub-stacks), distance to the nearest galaxy and the relative increase in brightness. Candidates arising from image subtraction artifacts at the cores of bright galaxies were revealed by low increase in brightness and rejected. Point source-like candidates
with high S/N and a small angular separation from the core of a galaxy
could come from AGN and were saved for further
investigation.

In order to identify possible \sne, a careful examination of the remaining candidates was needed. This included: 
\begin{itemize}
	\item   The image and subtraction stamps  from all our previous NIR and optical data were inspected. Light curves from aperture photometry on the subtractions of the archival data were built for the saved candidates and checked for previous activity.
	\item  High resolution {\it HST}/ACS images of the cluster were examined. The {\it HST}/ACS images have higher spatial resolution and S/N (in principle diffraction limited) compared to ground images. For this reason, faint background galaxies are often only visible in the {\it HST}/ACS images in which the transient could arise. In the case when the photometric/spectroscopic redshift of the presumable host galaxy was known, it was used to estimate the absolute magnitude of the transient. The lensing magnification was also taken into account.  
	
\end{itemize} 

After the survey was completed we repeated the transient search with
improved reference images for our \Jband-band data. Observations that
were obtained less than 14 days apart were co-added to obtain better
image depth. We were left with 15 combined images in total, and as
reference for image subtraction, we used a deep image from 2011. Since there is a gap of
observations between late 2009 and 2013, the reference image should not
contain light from any SNe present in the earlier epochs. We limited
the search area to the region with lensing magnification, $\mu \ge 2$,
for an object at $z_s=2$. We required at least two consecutive
detections and lowered the detection threshold to S/N$>4$. The initial number of
candidates was $\sim$$10000$ (number of objects detected by the
detection algorithm), of which with $\displaystyle\sim$$2/3$ were in the
magnified region. We required less than five pixel separation to be considered the same candidate appearing in the different
subtraction epochs. Given that our data span over five years, we rejected candidates that appeared in all the epochs, since we do not expect
a significant fraction of \sne to be visible for such a long time in the band and redshifts considered here. In that
way we discarded repetitive subtraction residuals. After the candidates ordering
by their occurrence and the cuts applying, 184 candidates remained. Most of the candidates were subtraction residuals, arising from several reasons such as
imperfect alignment between the \textit{new} and the \textit{ref}, or
imperfect matching of the PSF of these two images. To be able to find
real \sn candidates, members of our team (TP, RA, and AG) visually
inspected each of these remaining 184 candidates as described previously for the on-the-fly search. The candidates were ranked independently and the results were compared afterwards. Spurious detections appearing as candidates because of
repetitive residuals on the same locations were rejected by
all the inspectors. After the visual inspection, there were ten candidates that were rated highly. To exclude AGNs, we check the following: $i$) known AGN at
the position of this transient and the galaxy, $ii$) galaxy activity in the previous epochs, and $iii$) that the transient is not
located in the centre of the
galaxy and cross-match the position to the X-ray point source catalogue \footnote{http://cxc.cfa.harvard.edu/csc/}.  Two of the candidates were classified as AGNs.

\subsection{Results}\label{sec:typing}
The transient search yielded a detection of eight \sn candidates, six behind A1689, and two associated with cluster members, which are presented in Sect.\,\ref{sec:snebehind} and \ref{sec:sneinA1689} and summarized in Table~\ref{tab:SNe_found}.
The transients were photometrically typed with a prior of the host galaxy redshift (spectroscopic or photometric).  The only exception was
 CAND-1208 (\sn~2010lw) for which a spectrum was obtained.
\begin{table*}[t]
	\caption{Supernovae found in our VLT/HAWK-I+NOT/ALFOSC+VLT/ISAAC surveys.}
	\begin{center}
		\begin{tabular}{lcccccc}
			\hline \hline
			SN ID &RA & DEC & Type  & $z$ & Redshift info & Magnification \\ \hline 
			CAND-1208 & 197.83987&  -1.36145& Ia  & 0.189 & SN spectrum & -  \\ 
			CAND-802 & 197.88393& -1.37678&  Ia  & 0.214 & Host spectrum& - \\
			CAND-7169 &197.88005 & -1.35668 & Ia &  0.196 & Host spectrum & -\\ 

			 \hline 
			
			CAND-ISAAC & 197.86670& -1.32360 & IIP  & 0.637 & Host spectrum& $1.4\pm0.07$ \\
 			CAND-669 & 197.88937& -1.39524& IIL  & 0.671 & Host spectrum &$0.31\pm0.04$ \\ 
			CAND-821 & 197.86011& -1.35867& IIn   & 1.703 & Host spectrum &$1.58\pm0.07$\\ 
			CAND-1392 & 197.87369&-1.37000& IIP   & 0.944 &  Host spectrum& $1.09\pm0.07$\\
			CAND-10658 & 197.84296&-1.36251& IIn & $z=0.94^{+0.07}_{-0.27}$ & Host photo-$z$ & $0.58\pm0.06$ \\
			CAND-10662 & 197.85341& -1.32859& IIP  & $z=1.03^{+0.20}_{-0.17}$ & Host photo-$z$& $1.04\pm0.07$\\
			\hline
		\end{tabular}
		
	\end{center}
	\label{tab:SNe_found}
\end{table*}

The brightness of each transient was measured using PSF photometry, where the flux for each epoch was fitted simultaneously together with a host-galaxy model and the transient position as described in, for example, \citet{2008A&A...486..375A}, Sect\,3.1.  The transient fluxes were then calibrated to the tertiary standards mentioned in Sect.\,\ref{sec:reduction}.

We matched the coordinates of the candidate
with our galaxy catalogues to determine the redshift of the SN host galaxies. When the redshift of the candidate host galaxy was not available, we used deep multi-band images from our survey and the template-fitting \textit{hyperz} code \citep{2000A&A...363..476B} to obtain the photometric redshift. 

For the \sn typing we used the light curve templates used in G09 that are also listed in Table~\ref{tab:peak_v}.  The templates were tested against the observed data for a grid of redshifts around the host $z$ and its $1\sigma$ confidence limits. The synthetic light curves in the observer NIR filters for redshift $z$ were obtained by applying cross-filter $k$-corrections, distance modulus, and time dilation corrections. There is also the possibility to have different reddening parameters ($E(B-V)$ and $R_V$). When we estimated the absolute \Vband band magnitude $M_V$, we took the lensing magnification from the galaxy cluster into consideration.

\begin{figure*}[htbp]
	\begin{center}
		\includegraphics[width=\hsize]{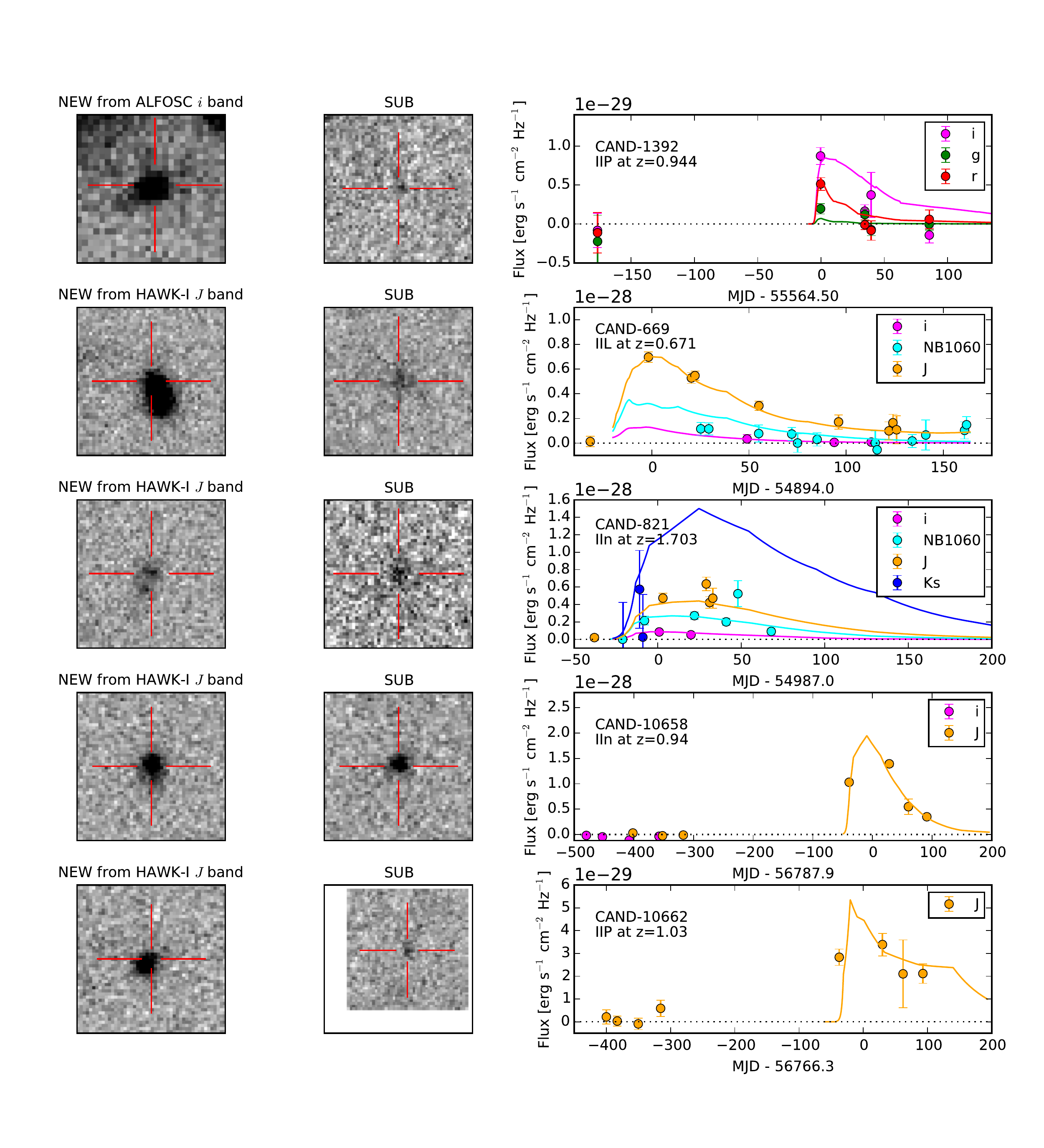}
		\caption{High-$z$ core-collapse supernovae behind A1689. The first two columns show $5\arcsec\times 5\arcsec$ image and subtraction stamps from epoch closest to maximum brightness. The location of the supernova candidates are indicated by the red markers. The third column contains the light curves of the supernova presented in the order as in the text: CAND-1392, CAND-669, CAND-821, CAND-10658 and CAND-10662.  The solid lines represent the best-fitted SN template which are IIP, IIL IIn, IIn and IIP for the five \sne, respectively. 
		}
		\label{lc_patches}
	\end{center}
\end{figure*}

The typing is not solely based on the best all-band $\chi^2$ fit, but on the best match to the fit parameters: the host redshift, the \sn peak absolute magnitude, and the duration of the transient.  The stamps, the multi-band photometry, and the best-fitted \sn type light curves are shown in Figures~\ref{lc_patches}~and~\ref{lc_patches_cluster} for the CC and \sneia, respectively.

\begin{table}[t]
	\caption{Supernova properties.}
	\begin{center}
		\begin{tabular}{l c c c}
			\hline \hline
			SN Type & $M_V \ (\rm mag)$ & $\sigma_{M_V} \ (\rm mag)$ & $f_{\rm CC}$ \\ \hline 
			Ia  & -19.30 & 0.30 &  \  \\  \hline 
			IIP & $-16.90\pm0.37$ & 0.97 & 0.524 \\ 
			IIL & $-17.98\pm0.34$ & 0.90 & 0.073 \\ 

			IIn & $-18.62\pm0.32$ & 1.48 & 0.064  \\ 
			Ib & $-17.54\pm0.33$ & 0.94 & 0.069  \\ 
			Ic & $-16.67\pm0.40$ & 1.04 & 0.176  \\ 
			faint CC SNe & $<15$ &  &   0.094 \\ 
			\hline 
			\multicolumn{4}{c}{{\bf Notes.} Second and third column indicate the peak \Vband band }\\
			\multicolumn{4}{l}{brightness $M_V$ and its one-standard-deviation $\sigma_{M_V}$, respecti-} \\ 
			\multicolumn{4}{l}{vely. The third column stems for the fractions of core-colla-} \\
			\multicolumn{4}{l}{pse supernova subtypes. Values adopted from} \\
			\multicolumn{4}{l}{  \citet{2014AJ....147..118R} and \citet{Li}.} \\
		\end{tabular}
		
	\end{center}
	\label{tab:peak_v}
\end{table}

\subsection{Core-collapse supernovae behind A1689}\label{sec:snebehind}
\textit{CAND-01392:} The transient was discovered on 2011-01-02 in the three optical ALFOSC bands, but no NIR epochs were obtained while it was active. The most probable host for this transient is a galaxy that is centred $0\farcs32$ from the transient. The  spectroscopic redshift of the host galaxy is $z=0.944$.
At this redshift, the candidate is at a projected distance of $\sim$$3$~kpc from the host galaxy centre and the estimated magnification from the galaxy cluster is $1.09\pm0.07$ mag. The transient is most likely a CC~\sn, where the best fit
is a Type~IIP~\sn with an absolute magnitude $M_V=-18.2\pm0.13$, which differs by $1.3\sigma$ from the mean peak brightness of this class. The light curve is also consistent as the fading part of a SN~IIn.

\textit{CAND-0669:} The transient was found in the \Jband band on 2009-03-01, and then observed later in the {\it NB1060} and $i$ bands. The core of the most probable host is located $0\farcs49$ from the transient. The spectroscopic redshift of the galaxy is $z=0.671$, which means that the projected separation corresponds to $\sim$$3.5$~kpc. The magnification from the galaxy cluster is $0.31\pm0.04$ mag. The best fit was obtained for the bright \sn~IIL with an absolute magnitude $M_V=-18.87\pm0.07$. It is also possible, given the absolute magnitude,  that the transient is a \sn~IIn. The absolute magnitude is also consistent with a \snia, however, taking into account the brightness in the {\it NB1060} band and the lack of the second maximum in the observed {\it J} band, which corresponds to the rest-frame {\it I} band, a thermonuclear \sn can be excluded \citep[see~e.g.][]{2005A&A...437..789N}.

\textit{CAND-0821:} Our most distant transient was detected on 2009-06-05 in the  $i$ band and confirmed a few days later in the \Jband band. The \sn has already been published in \citet{2011ApJ...742L...7A}. An accurate galaxy host redshift $z=1.703$ was determined from a VLT/X-SHOOTER spectrum taken almost a year later, making CAND-821 one of the most distant CC~SNe ever discovered. The \sn is located $\sim7$~kpc from the host galaxy centre. The \sn has a significant lensing magnification of $1.58\pm0.07$ mag.  The best fit for this transient is a SN~IIn with $M_V=-19.56\pm0.06$.  The detection in the rest-frame UV (observers $i$ band) excludes a thermonuclear SN. Further details are presented in \citet{2011ApJ...742L...7A}.

\textit{CAND-10658:} This \sn candidate was found in the NIR epochs from 2014 in the post-survey search. The transient is located $0\farcs38$ from the core of a galaxy for which we estimated the photometric redshift to be $z=0.94^{+0.07}_{-0.27}$. At this redshift, projected host galaxy separation corresponds to $\sim3$~kpc. The lensing magnification from the galaxy cluster is $0.58\pm0.06$ mag for the given redshift.
The light curve is consistent with a \sn~IIn with an absolute \Vband band magnitude of $M_{V} = -22.00\pm0.15$. \sne~IIn with this intrinsic brightness have been observed \citep{2009ApJ...695.1334S,2011ApJ...729..143C}, and recent discoveries confirm that the SNe~IIn class consists of objects that show the largest dispersion in peak magnitudes up to $M_R=-22.3$ \citep{2014ApJ...786...67A,2013A&A...555A..10T,2014AJ....147..118R}. The absolute magnitudes are usually reported in the $r/R$ band since most SNe~IIn have the best coverage in the red bands. The rapid colour evolution of SNe~IIn makes the conversion to \Vband band less straightforward \citep{2013A&A...555A..10T}. However, studies on individual \sne~IIn suggest a (V-R) colour index of $0.05\pm0.11$ at post-max epochs and  $0.46\pm0.05$ around maximum \citep{2000MNRAS.318.1093F,2002ApJ...573..144D}.  In conclusion, the absolute magnitude of the candidate is compatible with \sn~IIn, thus the most probable classification for the transient is a SN~IIn in the superluminous domain.

\textit{CAND-10662:} As CAND-10658, this transient was found in the NIR epochs from 2014 in the post-survey search. The S/N~$<5$ so it would not have been detected on the fly.  The most likely host galaxy has a photometric redshift of $z=1.03^{+0.20}_{-0.17}$ and is located $0\farcs27$ from the transient.  With the given redshift the magnification from A1689 at this position is $1.04\pm0.07$ mag and the projected separation corresponds to $\sim$$2.2$~kpc. The best light curve fit was obtained for a SN~IIP with $M_V=-17.46\pm0.21$.

\textit{CAND-ISAAC:} This transient was detected in the ISAAC pilot survey and typed as a SN~IIP at a photometric redshift of $z=0.59\pm0.05$ in G09. A spectrum of the host galaxy was published later \citep{2012ApJ...754...17F} placing it at $z=0.637$, which is consistent with the initial estimate. 
 
\subsection{Type Ia supernovae in and behind A1689}
\label{sec:sneinA1689}
\begin{figure*}[htbp]
	\begin{center}
		\includegraphics[width=6.5in]{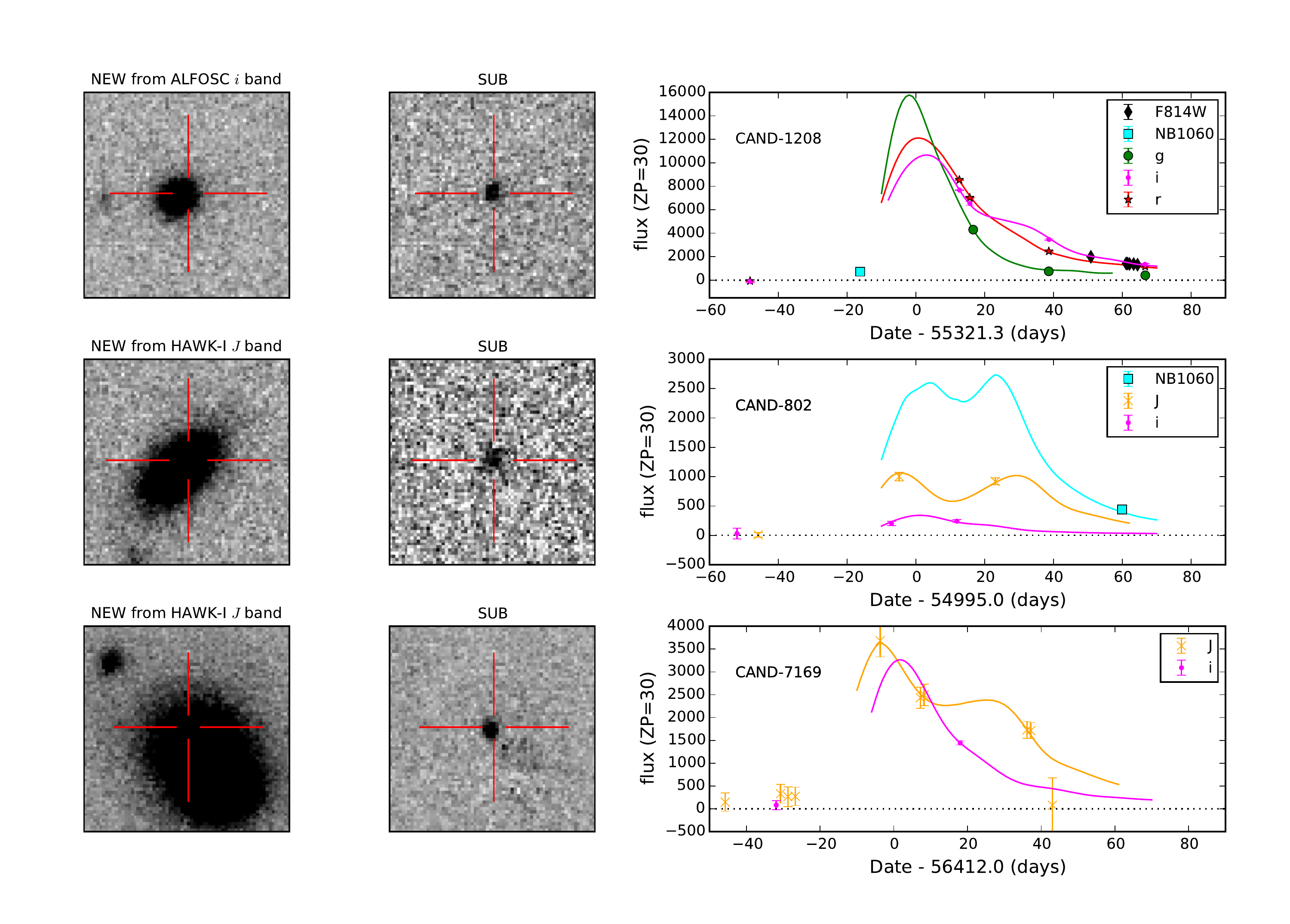}
		\caption{Supernovae Type Ia found in the surveys. The first two columns show 7$\arcsec \times 7\arcsec$ stamps from the epochs (image and subtraction image) closest to maximum brightness of the transient.
			 The location of the supernova candidate is indicated by the red markers. The third column contains the light curves of the \sne  presented in the order as in the text: CAND-1208, CAND-802 and CAND-7169. }
		\label{lc_patches_cluster}
	\end{center}
\end{figure*}

Three \sneia were found in the survey, of which two were found be associated with host galaxies that belong to A1689.  The host galaxies are shown in the first column of Figure~\ref{lc_patches_cluster}.   The observed light curves were used to classify the SNe using the method described in~\ref{sec:typing}.  Given this, we further use the SNooPy package \citep{2011AJ....141...19B} to fit light curve parameters, such as the date of {\it B} band maximum, $t_\mathrm{max}$, the brightness decline between peak and day $+15$ of the rest-frame \Bband-band light curve, $\Delta m_{15}$, the host galaxy extinction , $E(B-V)_\mathrm{host}$,
and the distance modulus, $\mu$. This process involves extrapolation from the multi-band observed photometry.  In order to fit $\mu$, SNooPy takes advantage of the fact that normal \sneia are a very homogeneous class of objects with a narrow distribution of absolute magnitudes.
The SNooPy package includes optical to NIR \snia light curve templates based on a \snia sample obtained by the Carnegie Supernova Project. The spectral energy density template from \citet{2007ApJ...663.1187H} is used for calculating cross-filter {\it k} corrections.

The best-fit light curve parameters for the three \sneia are presented in Table~\ref{tb:sneinA1689}, and the corresponding light curves are shown in Figure~\ref{lc_patches_cluster} together with the measured data.  In the last column of the table, we present the distance moduli calculated from the redshifts, assuming a flat Universe with $H_0=72\,$km/s/Mpc and $\Omega_{\rm M}=0.28$, which are the values used for the calibration of SNooPy. Details of the individual transients are given below.

\begin{table*}[htb]
\centering
\caption{Fitted light curve parameters obtained with SNooPy \label{tb:sneinA1689}. 
		The redshifts were adopted from Table~\ref{tab:SNe_found}, and the distance moduli in the last column 
		were calculated from the redshift assuming a flat Universe with $H_0=72\,$km/s/Mpc and $\Omega_M=0.28$.}
\begin{tabular}{r@{}l l r r l r r}
\hline\hline
\multicolumn{2}{c}{\sn} & \multicolumn{1}{c}{$z$}  & \multicolumn{1}{c}{$t_\mathrm{max}$} & 
  \multicolumn{1}{c}{$\Delta m_{15}$} & \multicolumn{1}{c}{$E(B-V)_\mathrm{host}$} & \multicolumn{1}{c}{$\mu$} &
  \multicolumn{1}{c}{$\mu_\mathrm{cosmo}$}\\
 & & &\multicolumn{1}{c}{(days)} & \multicolumn{1}{c}{(mag)} & \multicolumn{1}{c}{(mag)} &
   \multicolumn{1}{c}{(mag)} & \multicolumn{1}{c}{(mag)}\\
\hline
$1208$&$^{a}$ & $0.189$ & $55321.3(0.4)$ & $0.67(0.02)$ & $0.06(0.02)$ & $39.34(0.03)$ & 39.76\\
$802$&         & $0.214$ & $54995.0(4.5)$ & $1.57(0.27)$ & $2.12(0.80)$ & $40.97(0.42)$ & 40.06\\
$7169$&       & $0.197$ & $56412.0(1.6)$ & $1.67(0.01)$ & $0.39(0.07)$ & $39.66(0.05)$ & 39.86\\
\hline
\multicolumn{8}{l}{$^{a}$ SN~2010lw} \\
\end{tabular}
\end{table*}

\begin{figure}[htbp]
	\begin{center}
		\includegraphics[width=\hsize]{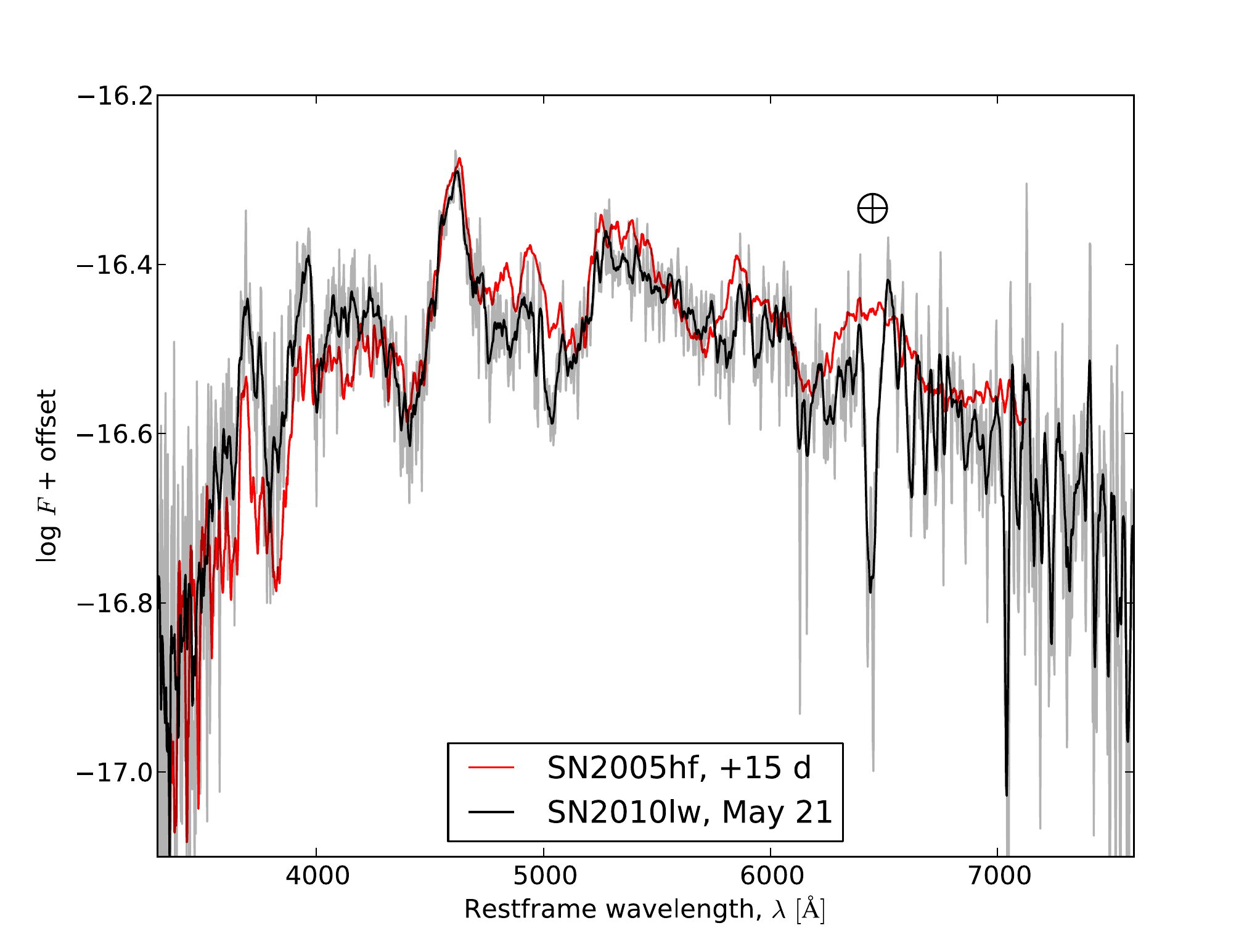}
		\caption{Spectrum of SN~2010lw (CAND-01208) obtained with ALFOSC at the NOT (black line). The Earth symbol indicates a strong telluric feature that is not corrected for. The red line shows a spectrum of the normal \snia~2005hf at 15 days past maximum brightness.}
		\label{SN2010lw_spec}
	\end{center}
\end{figure}

\textit{CAND-01208 (SN~2010lw):} The transient was discovered with the NOT on 2010 May 17.944 UT in the $i$ band and also observed in the $g$ and $r$ bands \citep{2011CBET.2642....1A}.  It was located  $0\farcs77$ west and $0\farcs35$ north of its host galaxy, which is a member of A1689 at redshift $z=0.189$. 
A spectrum of the candidate was obtained with NOT/ALFOSC on 2010-05-21.9 UT, using Grism 4 and a $1.3\arcsec$ slit. The spectrum was reduced
using standard IRAF routines. Using SNID \citep{2007ApJ...666.1024B}, we found several good matches to normal \sneia at +16 ($\pm8$) days after maximum brightness (see Figure~\ref{SN2010lw_spec} for a comparison between SN~2010lw and the normal \snia~2005hf). SNID also provides good matches to over-luminous 91T-like SNe and the peculiar SN 2007if \citep{2010ApJ...713.1073S}, which are also consistent with the parameters derived from the light curve fits.

\textit{CAND-0802:} The transient was discovered on 2009-06-07 in the \Jband band and later observed in the $i$ band and the {\it NB1060} band. The SN is located $4\farcs1$ from the nucleus of the host galaxy at a spectroscopic redshift of $z=0.214$ \citep{2007ApJ...664..761M}. Even though the galaxy is located close to the cluster core region in
projection space, most likely this galaxy does not belong to A1689. A1689 member galaxies have a relatively wide internal velocity spread and they extend to large radii  \citep{2009ApJ...701.1336L}. Moreover, there is evidence that there are two groups of galaxies merging \citep{2004ogci.conf..183C}. Velocity offsets from the mean cluster redshift are defined as $ v = c(z_{galaxy} - z_{cluster})/(1 + z_{cluster})$ \citep{1979ApJ...232...18H}. The redshift difference for this host galaxy is significant and it would require a velocity offset of $6300$~km~s$^{-1}$. \citet{2009ApJ...701.1336L} find maximum amplitude of $\mid$$4000$$\mid$ km~s$^{-1}$, which places this \sn host behind the cluster and not in it. 

 The best-fit light curve parameters suggest a significant host reddening, which is rarely found in transient searches at bluer wavelengths, and a steep decline rate. The low-resolution spectrum of the host galaxy shows strong emission lines that indicate ongoing star formation. The presence of dust is usual for these environments. Given the high reddening, the best-fitted distance modulus could be slightly biased by the assumption of the total-to-selective extinction, but despite this we note that the distance modulus is within $\sim2\sigma$ of the expected value based on the redshift.

\textit{CAND-7169:} The transient was discovered on 2013-05-06 in the \Jband band and later observed in the $i$ band. The host is an elliptical galaxy with spectroscopic redshift  $z = 0.1958$. The SN was located $1\farcs42$  from the core of its host galaxy.  The SNooPy fit results suggests a rapidly declining, reddened SN. 

\subsection{Discussion}
We compare our results to the expected number of events and their redshift distribution using existing rate models applied to our survey.  

We estimated the expected number of CC~\sne by following the same procedure as in G09.  We used CC~\sn rate as extrapolated in G09 and based on the results in \citet{2007MNRAS.377.1229M}.  The control times are calculated in the same way as for the volumetric rates estimation presented in the next sections. We assumed moderate overall reddening in the host galaxies with $E(B-V)=0.15$ and Milky Way-like extinction law $R_V=3.1$ \citep{1989ApJ...345..245C}.

\begin{figure}[htbp]
	{\centering
		\includegraphics[width=\columnwidth]{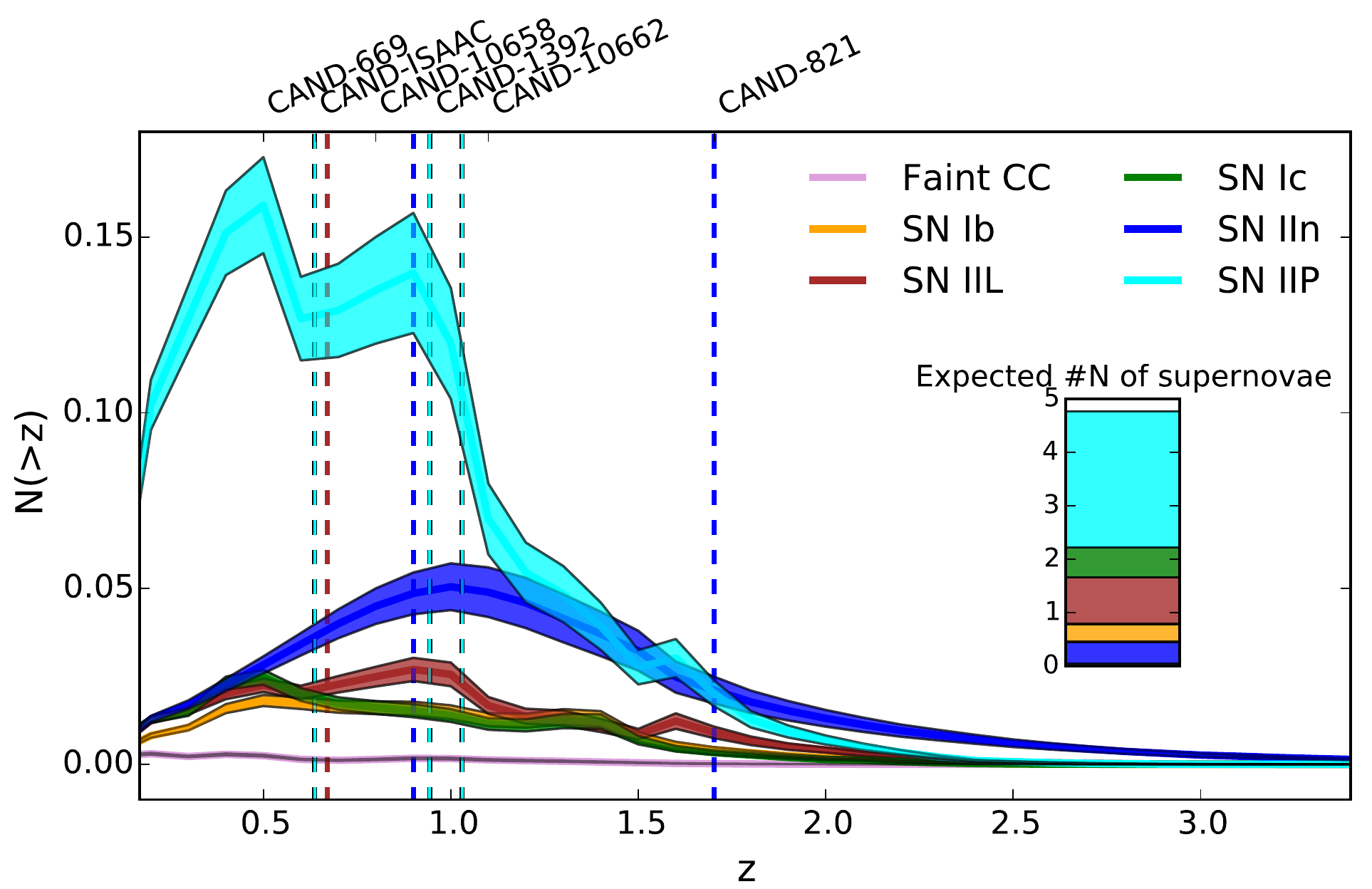}
		\caption{Redshift distribution of core-collapse supernova discoveries expected from the surveys with a rate model from \cite{2007MNRAS.377.1229M}. The redshift of the core-collapse supernova candidates from our surveys are indicated with dashed line. The total expected number of core-collapse supernovae is also shown. }
		\label{dN_dz}
	}
\end{figure}

The result is shown in Figure~\ref{dN_dz} and indicates that our survey was sensitive up to $z\approx2$ for most \sn subtypes and even to $z\approx3$ for the brightest CC~\sne. Moreover, the most expected subtypes were \sne~IIP around $z\approx0.5-1$ and \sne~IIn at $z\approx0.8-1.2$, thus reflecting the types and redshift distribution of our discoveries well.

Next, we investigate the consistency of detecting a single \snia outside A1689 with the expectations. We ran a similar simulation considering three \snia rate models based on the rates from the Supernova Legacy Survey (SNLS; \citealt{2006AJ....132.1126N}), the so-called "A+B model" from \citet{2005ApJ...629L..85S}, and  the best fit to the GOODS rates \citep{2008ApJ...681..462D, 2012ApJ...757...70D} corresponding to a delay time $\tau=1.0$ Gyr. This delay time refers to the time  between star formation and the \snia explosion and  was assumed to follow a Gaussian distribution.  The integrated number of \snia events are $2.2$, $3.7$, and $1.7$, respectively. We conclude that the detection of a single event is consistent with expectations.

\section{Volumetric SN rates}\label{sec:volumetric}
Next, we calculate the volumetric \sn rates. The volumetric SN rate, $r_V^j$, for a \sn type $j$ (in units  Mpc$^{-3}$yr$~^{-1}h^3_{70}$) is given by
\begin{equation}
r_V^j(z)=\frac{N_j(z)}{T_j(z)\cdotp V_C(z)}(1+z),
\end{equation}
where $T_j(z)$ is the visibility time (often called the survey control time), which indicates the amount of time the survey is sensitive to detecting a \sn candidate \citep{1938ApJ....88..529Z}.  The value $N_j$ is the number of \sne and $V_C$ is the co-moving volume. The factor $(1+z)$ corrects for the cosmological time dilation. The number of observed \sne, $N_{\rm raw}$, was corrected for the redshift migration effect to obtain the real debiased number, $N_{\rm debiased}$ (see Sect.~\ref{sec:migration}). We  calculated the volumetric rates for the search area as defined in the post-survey search, so we only considered the \sne inside this field (see Figure~\ref{lensing}).

The monitoring time above the detection threshold for a \sn of type $j$, $T_j$, is a function of the \sn light curve, absolute intrinsic SN brightness, $M$, detection efficiency (see Appendix~\ref{sec:det_eff}), $\epsilon$, extinction by dust, $\Delta m_{\rm ext}$, and the lensing magnification, $\Delta m_{lens}$. The probability distributions of the absolute intrinsic brightness $P(M)$ are assumed to be Gaussian with properties listed in Table~\ref{tab:peak_v}. Following a similar procedure as in G09, the control time is obtained as 
\begin{equation}
\begin{split}
T_j(z,\Delta m_{\rm ext}+\Delta m_{\rm lens}) =  \\
= \int{\epsilon(m){P(M)} \Delta t_j(z,M+\Delta m_{\rm  ext}+\Delta m_{\rm lens}) dM},
\end{split}
\end{equation}
where  $\Delta t_j$ is the time period when the \sn light curve is above the detection threshold. We combined the control time of the NIR and optical surveys and included the VLT/ISAAC survey from G09.

We calculated the control time for \sneia and CC separately. The control time depends on the properties of the light curves, so different subtypes of CC~\sne have different control times. The total CC control time was obtained by weighting the contribution from the various CC~\sn subtypes with their fractions (shown in Table~\ref{tab:peak_v}) and then summed.

The volumetric rates were measured in redshift bins chosen to match other surveys for easier comparison. There are five redshift bins in the range $0.4 \le z < 2.9$ with equal bin width of $\Delta z=0.5$.  We start at $z=0.4$  and postpone the discussion of the cluster rate to Sect.\,\ref{sec:cluster_rates}. We placed the measurements of the rates at the effective redshift of each bin, where the weighting is done with the control time.

The co-moving volume, $V_C$, contained in the redshift bin between $z_1 < z < z_2$ was computed as
\begin{equation}
V_C = \int\limits_{z1}^{z2}\frac{c \cdotp d_L^2(z)} {H(z)\cdotp (1+z)^2} \omega \cdotp  dz,  
\end{equation} 
where $d_L$ is the luminosity distance, $c$ is the speed of light in vacuum, $H(z)$ is the Hubble parameter  at redshift $z$ in units of $\rm km \cdotp s^{-1} Mpc^{-1}$, and $\omega$ denotes the solid angle, corrected by the lensing magnification $\mu$. The spatial variation of the magnification is accounted for by integrating over the field of view when calculating $\omega$ (see G09 for details). Figure~\ref{magnitude_limit} shows the \Jband band survey  volume summed in the line of sight for each \sn type. Given that the \Jband band is more sensitive to high-$z$ \sne, the corresponding volume for the $i$ band survey is between $\sim1.4$ and $\sim2.1$ times smaller, depending on SN type.

 Observations of \sne are likely to suffer from extinction in the host galaxy, which influences the rate estimate. Making an appropriate correction to account for this is very important, however, this correction represents one of the most uncertain assumptions in the rates analysis.  We assumed a Milky Way-like extinction law \citep{1989ApJ...345..245C} with $R_V=3.1$ for both \snia and CC~SN rates. In Sect. \ref{sec:ext_syst} we compare this assumption to the impact of using a starburst extinction law with $R_V=4.05$ \citep{2000ApJ...533..682C} for CC~\sne.

The values for the colour excess were drawn from a positive Gaussian distribution with a mean $E(B-V)=0.15$ and $\sigma_{ E(B-V)}=0.02$ similar to \citet{2012ApJ...757...70D}. Our choice of extinction correction is further justified by  \citet{2014ApJ...780..143A}, who studied $z$$\sim$$2$ galaxies behind A1689 and found an average reddening $E(B-V) = 0.15$ mag.

Extinction corrections for \sn rates were studied in  \citet{1998ApJ...502..177H} and \citet{2005MNRAS.362..671R}  with MCMC predictions that depend on several assumed parameters of the \sn, host galaxy, and dust properties. \citet{2012ApJ...757...70D} and \citet{Mattila12} compiled observed extinction properties of nearby CC~\sn sample with mean extinction $A_V=0.42\pm0.09$ mag. This is very close to the predicted mean value $A_V=0.44$ derived from a distribution of expected $A_V$ values from the model in \citet{2005MNRAS.362..671R}. It is difficult to measure $A_V$ directly for individual CC~\sne at higher $z$; we can only detect those with low $A_V$ and there is a relatively large spread of their absolute magnitudes.
Since a detailed understanding of the evolution of the dust content in galaxies with redshift is still lacking, we follow previous work \citep{Melinder, Graur11, Strolger15} and assume equal reddening over our entire redshift range.
 
To calculate the upper limits for \snia rates,  we used the high extinction dust model from \citet{2014AJ....148...13R}, where $A_V$ follows a Gaussian distribution with $\mu = 0.5$ and $\sigma = 0.62$. 
 
\subsection{Results}
Thanks to the magnification from A1689, our search was sensitive to high-redshift \sne (see Figure~\ref{dN_dz}). We present measurements  and pose upper limits on the \sn rates in five redshift bins from $0.4~\le~z~<~2.9$. Since we did not detect any confirmed \snia behind the galaxy cluster in these redshift bins and our survey was sensitive to these events, we only present the limits.

\begin{figure*}[htbp]
	\begin{center}
		\includegraphics[width= 5.8in]{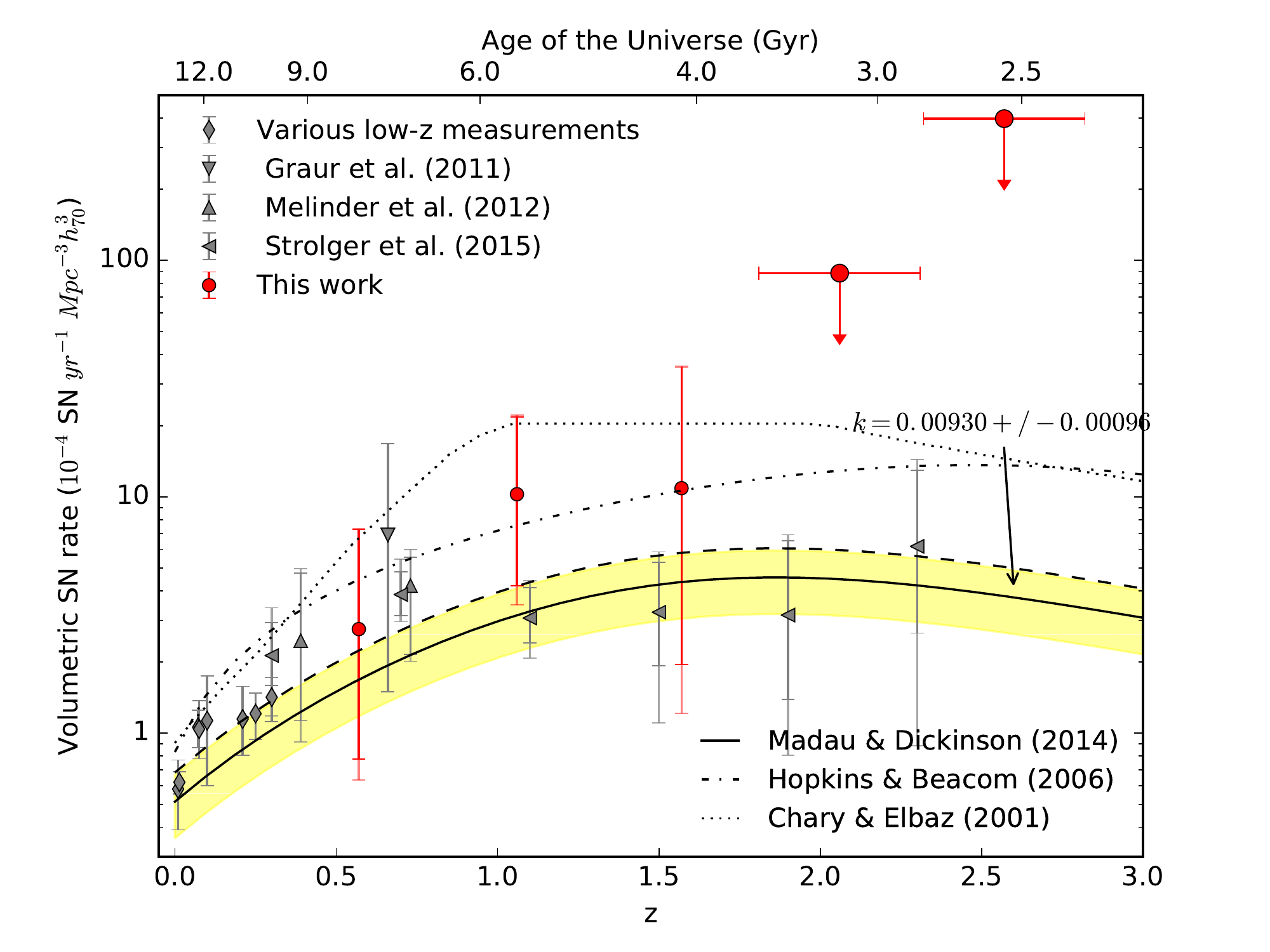}
		\caption{Measured core-collapse supernova rates with our NIR+optical (VLT/HAWK-I + NOT/ALFOSC) surveys presented here and VLT/ISAAC survey presented in \citet{Second}. Core-collapse supernova rate measurements from the literature are also shown; we plot the results with normal host galaxy extinction. Error bars are statistical with total errors (statistical and systematic added in quadrature) as a transparent/faded extra error bar for all surveys. The various lines represent cosmic star formation histories from several authors scaled with $k_{8}^{50}=0.007$. The best-fitted value to the rates data of the scale factor $k_{\rm CC}=0.00930\pm0.00096$ is also shown. }
		\label{cc_rates}
	\end{center}
\end{figure*}

Our rates are summarized in Table~\ref{table:rates} and shown in Figure~\ref{cc_rates}. For comparison, in the same figure we also plot measurements from other authors. At lower redshifts we include measurements from \citet{2012A&A...537A.132B, Mattila12, Li, Cappellaro, 2014ApJ...792..135T,2015MNRAS.450..905G,2009A&A...499..653B,2008A&A...479...49B,SUDARE}.
At the higher redshifts range, ($z \gtrsim0.4$), fewer results exists; these are Subaru Deep Field (SDF) at $\langle z \rangle=0.66$ \citep{Graur11}, the Stockholm VIMOS Supernova Survey (SVISS) at $\langle z \rangle=0.39$ and $\langle z \rangle=0.73$ \citep{Melinder}, and the combined rates from GOODS+CANDELS+CLASH survey at  $\langle z \rangle =0.3$, $\langle z \rangle=0.7$,  $\langle z \rangle =1.5$, $\langle z \rangle=1.9$ and $\langle z\rangle=2.3$ \citep{2004ApJ...613..189D,2012ApJ...757...70D, Strolger15}. 

Our rate measurements are slightly higher than the results from all other surveys at $\langle z \rangle =1.06$ and $\langle z \rangle =1.57$, but our rate measurements are consistent with all other measurements when taking statistical and systematic errors into account. We also plot the predictions from different star formation histories which we discuss in Sect.~\ref{SFH}.
 
\begin{table*}
	\center
	\caption{Field SN numbers and volumetric rates}\label{table:rates}
	\begin{tabular}{lccccc}
		\hline
		\hline
		Redshift bin  & $0.4\le z < 0.9 $ & $0.9\le z < 1.4$ & $1.4 \le z < 1.9$ & $1.9 \le z<2.4$ & $2.4 \le z<2.9$ \\
		\hline
		Effective redshift & 0.57 & 1.06 & 1.57 & 2.06 & 2.56 \\
		$\rm N_{\rm CC}$(raw) & 1 &  3 & 1 & 0 & 0 \\
		$\rm N_{\rm CC}$(debiased) & 1.51 &  2.48 & 1.01 & 0 & 0  \\
		$ r_{\rm V,CC}$ (no ext.)$^{a,b}$ & $\displaystyle2.6^{+4.2}_{1.8}\displaystyle$ & $8.7^{+9.8}_{-5.2}$  & $8.1^{+18.5}_{-6.7}$   & $<57$ & $<251$ \\
		$r_{\rm V,CC}$ (norm.ext.)  & $\displaystyle2.8^{+4.5}_{-2.0}$ & $10.3^{+11.5}_{-6.1}$  & $10.8^{+24.4}_{-8.9}$  & $ < 88$ & $< 398$ \\
		
		\hline
		Effective redshift & 0.61 & 1.09 & 1.59  & 2.07 & 2.54 \\
		$\rm N_{\rm Ia}$ (raw) &  0 & 0 & 0 & 0 & 0 \\
		$ r_{\rm V, Ia}$(no ext.)$^a$  & $<3.6$ &$<2.7$&$<3.5$ & $<5.8$ & $<20.4$ \\
		$ r_{\rm V, Ia}$(with high ext.)  & $<3.8$ &$< 3.0$ & $<4.5$ & $<9.6$ & $<62.1$ \\
		\hline

		\multicolumn{6}{l}{{\bf Notes.} The volumetric supernova rates $r_V$ are given in [10$^{-4}$ yr$^{-1}$ Mpc$^{-3} h^3_{70}$]. }\\
		\multicolumn{6}{l}{$^a$The errors, are the 68\% Poisson statistical uncertainties of the number of SNe propagated to the rates.} \\

		\multicolumn{6}{l}{$^b$The highest two redshift bins for the  CC~\sn rates and all the \snia rates are upper limits with $2\sigma$ Poisson} \\
		\multicolumn{6}{l}{uncertainty.}
	\end{tabular}
\end{table*}

\subsection{Sources of uncertainties in the CC~SN rates}  
\label{systematics}
Our error budget is dominated by the Poisson uncertainties from the low number of events in our survey.  To assess the impact of the many assumptions we made, we estimate of the systematics errors in detail and discuss them below. The resulting systematic uncertainties per redshift bin are summarized in Table~\ref{table:systerrs} along with the statistical uncertainties for comparison. In order to quantify the contribution  to the systematic uncertainties of the rate calculation, we re-calculated the control time with several assumptions and derived the volumetric rate. The uncertainties in the detection efficiency function are small in comparison with the sources of error. Since we only provide upper limits for \sneia rates at redshift bins where there are ample existing measurements \citep[see \eg][for a recent volumetric \snia rate compilation]{2014AJ....148...13R}, we did not engage in detailed estimation of the systematic uncertainties.

\textit{Mis-typing:} Incorrectly classified candidates as CC~\sne instead of \snia or AGN, introduces an error in the rates. For this reason, we discuss the \sn typing thoroughly in Sect.~\,\ref{sec:snebehind}. As an attempt to quantify the possible systematic offset caused by misclassification,  we excluded one of the \sne in the rate calculation at $z=0.94$. In this way, we obtained lower rates of $\sim$$20\%$ and $\sim$$30\%$ at  $\langle z \rangle=0.57$ and $\langle z \rangle=1.06$, respectively which is  smaller than the statistical uncertainty.

\textit{Redshift migration:}\label{sec:migration}
Uncertainty in the \sn redshift introduces uncertainty in the rate determination. Using Monte Carlo simulations we redistributed the \sne with a Gaussian distribution where the mean and standard deviations are the \sn redshift and its error, respectively. The derived errors are then propagated into the \sn rates.

The effect of the redshift migration can have a significant contribution on the systematics if the \sn typing only relies on the photometric redshift of the host galaxy. This highlights the importance of having a spectroscopic redshift measurement of the galaxy or, even better, spectroscopic confirmation of the \sn candidate.

\textit{Host galaxy extinction:}\label{sec:ext_syst}
To estimate the extinction correction uncertainties in the CC~\sn rates, we tested several assumptions for the extinction parameters. \citet{2014AJ....148...13R} (lower panel of their Figure~7) show three dust models for CC~\sne population based on observational evidence.
The different distributions for $A_V$ are generated from the positive half of a Gaussian distribution centred at $A_V=0$ with dispersion $\sigma$, plus an exponential distribution of the form $e^{-A_V/\tau}$. Their low extinction model with $\sigma=0.15,\tau=0.5$ matches well with our assumption made for the rate calculation. Their high extinction model with $\sigma=0.8,\tau=2.8$ yields higher CC~\sn rates by $15\%$ at $\langle z \rangle=0.57$, $37\%$ at $\langle z \rangle=1.08$, and $72\%$ at $\langle z \rangle=1.57$.

We also calculated the rates with negligible extinction, $E(B-V)=0$ (also shown in Table~\ref{table:rates}). The CC~\sn rates decreased by $7\%$ at $\langle z \rangle=0.57$, $\langle z \rangle=1.08$, $16\%$ at $\langle z \rangle=1.08$, and $30\%$ at $\langle z \rangle=1.57$, which provides the extreme lower limit to the estimate.

 \citet{2007MNRAS.377.1229M} and \citet{Mattila12} showed that a large fraction of the \sne exploding in galaxies can be missed because of severe dust obscuration. In particular, this should happen in galaxies known as  luminous IR galaxies (LIRGs) and ultra-luminous IR galaxies (ULIRGs), where most of the UV light is processed into thermal IR heating by dust. We used the correction factors to account for the missing \sn fraction from \citet{Mattila12} as an extreme upper limit to the uncertainty estimate of the host extinction correction. The resulting missing \sn fractions weighted by our control time are $47\%$ at $\langle z \rangle=0.57$,  $56\%$ at $\langle z \rangle=1.06$ and $58\%$ at $\langle z \rangle=1.57$. These  correction factors are based on the observations of the nearby LIRG Arp 299 and the missing fraction of \sne in high-$z$ U/LIRGs remains uncertain. In particular the assumption that the properties of the U/LIRGs do not change significantly with time might not hold.

 We also quantified the impact of  the  assumption regarding the extinction law for CC rates, by assuming attenuation law with $R_V=4.05$ that is  appropriate for starburst galaxies  \citep{2000ApJ...533..682C}.We found a small impact on our rates results; they become higher  by $1\%$ at $\langle z \rangle=0.57$, $1\%$ at $\langle z \rangle=1.08$, and $2\%$ at $\langle z \rangle=1.57$.

\textit{CC~\sn fractions and peak magnitudes:} Throughout the CC~\sn rates calculations, we have used the \sn properties from the Asiago Supernova Catalogue (ASC) as compiled by \citet{2014AJ....147..118R} and the fraction of the CC~SN subtypes from the Lick Observatory Supernova Search (LOSS) \citep{Li}. If we instead use the \sn fractions and properties as compiled in Table~1 in G09 based on older ASC results \citep{2002AJ....123..745R}, we obtained lower rates, $\sim$$17\%$ at $\langle z \rangle=1.08$ and $\sim$$19\%$ at $\langle z \rangle=1.57$.  However,  the uncertainty in the relative fractions  is large. For instance, in the newer estimate,  \citet{2014AJ....147..118R} found a much lower fraction of Type IIL of 7.3\% than the  20\% in  \citet{2002AJ....123..745R}. Also, measured CC~\sn fractions are based on a nearby sample study and might not be representable at higher redshifts where CC~\sn fractions are quite unknown.

Unlike \sneia, CC~\sne show large spread in the values of the mean peak absolute magnitudes. To estimate the effect on the rates, we propagated the standard error of the mean peak absolute magnitudes for each subtype of CC~\sne (shown in Table~\ref{tab:peak_v}, values taken from \citet{2014AJ....147..118R}). We obtained $-12/\!+\!14\%$ at $\langle z \rangle=0.57$, $-20/\!+\!23\%$ at $\langle z \rangle=1.08$, and $-25/\!+\!29~\%$ at $\langle z \rangle=1.57$. 

\textit{Magnification maps:} The uncertainties in the magnification maps of A1689 propagate to uncertainties of the estimated rates since they affect both the control time and the probed co-moving volume. A larger (smaller) magnification would result in an increased (decreased) control time, but at the same time a smaller (larger) volume is probed. The  uncertainties originating from the magnification maps are insignificant at $\langle z \rangle=0.57$, $\pm3\%$ at $\langle z \rangle=1.08$ and $-3/\!+\!1\%$ at $\langle z \rangle=1.57$.

\textit{Cosmic variance:} We used a relatively small field of view around A1689 to measure \sn rates, which can lead to uncertainty due to the local inhomogeneity. To account for this, we calculated our systematic errors coming from cosmic variance based on the recipe from \citet{2008ApJ...676..767T}\footnote{http://casa.colorado.edu/~trenti/CosmicVariance.html} and assumed that the variances of the \sne follow the overall galaxy population. We estimate that the uncertainties are $\sim$$14\%$ at $\langle z \rangle=0.57$, $\sim$$15\%$ at $\langle z \rangle=1.08$, and $\sim$$17\%$ at $\langle z \rangle=1.57$.

\begin{table*}
	\caption{Relative systematic error budget}\label{table:systerrs} \centering \begin{tabular}{l ccc} \hline\hline Error source &\multicolumn{2}{c}{CC Supernovae} \\
		\hline 
		& $0.4\le z<0.9$ & {$0.9\le z<1.3$} & {$1.4 \le z<1.8$} \\
		\hline
		Redshift migration & $\pm0.68$ & $\pm1.74$ & $\pm0.11$ \\
		Subtype fractions    &$\pm0.09$& $\pm1.38$ & $\pm2.35$ \\
		Peak magnitudes & $-0.32/+0.38$ & $-2.01/+2.35$&$-2.78/+2.98$ \\
		Extinction  law    & $\pm0.03$ & $\pm0.01$ & $\pm0.13$\\
		Dust extinction    & $-0.05/+0.04$ & $-0.25/+0.35$ & $-0.49/+0.48$\\
		Extreme dust extinction limits   & $-0.19/+1.21$ & $-1.55/+4.90$ &  $-2.62/+4.73$\\
		Magnification maps & $\pm0.01$&  $-0.28/+0.35$   &$-0.37/+0.14$ \\
		Cosmic variance & $\pm0.38$&  $\pm1.56$   &$\pm1.83$ \\
		\hline
		Total systematic & $-0.76/+0.79$ & $-3.01/+3.27$ & $-3.70/+3.83$\\
		Statistical & $-1.98/+4.51$ & $-6.06/+11.48$ & $-9.56/+26.36$\\
		\hline
	\end{tabular} 
	\tablefoot{The total errors have been computed by co-adding the individual errors in quadrature. We do not add the extreme extinction limits and the cosmic variance since they do not represent a measurement error.}
\end{table*}

\section{Core-collapse supernova rate and cosmic star formation history}
\label{SFH}
Since CC~\sne have massive and short-lived stars as progenitors, their rates, $r^{\rm CC}_{V}(z)$, should reflect the ongoing star formation rate $\psi(z)$ (SFR) with a simple relation, 
\begin{equation}\label{eq_rate}
	r^{\rm CC}_{V}(z)=k_{\rm CC} \cdot \:h^2 \cdot\psi(z),
\end{equation}
where the scale factor $k_{\rm CC}$ is the number of stars per unit mass that explode as \sne. The constant is obtained as the ratio of the two integrals
\begin{equation}
	k_{\rm CC}^{}=\frac{\int_{M_{min}}^{M_{max}}{\phi(M)dM}}{\int_{0.1\msun}^{125\msun}{M\phi(M)dM}},
\end{equation}
where $\phi(M)$ is the initial mass function (IMF), which is the empirical function that describes the mass distribution of a population of stars. We assumed that $k_{\rm CC}$  does not evolve with redshift. Similar to previous high-$z$ CC~\sn rate studies, we used a \citet{1955ApJ...121..161S} IMF and $M_{min}=8 \: \rm \msun$ and $M_{max}= 50 \: \rm \msun$ as range of stellar masses that explode as CC~\sne. With these assumptions, the scale factor is $k_{8}^{50}=0.0070^{+0.0019}_{-0.0022} \: \rm \msun ^{-1}$. The uncertainty comes from the assumed IMF and the extremes of integration i.e. the stellar mass ranges that end up as CC~\sne \citep[see \eg][for a detailed discussion]{Melinder, SUDARE, Strolger15}.

The estimate of the SFH is based on several SF tracers, which depend on the parametrization of the SF function and the applied dust extinction. In fact, only recently has a clear picture of the SFH emerged \citep{2014ARA&A..52..415M} in which these authors found a consistent SFR density which peaks at $z\approx1.9$ and then declines exponentially. By using a scale factor of $k_{\rm CC}=0.007$, Madau \& Dickinson predicted CC~\sn rates that were in good agreement with the results in  \citet{2012ApJ...757...70D}. Before the revised version of cosmic SFH presented by \citet{2014ARA&A..52..415M}, it was suspected that the cosmic CC~\sn rate did not match the massive SF rate and that the CC~\sn rate  was lower by factor two \citep{2011ApJ...738..154H}. From the overall CC~\sn rate available and the predictions from the \citet{2014ARA&A..52..415M} SFH, that problem is not evident anymore.

Instead of using the scale factor $k$ from a Salpeter IMF and the above-mentioned mass ranges, we can empirically obtain  the value for $k$ by fitting the $\psi(z)$ to our CC rates and all available literature values. We used the weighted least-squares fit of the SFH parametrization from \citet{2014ARA&A..52..415M} as follows:
\begin{equation}
	\psi(z) = A\,\frac{(1+z)^C}{1+((1+z)/B)^D}\quad[\rm\msun \,year^{-1}Mpc^{-3}]
\end{equation}
with their best-fit values $A=0.015$, $B=2.9$, $C=2.7$, and $D=5.6$.
We used the Levenberg-Marquardt least-squares algorithm and we obtained the best fit for $k_{\rm CC}=0.00930\pm 0.00096 $, which is similar to what was found in \citet{Strolger15}. The result with its uncertainty is shown in Figure~\ref{cc_rates}. This value is within $\sim$$1\sigma$ of $k_{8}^{50}$.

\citet{Strolger15} attempted to predict the shape of $\psi(z)$ by using re-binned measured values in five redshift bins from the literature and the combined GOODS+CANDELS+CLASH surveys. We added our results and the newly published measurements at $\langle z \rangle=0.10$ and $\langle z \rangle=0.25$ from the Supernova Diversity and Rate Evolution survey (SUDARE; \citealt{SUDARE}) to obtain new comprehensive weighted average rates in the same redshift bins. We obtained $A=0.016\pm 0.004$, $B=1.9\pm0.7$, $C=4.3\pm 1.7$, and $D=6.2\pm2.5$. However, in addition to the comparable number of points with the number of parameters, this procedure is sensitive to the choice of binning and anchoring low-$z$ point. In conclusion, a robust prediction of the SFH from the CC~\sn rate remains difficult at present.

\section{Supernovae in the resolved strongly lensed multiply-imaged galaxies }
\label{sec:multi}

Based on photometric studies on {\it HST}/ACS images, there are more than 100  multiple images of $40$ strongly lensed background galaxy sources with redshifts ranging from $z=1$ to $z=4.9$ behind A1689 (\citealt{2005ApJ...621...53B}, \citealt{2007ApJ...668..643L}, \citealt{2015MNRAS.446..683D}). Our survey was sensitive to \sne that could have exploded in these galaxies. For example, the multiple images (labelled 1.1 and 1.2) of galaxy source 1 at $z$=3.04, are magnified by $3.80^{+0.30}_{-0.29}$ and $5.58^{+1.21}_{-0.74}$ mag, respectively \citep{Third}. With this magnification, most subtypes of \sne were detectable with our NIR survey. Furthermore, the time delay between these two images is $83^{+77}_{-69}$ days \citep{Third}, which is well within the total length of our programme.  In other words, if a SN had exploded in this galaxy, the SN could have been detected in both 1.1 and 1.2 during the survey (see Figure~\ref{WFIRST_multiple}).

Because of their standard candle nature, observations of  \sneia through lensing clusters are particularly interesting. By estimating the absolute magnification of \sneia, it is possible to break the so-called mass-sheet degeneracy of gravitational lenses. Thus, they could be used to put constraints on the lensing potential, if the background cosmology is assumed to be known \citep[see~e.g.][]{2014MNRAS.440.2742N}. This can be further improved by observations of strongly lensed \sneia, through the measurements of time delays between the multiple images. In contrast, if the lensing potential is well known instead, measurements of time delays of any transient source can be used to measure the Hubble constant and, to a lesser degree, the density of the cosmic fluids \citep{1964MNRAS.128..307R,2001ApJ...556L..71H,2003ApJ...592...17B, 2003MNRAS.338L..25O,2007ApJ...660....1O, 2014ApJ...789...51Z, Third}.
At the high redshifts where these galaxies are located, the \sn rates are dominated by CC~\sne and the \snia rates are expected to be lower owing to the delay between the star formation and explosion. 

Here, we calculate the expected number of \sne in the multiply lensed background galaxies from our NIR survey. As in \citet{Third}, we only considered those that have a spectroscopic redshift and a predicted time delay of less than five years. There are 17 systems with 51 images that satisfy these criteria. These are shown in Figure~\ref{multiple}. To obtain the expected number of \sne in each galaxy $N_i$, we multiply the \sn rate $R_i$ and the control time $T_i$,
\begin{equation}
N_i=R_i\cdot T_i,
\end{equation}
where $i$ indicates the individual galaxies. To estimate the \sn rate in each galaxy, we used the SFR estimated previously in \citet{Third} and G09, using the rest-frame UV luminosity at $2800\, \AA$ as a tracer. The expected CC~SN rate for each galaxy in units yr$^{-1}$ was then calculated from
\begin{equation}
R_{\rm CC} = k_8^{50} \cdot SFR,
\end{equation} 
where the SFR is given in in units $[\msun$yr$^{-1}$].  The \snia rates were obtained from the \citet{2005ApJ...629L..85S} two-component model
\begin{equation}
R_{Ia} = A \cdot SFR + B \cdot M_*,
\end{equation} 
where the stellar mass $\rm M_*$ of the individual galaxies was obtained from the rest-frame \Bband band luminosity via the relations from \citet{2003ApJS..149..289B}. The total expected number of SNe over all the systems are then simply summed. The result is  $N_{\rm CC}=0.23 \pm 0.11$ and   $N_{\rm Ia}=0.14 \pm 0.09$ for CC~SNe and  \sneia, respectively. This means that the chance of detection a SN in the strongly lensed galaxies was rather low.

\begin{figure*}[htbp]
	\begin{center}
		\includegraphics[width=\textwidth]{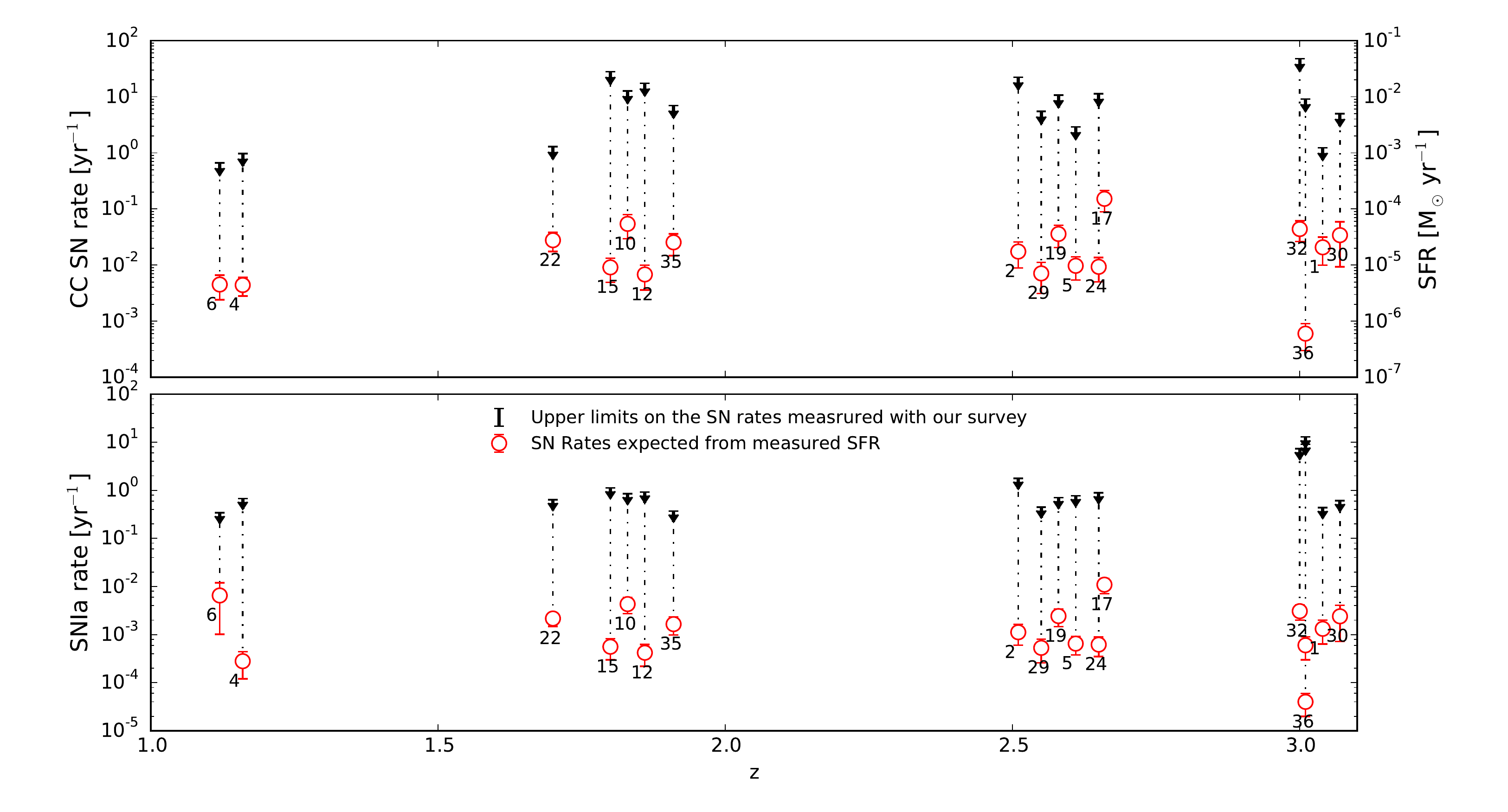}
		\caption{Comparison of the supernova rates limits measured from our survey and those obtained from the SFR published in \citet{Third}. The limits represent 90 \% C.L. The numbers label the background galaxies shown in Figure~\ref{multiple}. Only the galaxies with multiple images and delay time less than five years (i.e less than the duration of the survey) are considered here.} 
		\label{multi_rates}
	\end{center}
\end{figure*}

\begin{figure*}[htbp]
	\begin{center}
		\includegraphics[width=\textwidth]{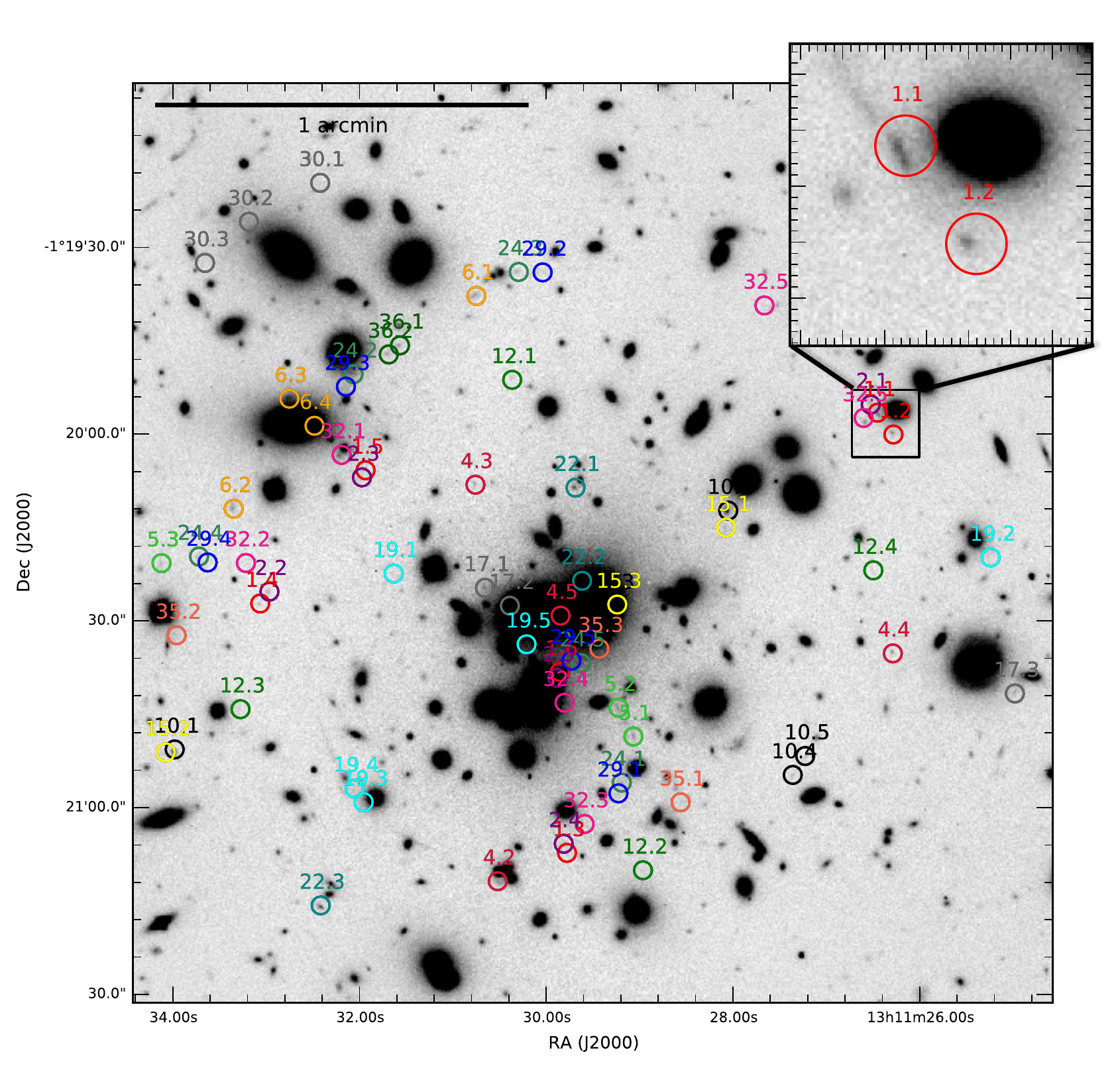}
		\caption{Near-infrared VLT/HAWK-I image overplotted with the 17 background galaxies with 51 multiply images used in Figure \ref{multi_rates}, Sect.\,\ref{sec:multi} and \ref{expectation}. The predicted magnifications for these galaxies are of order of few magnitudes. As an example, a magnified view of system 1 at $z=3.04$ is shown, in which images 1.1 and 1.2 are magnified by $3.80^{+0.30}_{0.29}$ and $5.58^{+1.21}_{-0.74}$ magnitudes, respectively. }
		\label{multiple}
	\end{center}
\end{figure*}

\begin{figure}[htbp]
	\begin{center}
		\includegraphics[width=\hsize]{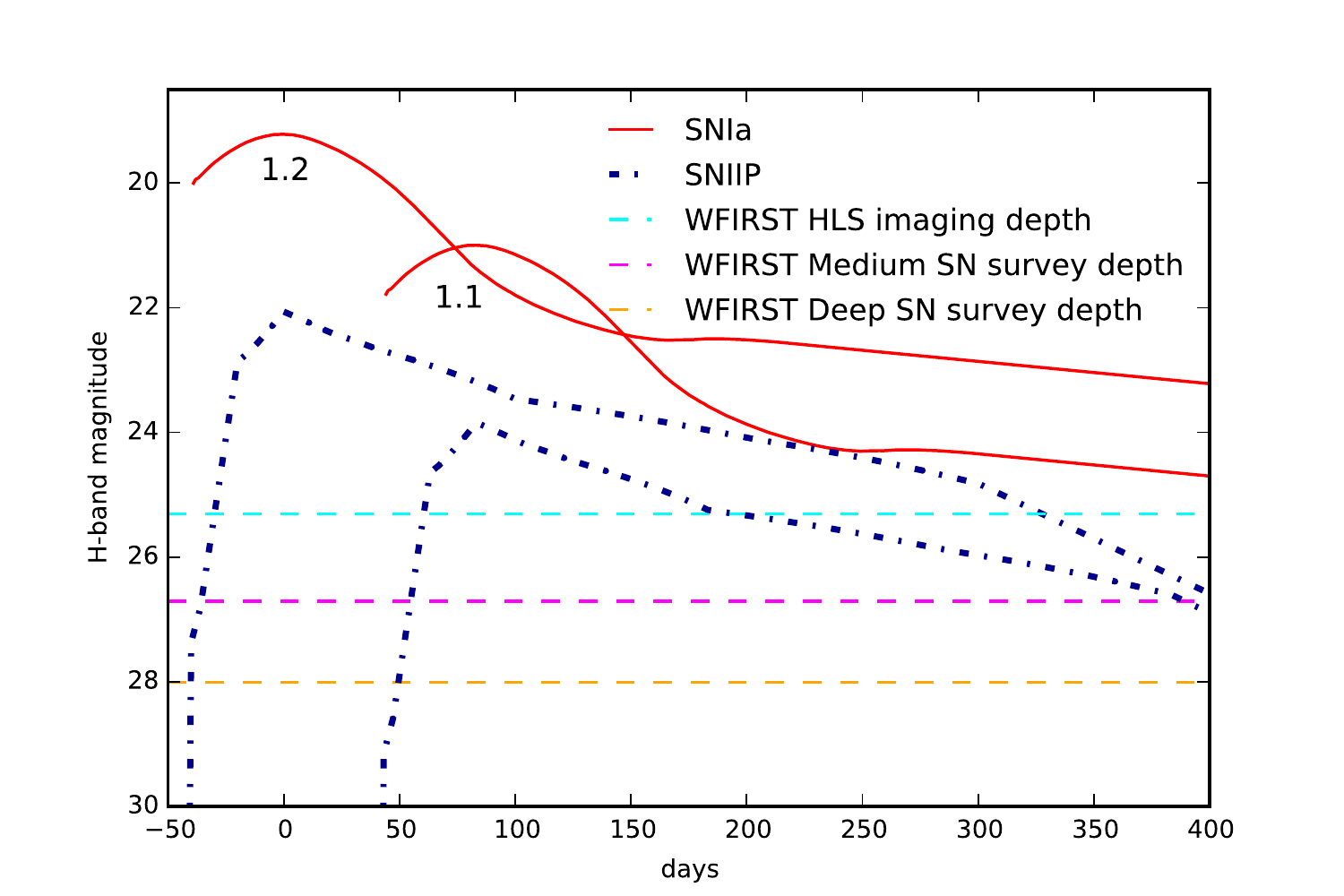}
		\caption{Simulated light curves for \snia and IIP in the \Hband band for the multiply-imaged system 1 at $z=3.04$; images 1.1 and 1.2 are magnified $3.80^{+0.30}_{0.29}$ and $5.58^{+1.21}_{-0.74}$ mag, respectively. The time delay between the images is $83^{+77}_{-69}$ days \citep{Third}.  Also shown are the magnitude limits for the three main survey modes planned for WFIRST \citep{2013arXiv1305.5422S}. Without the magnification from A1689, the SN~IIP would hardly be observable. }
		\label{WFIRST_multiple}
	\end{center}
\end{figure}

Since we have not detected any \sn in these galaxies with multiply lensed images, we calculated the upper limit of the rates for both \sneia and CC. The control time of the  galaxies belonging to the same system (indicated with the same first number in Figure \ref{multiple}) was summed. The result of this comparison is shown in Figure~\ref{multi_rates}. 

\section{A1689 cluster SN rates}\label{sec:cluster_rates}
In this section, we use the \sne detected in galaxies in A1689 to measure the SN rate in the galaxy cluster. Only $\le15\%$ of the galaxies in A1689 show traces of star formation \citep{2008A&A...485..633L}, which confirms that most of the members are galaxies with old stellar populations. Hence, it is not surprising, that the \sne we found are \snia. 

It is not meaningful to discuss the volumetric rate for galaxy clusters, instead the rates are normalized with the cluster total stellar luminosity in one specific band. Here, we used the total \Bband band luminosity and units defined as SNuB$ \equiv 10^{-12} \,\rm{SNe} \, L^{-1}_{\odot,B} yr^{-1}$. Furthermore, when \sn rates in clusters are compared at different redshifts, it is more useful to normalize by the stellar mass of the galaxy cluster, since the luminosity changes with the stellar population age \citep{2012ApJ...745...32B}. The units are then SNuM$ \equiv 10^{-12} \,\rm{SNe} \, \msun^{-1} yr^{-1}$.

To determine the \sn rate in the  galaxy cluster in units SNuB, we used the following relation:
\begin{equation}
\label{cl_rate}
R_{Ia}=\frac{N}{T\cdot L_B}
\end{equation}
where $N$ is the observed number of cluster \sne, $T$ is the total control time and $L_{B}$ is the luminosity of each galaxy member of the cluster in the \Bband band. The total control time represents the amount of time a \sn is above the survey detection limit at the cluster redshift. This control time was calculated in an analogous manner as for the volumetric rates. When normalizing by the total stellar mass of the galaxy cluster, the  $L_{B}$ is replaced by  $M_{*}$ in Eq. \ref{cl_rate}.

The total stellar luminosity and mass of A1689 within $r_{500}$ (the radius within which the mean cluster density exceeds the critical density by a factor of 500), are adopted from \citet{2008A&A...485..633L}  $L_{B}=5.99\pm 0.23\cdot10^{12} L_{\odot,B}$  and $M_{*}=9.0\pm2.0\cdot10^{12} \msun$. They obtained the total luminosity with good precision by combining X-ray and optical data, excluding the background galaxies by using their colours. For the total mass, they used empirical relations between the stellar mass and luminosity of the cluster galaxies. 

\begin{table*}
	\caption {Compilation of \snia rates in galaxy clusters} \label {cluster_rates_table}
	\vspace{0.2cm}
	\begin{centering}

		\begin{tabular}{lccccc}
			\hline\hline
			Survey & Number of & Mean  & $N_{\rm Ia}$ &  \snia Rate$^a$ & Reference\\ 
			& clusters   &    redshift & & [SNuB$\:h^{2}$] &  \\    
			\hline
			Clusters in Cappellaro et al. (1999) SN sample& $\cdots$  & 0.02  & 12.5 & $0.57^{+0.22}_{-0.16}$ &  \citet{2008MNRAS.383.1121M} \\
			SDSS-II   & 71 & 0.084   & 9 & $0.46^{+0.21}_{-0.15} \pm 0.01$& \citet{2010ApJ...713.1026D} \\
			 WOOTS   & 140& 0.15  & 6 & $0.73^{+0.45}_{-0.29} \pm 0.04$ & \citet{2007ApJ...660.1165S} \\
			 
			\textbf{A1689}   & 1& 0.18  & \Ncluster & $0.14^{+0.19}_{-0.09}\pm0.01$  & \textbf{This work} \\
			
			SDSS-II  &492  & 0.225  & 25 & $0.68^{+0.17}_{-0.14} \pm 0.02$ & \citet{2010ApJ...713.1026D} \\
		{\it HST} archival images  & 6 & 0.25   & 1 & $0.80^{+1.84}_{-0.65}$ & \cite{2002MNRAS.332...37G} \\
			CFHT SNLS & $\cdots$ &0.45 &3 &$0.63^{+1.04}_{-0.33}$  & \citet{2008AJ....135.1343G}\\
			 {\it HST}/ACS survey  & 15& 0.6   & 6 & $0.71^{+0.35}_{-0.24}\pm0.26$ & \citet{2010ApJ...718..876S} \\
			 {\it HST} archival images  & 3 &0.9   & 1  & $1.63^{+2.16}_{-1.06}$ & \cite{2002MNRAS.332...37G} \\
			
			{\it HST} Cluster SN survey  & 25  &1.14 & 8 & $0.50^{+0.23}_{-0.19}$ $^{+0.10}_{-0.09}$  & \citet{2012ApJ...745...32B} \\
			
			\hline
		\end{tabular}
		
		Note: $^a$ All quoted errors are $1\sigma$ statistical and systematic, where available.

	\end{centering}
\end{table*}

The cluster rate we measure is $0.14^{+0.19}_{-0.09}\pm0.01$ $\rm SNuB\,$$h^2$, where the error bars indicate $1\, \sigma$ confidence intervals, statistical and systematic, respectively. The cluster rate normalized by the stellar mass is  $0.10^{+0.13}_{-0.06}\pm0.02$ in $\rm SNuM\,$$h^2$.

Our measurement is based on two \sneia detected in one galaxy cluster, and as with most of previous studies, the errors are dominated by small number statistics. In Table~\ref{cluster_rates_table} we make a comparison of our A1689 \snia rate  per luminosity in the \Bband band with the existing values from the literature  \citep{2008MNRAS.383.1121M,2010ApJ...713.1026D,2007ApJ...660.1165S,2002MNRAS.332...37G,2008AJ....135.1343G,2010ApJ...718..876S,2012ApJ...745...32B}

Thermonuclear SNe have long-lived progenitors, so their rate does directly not reflect the cosmic SFH. Instead, there is an unknown time delay between the formation and explosion of the progenitors. Models  are typically parameterized by a delay time distribution (DTD), where a convolution with the SFH gives the \snia rate. Cluster galaxies should typically have more uniform stellar populations compared to their counterparts in the field, which in turn, are reflected in a simpler SFH.  \citet{2012ApJ...745...32B} parametrized the \snia DTD with a power law $\Psi \propto t^s$ and used the approximation of a single burst of star formation at $z=3$. These authors obtained $s=1.41^{+0.47}_{-0.40}$, which is consistent with the DTD estimates in field galaxies. For a detailed discussion, see \citet{2012ApJ...745...32B} and \citet{2010ApJ...722.1879M}.

\begin{figure}[htbp]
	\begin{center}
		\includegraphics[width=\hsize]{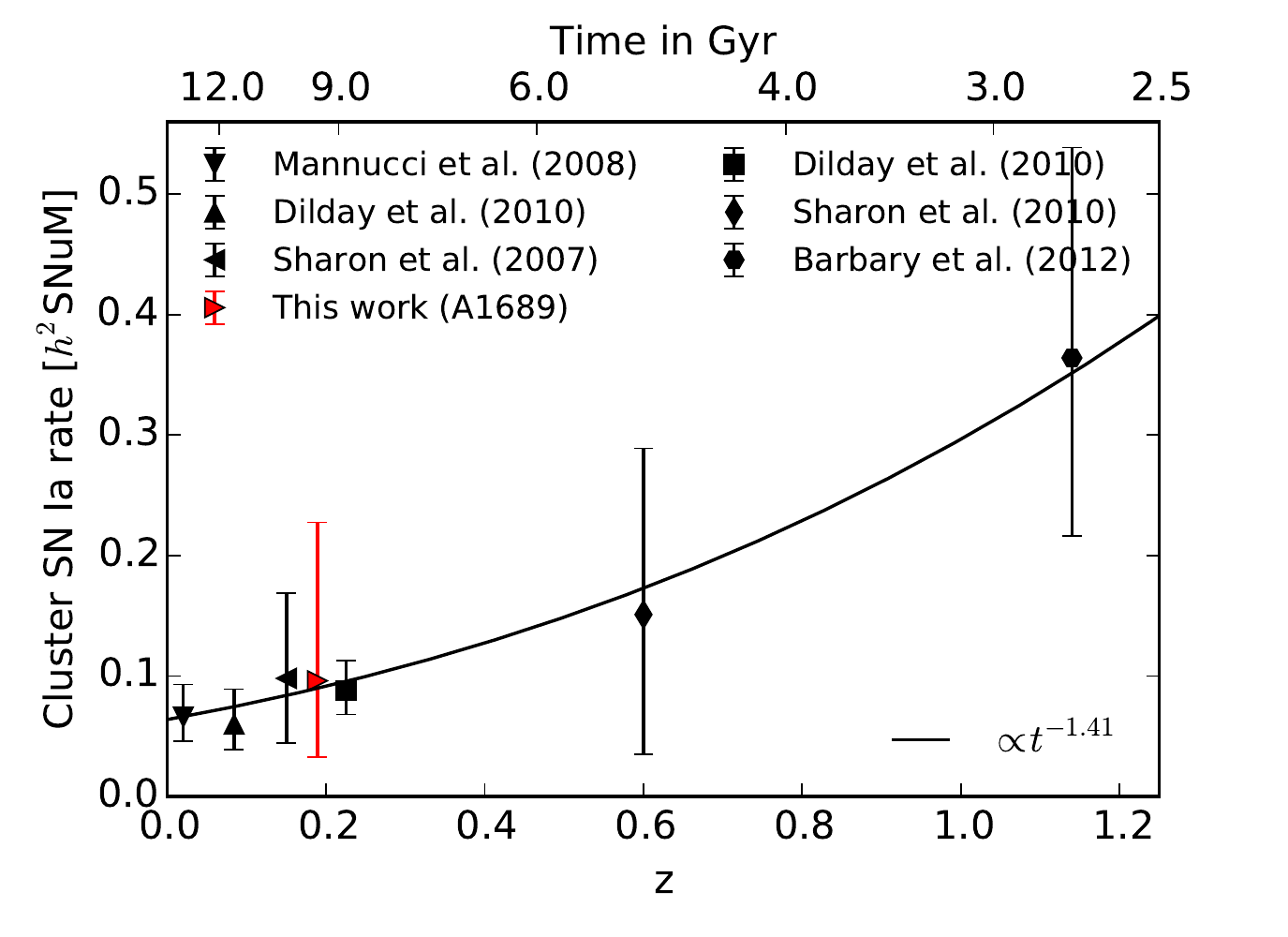}
		\caption{Cluster \snia rates in units SNuM\,$h^2$, where SNuM$ \equiv 10^{-12} \,\rm{SNe} \, \msun^{-1} yr^{-1}$. The line shows the best-fit power-law DTD from \citet{2012ApJ...745...32B}, where $\Psi(t) \propto t^s$ with $s=1.41$.}
		\label{cluster_rates}
	\end{center}
\end{figure}

In Figure~\ref{cluster_rates} we plot our result together with literature values of the \snia cluster rates per stellar mass unit (SNuM). We also show the best-fit power-law DTD from \citet{2012ApJ...745...32B}, where $\Psi(t) \propto t^s$ with $s=1.41$. Our measured value is consistent with the power law.

\section{Expectations for future transient surveys}
\label{expectation}
Here, we discuss the feasibility of finding \sne at $z\gtrsim2$, where the SN rates are poorly constrained, using current facilities. At these very high redshifts, it is particularly interesting to compare CC rates with those expected from the SFR. These two independent methods can serve as an additional test for the SFH at unprobed redshifts. Moreover, high-$z$ CC~SNe can be used for probing the IMF in the early Universe and for studying the possible evolution of the intrinsic properties of CC~\sne subtypes.

\begin{figure*}[htbp]
	\begin{center}
		\includegraphics[width=6.5in]{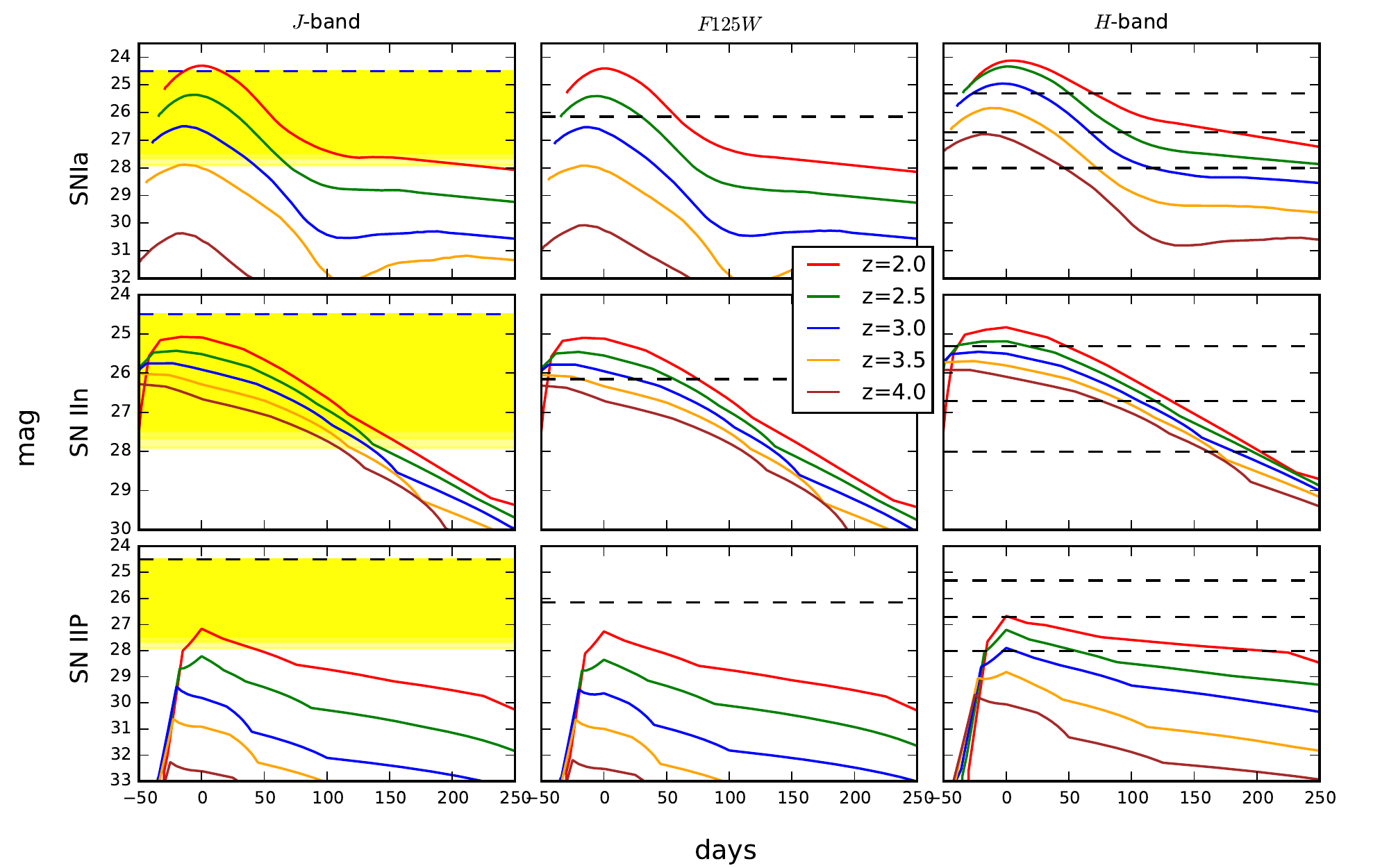}
		\caption{Observer frame light curves for a \snia (upper panels), SN~IIn (middle panel) and SN IIL (lower panels) in the HAWK-I \Jband, {\it HST} {\it F125W} and WFIRST \Hband bands, respectively. The horizontal dashed lines indicate the depths assumed in the calculation, 24.5 for the \Jband band and 26.15 for {\it F125W}. In the \Hband band the three depths represent the three main survey modes planned for WFIRST \citep{2013arXiv1305.5422S}. The yellow band denotes the average magnification from the galaxy cluster A1689 at the same redshifts in the central region with $2.7\arcmin\times3.3\arcmin$.}
		\label{future}
	\end{center}
\end{figure*}

To assess the necessary depth of a survey, in Figure~\ref{future} we plot simulated light curves of \sne Ia, IIn, and IIP in observers frame at redshifts $2.0, 2.5, 3.0, 3.5$, and $4.0$. The yellow band indicates the typical magnifications from a massive galaxy clusters for background objects at those redshifts. Assuming a depth of 24.5 mag in the \Jband band that is suitable for a ground-based survey, it is impossible to observe the light curves at any epoch, without the magnification provided by a cluster lens. For a {\it HST}-like survey with a depth of $26.1$ in the {\it F125W} and \Hband bands (similar to {\it F160W}), \sne~IIn and \sneia are observable, but the fainter IIP are already too challenging. For this reason, we continue making predictions  using the magnification help from the galaxy cluster.

 To predict the number of discoverable CC~\sne, we used the \citet{2007MNRAS.377.1229M} volumetric SN rate 
 as extrapolated in G09.  First, we considered a limiting depth of 24.5 mag in the \Jband band, duration of five years with seven visits per year with a monthly cadence. Second, we considered a limiting depth of 26.15 mag in the {\it F125W} band, which is easily obtainable with {\it HST}. The results are summarized in Table~\ref{table:future_surveys_volumetric}. For the possibility of measuring the CC~\sn rate,
  we can assess how many clusters are required to find at least $\sim$$3$ \sne per redshift bin.  
Even at the redshift bin $\langle z \rangle = 2.06$, $\sim$$45$ clusters are required to reach this goal. For the space-based set-up, the number of clusters decreases: at $\langle z \rangle = 2.06$, $\sim$$11$ clusters are needed, while at $\langle z \rangle = 2.56$, $\sim$$60$. From these estimates, we can conclude that it would be difficult to construct a realistic survey with the existing ground- and space-facilities that would significantly improve  the measurement of SN rates at $\langle z \rangle > 2.0$, even with the help of a gravitational telescope.  This will most likely not be possible
before the forthcoming next generation facilities, such as the space-based Wide Field Infrared Survey Telescope (WFIRST), are in operation. The WFIRST will be able to detect \sne to very high-$z$ in the \Hband band (see Figure~\ref{future}). Without considering the contribution from gravitational lensing, we estimated the expected number of CC~\sne to be discovered with WFIRST. During its planned six-years duration, it will dedicate six months spread over two years for a \sn search, focusing mainly on \sneia to study the possible evolution of the dark energy equation of state parameter \citep{2013arXiv1305.5422S}. However, even a larger number of CC~\sne can be expected and the results are shown in Table~\ref{table:WFIRST_volumetric}.

\begin{table*}
	\caption{Expectations for high-$z$ SNe in A1689 FOV}\label{table:future_surveys_volumetric} \centering \begin{tabular}{l cccccccccc} \hline\hline  Depth &Filter & Years & N$_{epochs}$& Cadence & N$_{\rm CC}^a$ & N$_{\rm CC}^a$ & N$_{\rm CC}^a$ & N$_{\rm CC}^a$ \\
		(mag)&  &&  /yr &(days)  & $1.9\leq z<2.3$  &$2.4\leq z<2.8$  &$2.9\leq z<3.3$  & $3.3\leq z<3.7$ \\
		\hline
		 $24.5 $& \Jband & $5$ &7  &30  & $0.07\pm0.01$ & $0.010\pm0.002$ & $0.0021\pm0.0004$ & $0.0012\pm0.0005$ \\
 		 $26.15 $& {\it F125W} & $5$ &7  &30  & $0.30\pm0.05$ & $0.04\pm0.01$ & $0.011\pm0.001$ & $0.0028\pm0.0005$   \\
		\hline
	\end{tabular} 
	\tablefoot{The errors in the N$_{\rm CC}$ originate from the propagated uncertainty in the \sn rate model. \\
		$^a$ The number of expected CC~\sne in the redshift bin.}
\end{table*}

\begin{table}
	\caption{Expectations for high-$z$ SNe in WFIRST, \Hband band, for both medium and deep supernova Survey. }\label{table:WFIRST_volumetric} \centering \begin{tabular}{l ccccccccc} \hline\hline   N$_{\rm CC}^a$ & N$_{\rm CC}$ & N$_{\rm CC}$ & N$_{\rm CC}$ \\
		  $1.9\leq z<2.3$  &$2.4\leq z<2.8$  &$2.9\leq z<3.3$  & $3.3\leq z<3.7$ \\
		\hline
		   $341 \pm 62$ & $58 \pm 10$ & $20\pm3 $ & $7.5\pm1.5$ \\
		  $1809 \pm 327$ & $898 \pm160$ & $ 360 \pm 63$ & $13 \pm2 $   \\
		\hline
	\end{tabular} 
	\tablefoot{The errors in the N$_{\rm CC}$ originate from the propagated uncertainty in the \sn rate model. The first row shows the expectations for the Medium Supernova survey with WFIRST,  planned to have depth of 26.71 mag and a FOV 8.96 deg$^2$, while second row shows the Deep Supernova survey, depth of 28.01 mag and a FOV 5.04 deg$^2$. The duration is 6 months over 2 years. Magnification from  lensing is not taken into account.\\
		$^a$ N$_{\rm CC}$ indicated the number of expected CC~\sne in the redshift bin.}
\end{table}

Next, we consider the expectations from the multiply-imaged galaxies behind A1689 (presented in Sect.\,\ref{sec:multi}) for future surveys.
Instead of using the volumetric \sn rate, we adopt the \sn rate in the resolved galaxies for which we use the predictions derived from the SFR. We simulated three upcoming transient surveys: the Zwicky Transient Facility (ZTF; \citealt{2014htu..conf...27B}), the Large Synoptic Survey Telescope (LSST; \citealt{2009arXiv0912.0201L}) and WFIRST. The ZTF and LSST are ground-based surveys with 1.2 and 8.4 m telescopes, respectively.  The results are shown Table~\ref{table:future_surveys}. The shallowness makes ZTF not  useful for this task. The situation is different for LSST with its improved depth, more suitable filters, and considering the length of the survey. The LSST goal is to revisit the same field $\sim$$164$ and $\sim$$180$ times in the $z$ and $i$ bands, respectively, over ten years, so close epochs can be combined for a better image depth. Around $70$ galaxy clusters with Einstein radii larger than $\theta_E>20''$ are estimated to be visible to LSST, which would amount to $\sim$$1000$ strongly lensed  multiply-imaged galaxies that are detectable with the LSST \citep{2009arXiv0912.0201L}. Thus, extrapolating from our expectation result from A1689 with 17 systems, we may expect roughly $\sim$$40$ strongly lensed \sneia with the possibility of measuring the time delay between the multiply images. The WFIRST all-sky survey will also be excellent for this task, since it will offer even more suitable filters for higher-$z$ \sne.

The James Webb Space Telescope (JWST; \citealt{2006SSRv..123..485G}) with 6.5~m aperture will have an  unmatched resolution and sensitivity reaching up to $\sim$$31$ magnitude from the optical to the mid-IR. With this depth, the aid of gravitational telescopes are not necessarily needed to discover \sne at $z\lesssim 4$. Using the magnification of the galaxy clusters, however, the JWST will be able to discover lensed CC~\sne at redshifts exceeding $z=6$ \citep{2013MNRAS.435L..33P}. To achieve this, it will require a dedicated multi-year search towards galaxy clusters with its relatively small field of view of $2.2\arcmin \times 2.2\arcmin$. To improve the lens models of galaxy clusters that will be used as gravitational telescopes by JWST, there is an ongoing 190-orbits {\it HST} programme targeting 41 massive clusters named RELICS\footnote{Reionization Lensing Cluster Survey, RELICS;	https://relics.stsci.edu/}.

\begin{table*}
	\caption{Expectations for SNe in the multiply-imaged galaxies in A1689 for future surveys}\label{table:future_surveys} \centering \begin{tabular}{l ccccccccccc} \hline\hline Survey & Filter & Depth & Years & N$_{epochs}$& Cadence & N$_{\rm CC}^a$ & N$_{\snia}^a$ & $z_{max}^b$ & $z_{max}^b$ & N$_{gal}^c$ & N$_{gal}^c$\\

		&  & (mag) & &/yr &(days)  & &&CC &\snia &CC &\snia \\
		\hline
		ZTF & R & $22.5^d$ & $3$ &7  &15  & $0.017\pm0.007$ & $0.04\pm0.03$ & 1.15 & 1.15  & 6& 6\\
		ZTF    &$i$ & $22.5^d$ & $3$ &7  &15  & $0.03\pm0.01$ & $0.06\pm0.05$ & 1.15 & 1.15 &6 & 6  \\
		LSST & $i$ & $24.0^e$& $10$ &7  &30  & $0.18\pm0.09$ & $0.21\pm0.17$ & 3.05 & 1.70 &14 &7  \\
		LSST    & $i$ & $25.0^d$ & $10$ &7  &30  & $0.38\pm0.18$ & $0.26\pm0.20$ & 3.05 & 1.70 &23 &7 \\
		LSST   & $z$ & $22.76^e$ & $10$ &7  &30  & $1.14\pm0.61$ & $0.68\pm0.40$ & 3.05 & 3.05 &25 &33 \\
		WFIRST   & H & $28.01^f$ & $2$ &3  &30  & $1.74\pm0.82$ & $0.17\pm0.08$ & 3.05 & 3.05 &48 &48 \\
		\hline
	\end{tabular} 
	\tablefoot{The errors in the N$_{\rm CC}$ and N$_{\snia}$ originate from the propagated uncertainty in the SFR. \\
		$^a$ The number of expected \sne in the background galaxies with resolved multiply images. \\
		$^b$ The maximum redshift of the expected \sne. \\
		$^c$ Number of galaxies that can host observable \sne. \\
		$^d$ Limiting depth of a co-add of several images. \\
		$^e$ Limiting depth ($5\sigma$ for a single visit with 2x15s exposure \citep{2009arXiv0912.0201L}. \\
		$^f$ $5\sigma$ limiting depth for the WFIRST Deep Supernova Survey \citep{2013arXiv1305.5422S}.		
		 }
\end{table*}

\section{Summary and conclusions}
In this work we present the results of a dedicated ground-based NIR rolling search, accompanied with an optical programme, which is aimed at discovering high-$z$ lensed \sne behind the galaxy cluster Abell~1689. 
During 2008--2014, we obtained a total of 29~and 19~epochs in the \Jband and  $i$ bands, respectively, and discovered five~CC~\sne behind the cluster and two\sneia associated with A1689. Notably, one of the most distant CC~\sn ever discovered was found at $z=1.703$ with significant magnification.

Using these discoveries, we compute the volumetric CC~\sn rates in three redshift bins in the range $0.4<z<1.9$, and put upper limits on the CC~\sn rates in two additional redshift bins in $1.9<z<2.9$. Upper limits of the volumetric \snia rate are also calculated for the same redshift bins.  All the measured rates are found to be consistent with previous studies at the corresponding redshifts. We further emphasize the comparably high statistical precision at high redshift, given the modest investment in observing time, which can be obtained using gravitational telescopes.

The impact of systematic uncertainties were calculated for the CC~\sn rates, which will become increasingly important for upcoming wide-field surveys such as the LSST which are expected to discover a large number of \sne \citep{2009JCAP...01..047L}.  We highlight the need for better understanding CC~SN properties, such as the subtype fractions and peak magnitudes, which will affect the systematics budget.

Since the CC~SN rate traces the cosmic SFH, we compare our results and literature values to the expected rates calculated from the \citet{2014ARA&A..52..415M} SFH.  We find that the measured and predicted rates are in good agreement with a scale factor of, $k_{\rm CC}=0.0093\pm0.0010$~$\msun^{-1}$.

By simulating ground- and space-based five-year surveys, we explore the possibility of finding lensed \sne at $z\gtrsim 2.0$. We find that there is very little room for improvement with the current facilities and that the next generation of telescopes, for example WFIRST, are needed.

Finally, we estimate the number of strongly lensed \sne with multiple images that can be expected to be discovered behind A1689 by upcoming transient surveys. We find that LSST, and in particular WFIRST, can be expected to find tens of strongly lensed SNe that would allow the time delays between the multiple images to be measured. Until the first light of these surveys, using gravitational telescopes is the only way to find high-$z$ \sne using available ground-based instruments.

\begin{acknowledgements}
The work is based, in part, on observations obtained at the ESO Paranal Observatory. It is also based, in part, on observations made with the Nordic Optical Telescope, operated by the Nordic Optical Telescope Scientific Association at the Observatorio del Roque de los Muchachos, La Palma, Spain, of the Instituto de Astrofisica de Canarias. TP thanks Brandon Anderson for proofreading this manuscript and Doron Lemze, Jens Melinder, and Kyle Barbary for useful discussions. The authors thank  John Blakeslee for making HST data available. RA and AG acknowledge support from the Swedish Research Council and the Swedish Space Board. The Oskar Klein Centre is funded by the Swedish Research Council. ML acknowledges the Centre National de la Recherche Scientifique (CNRS) for its support. JR acknowledges support from the ERC starting grant CALENDS (336736).
\end{acknowledgements}

\appendix
\section{Supernova detection efficiency}\label{sec:det_eff}
\begin{figure}[htbp]
	\begin{center}
		\includegraphics[width=\hsize]{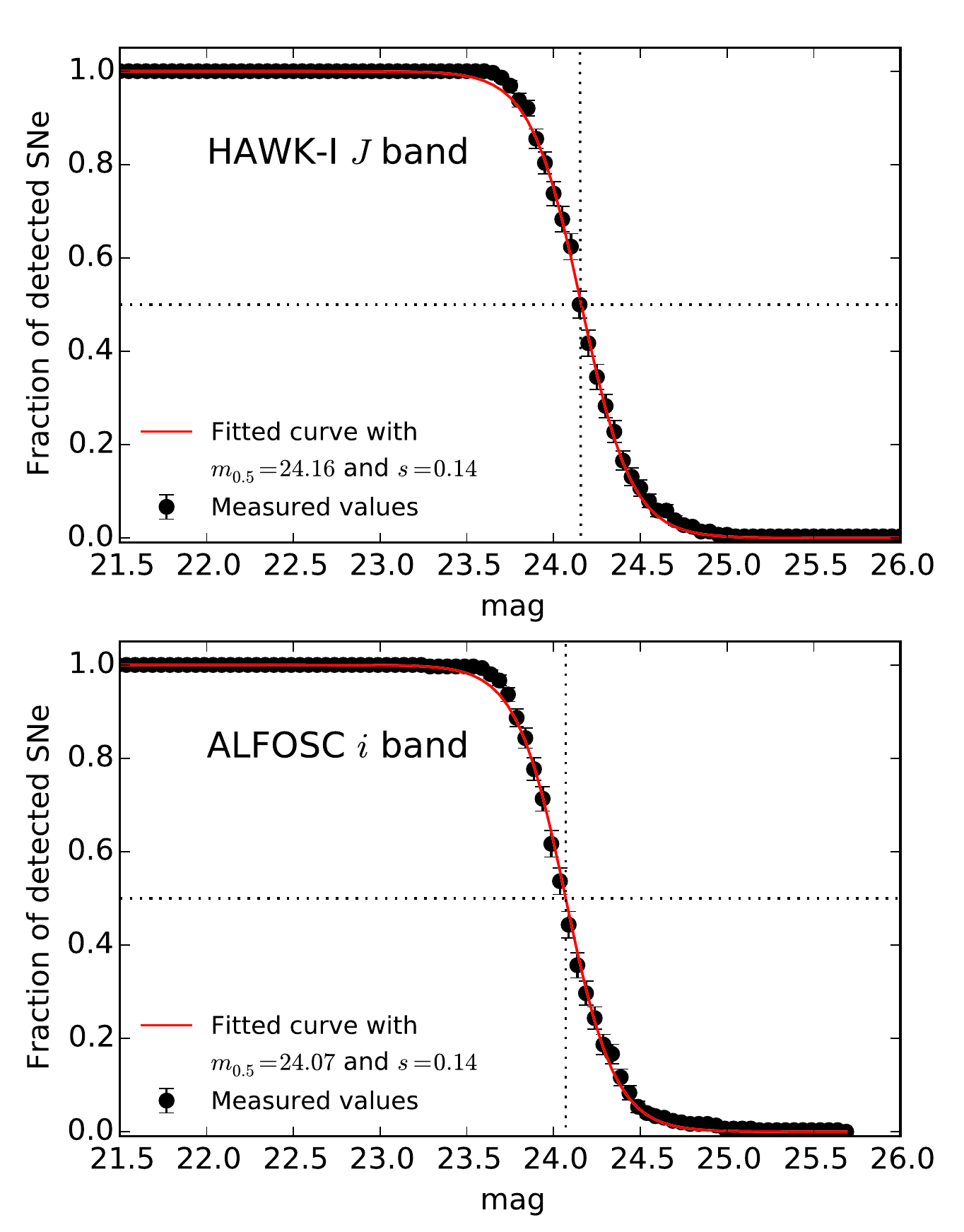}
		\caption{Example of efficiency curves of a detection of point-source transient in the HAWK-I \Jband-  and ALFOSC $i$ band data. Error bars represent $1 \sigma$ binomial uncertainties. The dotted lines indicate the magnitude at $50\%$ detection efficiency. The magnitudes are given in the Vega system. }
		\label{evaluation}
	\end{center}
\end{figure}
We estimated the efficiency of detecting transients in our surveys as a function of magnitude. In order to that we carried out
simulations by blindly adding 300 artificial \sne for each of the
combined images used in the search. The \sne were modelled with a
Moffat PSF fitted to isolated field stars in the images. We used three
different approaches to generate the positions of the artificial SNe to study how the detection efficiency is affected by the host
brightness: first, we used random positions with low background
values; second, close to galaxies; and the third in the coordinates
close to the strongly lensed galaxies. The resulting detection
efficiency did not change significantly depending on this choice,
which is most likely due to the overall bright sky foreground across the
image. After that, the simulated data went through the normal image
differencing and analysis.  For each magnitude bin, we counted the
number of successfully detected events. We performed the efficiency
simulations for each of the stacked images as used in the HAWK-I
(post-survey) and ALFOSC search, and the results are presented in
Table~\ref{HAWKI_data}~and~\ref{ALFOSC_data}, respectively.  The
efficiency curve as a function of magnitude $m$ can be parameterized
with the function
\begin{equation}
\epsilon(m) = \frac{1}{1 + e^{(m-m_{0.5} )/S}} ,
\end{equation}
where $m_{0.5}$ is the magnitude at $50\%$ detection efficiency and $S$ is the parameter that determines the exponential decline rate.  An example of an average efficiency curve for both instruments is shown in Figure~\ref{evaluation}. The mean value for $s$  is 0.14 for both instruments with a dispersion of 0.01 and 0.03 for HAWK-I and ALFOSC respectively.

The magnitude at $50\%$ detection efficiency correlates with the image depth, which is expected given that the sensitivity of the image determines the magnitude at which detection efficiency drops rapidly as follows:
\begin{equation}
\begin{split}
m_{0.5,HAWK\,I-J} =  m_{lim}-0.14, \\
m_{0.5,ALFOSC-i} =  m_{lim}-0.41.
\end{split}
\end{equation}
Here, we define image depth as the magnitude of a point source with a $5\sigma$ signal over the background level. In the HAWK-I mosaic images depth varies slightly due to the cross-looking gap between the four chips. HAWK-I has  average image depth of  $m_{\rm lim} = 23.8\pm 0.4$ and $m_{05} = 23.7\pm 0.5$, while  ALFOSC has $m_{\rm lim} = 24.0\pm0.6$ and $m_{05} = 23.4\pm 0.6$.

\

\section{Photometry}
Here we present the multi-band photometry from several instruments for the \sne presented in this work. For each observation, we list the modified Julian dates, the filter, and the measured flux with its error. For conversion to magnitude, the zero point is given. We note that the photometry of CAND-821 and CAND-ISAAC are provided in \citet{2011ApJ...742L...7A} and G09, respectively. 
\begin{table}
	\caption{Photometry of the CC~SN candidates behind A1689.}
\begin{center}
\begin{tabular}{lcccc}

	\hline	\hline
	{\small MJD} & {\small Filter}  & {\small Flux}  &{\small Flux} & {\small Zero point}\\
	
	{\small (days)} &  &  & error&{\small(mag)}\\

	\hline
	\multicolumn{4}{c}{\small CAND-1392}\\
	\hline
	{\small $ 55387.9 $} &$i$& {\small $ -320 $}&840  & 32.51 \\ 
	{\small $ 55564.17 $} & $i$ & {\small $ 3300$}& 400& $\cdots$ \\ 
	{\small $ 55599.12 $} & $ i $ & {\small $ 630 $}&310  & $\cdots$ \\ 
	{\small $ 55604.11 $} & $ i $ & {\small $ 1400 $}&1000  & $\cdots$ \\ 
	{\small $ 55650.11 $} & $ i $ & {\small $ -550 $}&380  & $\cdots$ \\ 
	 
	{\small $ 55387.98 $} & $ g $ & {\small $ -690 $}&1040 & 32.70 \\ 
	{\small $ 55564.22 $} & $ g $ & {\small $ 600 $}&200  & $\cdots$ \\ 
	{\small $ 55599.19 $} & $ g $ & {\small $ 370 $}&210  & $\cdots$ \\ 
	{\small $ 55650.03 $} & $ g $ & {\small $ -10 $}&230  & $\cdots$ \\ 
	
	{\small $ 55387.95 $} & $ r $ & {\small $ -270 $}&630  & 32.20 \\ 
	{\small $ 55564.24 $} & $ r $ & {\small $ 1200 $}&200  & $\cdots$ \\ 
	{\small $ 55599.17 $} & $ r $ & {\small $ -20 $}&150  & $\cdots$ \\ 
	{\small $ 55604.16 $} & $ r $ & {\small $ -200 $}&300  & $\cdots$ \\ 
	{\small $ 55650.08 $} & $ r $ & {\small $ 140 $}&290  & $\cdots$ \\ 
	
	\hline
	\multicolumn{4}{c}{\small CAND-669}\\
	\hline
{\small $ 54943.0 $} & $ i $ & {\small $ 1300 $}&1300  & $32.86$ \\ 
{\small $ 54987.9 $} & $ i $ & {\small $ 240 $}&600  & $\cdots$ \\ 
{\small $ 55006.92 $} & $ i $ & {\small $ 330 $}&490  & $\cdots$ \\ 

{\small $ 54862.26 $} & $ J $ & {\small $ 30 $}&90 & $30.67$ \\ 
{\small $ 54892.23 $} & $ J $ & {\small $ 1540 $}&90  & $\cdots$ \\ 
{\small $ 54914.32 $} & $ J $ & {\small $ 1160 $}&100  & $\cdots$ \\ 
{\small $ 54916.19 $} & $ J $ & {\small $ 1210 $}&80  & $\cdots$ \\ 
{\small $ 54949.1 $} & $ J $ & {\small $ 670 $}&80  & $\cdots$ \\ 
{\small $ 54990.11 $} & $ J $ & {\small $ 380 $}&130  & $\cdots$ \\ 
{\small $ 55016.05 $} & $ J $ & {\small $ 220 $}&160  & $\cdots$ \\ 
{\small $ 55018.1 $} & $ J $ & {\small $ 360 $}&150  & $\cdots$ \\ 
{\small $ 55019.97 $} & $ J $ & {\small $ 240 $}&250  & $\cdots$ \\ 

{\small $ 54919.16 $} & $ NB1060 $ & {\small $ 250 $}&110 & $30.37$ \\ 
{\small $ 54923.3 $} & $ NB1060 $ & {\small $ 250 $}&100  & $\cdots$ \\ 
{\small $ 54948.99 $} & $ NB1060 $ & {\small $ 170 $}&150  & $\cdots$ \\ 
{\small $ 54966.0 $} & $ NB1060 $ & {\small $ 160 $}&110  & $\cdots$ \\ 
{\small $ 54968.99 $} & $ NB1060 $ & {\small $ 0 $}&160  & $\cdots$ \\ 
{\small $ 54979.07 $} & $ NB1060 $ & {\small $ 60 $}&120  & $\cdots$ \\ 
{\small $ 55008.98 $} & $ NB1060 $ & {\small $ 0 $}&210  & $\cdots$ \\ 
{\small $ 55009.96 $} & $ NB1060 $ & {\small $ -110 $}&110  & $\cdots$ \\ 
{\small $ 55028.0 $} & $ NB1060 $ & {\small $ 40 $}&110 & $\cdots$ \\ 
{\small $ 55035.01 $} & $ NB1060 $ & {\small $ 140 $}& 260  & $\cdots$ \\ 
{\small $ 55054.97 $} & $ NB1060 $ & {\small $ 220 $}&150  & $\cdots$ \\ 
{\small $ 55056.0 $} & $ NB1060 $ & {\small $ 320 $}&140  & $\cdots$ \\ 
		\hline
		\multicolumn{3}{c}{\small CAND-10658}\\
		\hline
{\small $ 56366.26 $} & $ J $ & {\small $ 30 $}&30  & $29.87$ \\ 
{\small $ 56415.91 $} & $ J $ & {\small $ -30 $}&30  & $\cdots$ \\ 
{\small $ 56450.79 $} & $ J $ & {\small $ -10 $}&40  & $\cdots$ \\ 
{\small $ 56728.78 $} & $ J $ & {\small $ 1070 $}&40  & $\cdots$ \\ 
{\small $ 56796.06 $} & $ J $ & {\small $ 1450 $}&50 & $\cdots$ \\ 
{\small $ 56828.10 $} & $ J $ & {\small $ 570 $}&160  & $\cdots$ \\ 
{\small $ 56858.99 $} & $ J $ & {\small $ 370 $}&40  & $\cdots$ \\  
	\hline
	
	\multicolumn{3}{c}{\small CAND-10662}\\
	\hline
{\small $ 56366.26 $} & $ J $ & {\small $ 20$}& 20  & $29.60$ \\ 
{\small $ 56383.26 $} & $ J $ & {\small $ 0 $}&20  & $\cdots$ \\ 
{\small $ 56415.91 $} & $ J $ & {\small $ -10 $}&20  & $\cdots$ \\ 
{\small $ 56450.79 $} & $ J $ & {\small $ 50 $}&30  & $\cdots$ \\ 
{\small $ 56728.78 $} & $ J $ & {\small $ 230 $}&30  & $\cdots$ \\ 
{\small $ 56796.06 $} & $ J $ & {\small $ 280 $}&40  & $\cdots$ \\ 
{\small $ 56828.1 $} & $ J $ & {\small $ 170 $}&120  & $\cdots$ \\ 
{\small $ 56858.99 $} & $ J $ & {\small $ 170 $}&40  & $\cdots$ \\ 
	
		\hline

	\end{tabular}
\end{center}
\end{table}	

\begin{table}
\caption{Photometry of the \snia candidates.}
\begin{center}
\begin{tabular}{lcccc}
	\hline
	\hline
	{\small MJD} & Filter & {\small Flux} &{\small Flux}  & Zero point \\
	
	\multicolumn{1}{l}{\small (days)} & \multicolumn{1}{c}{\small }  &&{\small error}  &\multicolumn{1}{c}{\small (mag)}\\
	\hline
	\multicolumn{4}{c}{\small CAND-1208}\\
	\hline
{\small $ 55273.14 $} & $ i $ & {\small -1500 } & 1500  & $32.50$ \\ 
{\small $ 55333.98 $} & $ i $ & {\small $ 108800 $}&500 & $\cdots$ \\ 
{\small $ 55337.0 $} & $ i $ & {\small $ 92600 $}&2100  & $\cdots$ \\ 
{\small $ 55359.97 $} & $ i $ & {\small $ 49000 $}&1000  & $\cdots$ \\ 
{\small $ 55387.9 $} & $ i $ & {\small $ 19100 $}&1200  & $\cdots$ \\ 
{\small $ 55564.17 $} & $ i $ & {\small $ -350 $}&430  & $\cdots$ \\ 

{\small $ 55337.96 $} & $ g $ & {\small $ 46800 $}&400 & $32.67$ \\ 
{\small $ 55359.92 $} & $ g $ & {\small $ 8200 $}&200  & $\cdots$ \\ 
{\small $ 55387.98 $} & $ g $ & {\small $ 4500 $}&1200  & $\cdots$ \\ 

{\small $ 55273.09 $} & $ r $ & {\small $ -620 $}&370  & $32.22$ \\ 
{\small $ 55333.94 $} & $ r $ & {\small $ 76400 $}&400  & $\cdots$ \\ 
{\small $ 55336.95 $} & $ r $ & {\small $ 62800 $}&400  & $\cdots$ \\ 
{\small $ 55359.94 $} & $ r $ & {\small $ 22000 $}&500  & $\cdots$ \\ 
{\small $ 55387.95 $} & $ r $ & {\small $ 10000 $}&1000  & $\cdots$ \\ 
{\small $ 55564.24 $} & $ r $ & {\small $ -210 $}&430  & $\cdots$ \\ 
	
{\small $ 55382.41 $} & $ F814W $& {\small $ 1420 $}&30  & $30.0$ \\ 
{\small $ 55382.82 $} & $ F814W $& {\small $ 1410 $}&30  & $\cdots$ \\ 
{\small $ 55383.40 $} & $ F814W $& {\small $ 1380 $}&30  & $\cdots$ \\ 
{\small $ 55384.59 $} & $ F814W $& {\small $ 1370 $}&30  & $\cdots$ \\ 
{\small $ 55372.16 $} & $ F814W $& {\small $ 1990 $}&40  & $\cdots$ \\ 
{\small $ 55385.73 $} & $ F814W $& {\small $ 1310 $}&30  & $\cdots$ \\ 
	\hline
	\multicolumn{4}{c}{\small CAND-802}\\
	\hline

{\small $ 54949.16 $} & $ J $ & {\small $ 10 $}&30  & $29.58$ \\ 
{\small $ 54990.13 $} & $ J $ & {\small $ 680 $}&50  & $\cdots$ \\ 
{\small $ 55018.06 $} & $ J $ & {\small $ 620 $}&40  & $\cdots$ \\ 
{\small $ 56366.26 $} & $ J $ & {\small $ -20 $}&30  & $\cdots$ \\ 

{\small $ 54943.0 $} & $ i $ & {\small $ 260 $}&780  & $31.96$ \\ 
{\small $ 54987.9 $} & $ i $ & {\small $ 1700$}& 300  & $\cdots$ \\ 
{\small $ 55006.92 $} & $ i $ & {\small $ 2100 $}&200  & $\cdots$ \\ 

{\small $ 55054.97 $} & $ NB1060 $ & {\small $ 640 $}&80  & $30.41$ \\ 

	\hline
	\multicolumn{4}{c}{\small CAND-7169}\\
	\hline
{\small $ 56366.26 $} & $ J $ & {\small $ 100 $}&140  & $29.60$ \\ 
{\small $ 56381.26 $} & $ J $ & {\small $ 230 $}&140  & $\cdots$ \\ 
{\small $ 56383.27 $} & $ J $ & {\small $ 180 $}&150  & $\cdots$ \\ 
{\small $ 56385.26 $} & $ J $ & {\small $ 190 $}&140  & $\cdots$ \\ 
{\small $ 56408.29 $} & $ J $ & {\small $ 2600 $}&300  & $\cdots$ \\ 
{\small $ 56419.21 $} & $ J $ & {\small $ 1700 $}&200  & $\cdots$ \\ 
{\small $ 56420.24 $} & $ J $ & {\small $ 1700 $}&200  & $\cdots$ \\ 
{\small $ 56448.14 $} & $ J $ & {\small $ 1200 $}&130  & $\cdots$ \\ 
{\small $ 56449.17 $} & $ J $ & {\small $ 1190 $}&120 & $\cdots$ \\

	\hline

\end{tabular}
\end{center}
\end{table}

\clearpage


\begin{thebibliography}{108}
\expandafter\ifx\csname natexlab\endcsname\relax\def\natexlab#1{#1}\fi

\bibitem[{{Alard}(2000)}]{Alardi2}
{Alard}, C. 2000, \aaps, 144, 363

\bibitem[{{Alard} \& {Lupton}(1998)}]{Alard1}
{Alard}, C. \& {Lupton}, R.~H. 1998, \apj, 503, 325

\bibitem[{{Alavi} {et~al.}(2014){Alavi}, {Siana}, {Richard}, {Stark},
  {Scarlata}, {Teplitz}, {Freeman}, {Dominguez}, {Rafelski}, {Robertson}, \&
  {Kewley}}]{2014ApJ...780..143A}
{Alavi}, A., {Siana}, B., {Richard}, J., {et~al.} 2014, \apj, 780, 143

\bibitem[{{Amanullah} {et~al.}(2011{\natexlab{a}}){Amanullah}, {Goobar},
  {Cl{\'e}ment}, {Cuby}, {Dahle}, {Dahl{\'e}n}, {Hjorth}, {Fabbro},
  {J{\"o}nsson}, {Kneib}, {Lidman}, {Limousin}, {Milvang-Jensen},
  {M{\"o}rtsell}, {Nordin}, {Paech}, {Richard}, {Riehm}, {Stanishev}, \&
  {Watson}}]{2011ApJ...742L...7A}
{Amanullah}, R., {Goobar}, A., {Cl{\'e}ment}, B., {et~al.} 2011{\natexlab{a}},
  \apjl, 742, L7

\bibitem[{{Amanullah} {et~al.}(2011{\natexlab{b}}){Amanullah}, {Goobar},
  {Johansson}, {Joensson}, {Moertsell}, {Nordin}, {Paech}, {Riehm}, \&
  {Stanishev}}]{2011CBET.2642....1A}
{Amanullah}, R., {Goobar}, A., {Johansson}, J., {et~al.} 2011{\natexlab{b}},
  Central Bureau Electronic Telegrams, 2642, 1

\bibitem[{{Amanullah} {et~al.}(2008){Amanullah}, {Stanishev}, {Goobar},
  {Schahmaneche}, {Astier}, {Balland}, {Ellis}, {Fabbro}, {Hardin}, {Hook},
  {Irwin}, {McMahon}, {Mendez}, {Mouchet}, {Pain}, {Ruiz-Lapuente}, \&
  {Walton}}]{2008A&A...486..375A}
{Amanullah}, R., {Stanishev}, V., {Goobar}, A., {et~al.} 2008, \aap, 486, 375

\bibitem[{{Anderson} {et~al.}(2014){Anderson}, {Gonz{\'a}lez-Gait{\'a}n},
  {Hamuy}, {Guti{\'e}rrez}, {Stritzinger}, {Olivares E.}, {Phillips},
  {Schulze}, {Antezana}, {Bolt}, {Campillay}, {Castell{\'o}n}, {Contreras}, {de
  Jaeger}, {Folatelli}, {F{\"o}rster}, {Freedman}, {Gonz{\'a}lez}, {Hsiao},
  {Krzemi{\'n}ski}, {Krisciunas}, {Maza}, {McCarthy}, {Morrell}, {Persson},
  {Roth}, {Salgado}, {Suntzeff}, \& {Thomas-Osip}}]{2014ApJ...786...67A}
{Anderson}, J.~P., {Gonz{\'a}lez-Gait{\'a}n}, S., {Hamuy}, M., {et~al.} 2014,
  \apj, 786, 67

\bibitem[{{Astier} {et~al.}(2006){Astier}, {Guy}, {Regnault}, {Pain},
  {Aubourg}, {Balam}, {Basa}, {Carlberg}, {Fabbro}, {Fouchez}, {Hook},
  {Howell}, {Lafoux}, {Neill}, {Palanque-Delabrouille}, {Perrett}, {Pritchet},
  {Rich}, {Sullivan}, {Taillet}, {Aldering}, {Antilogus}, {Arsenijevic},
  {Balland}, {Baumont}, {Bronder}, {Courtois}, {Ellis}, {Filiol}, {Gon{\c
  c}alves}, {Goobar}, {Guide}, {Hardin}, {Lusset}, {Lidman}, {McMahon},
  {Mouchet}, {Mourao}, {Perlmutter}, {Ripoche}, {Tao}, \&
  {Walton}}]{2006A&A...447...31A}
{Astier}, P., {Guy}, J., {Regnault}, N., {et~al.} 2006, \aap, 447, 31

\bibitem[{{Barbary} {et~al.}(2012){Barbary}, {Aldering}, {Amanullah},
  {Brodwin}, {Connolly}, {Dawson}, {Doi}, {Eisenhardt}, {Faccioli}, {Fadeyev},
  {Fakhouri}, {Fruchter}, {Gilbank}, {Gladders}, {Goldhaber}, {Goobar},
  {Hattori}, {Hsiao}, {Huang}, {Ihara}, {Kashikawa}, {Koester}, {Konishi},
  {Kowalski}, {Lidman}, {Lubin}, {Meyers}, {Morokuma}, {Oda}, {Panagia},
  {Perlmutter}, {Postman}, {Ripoche}, {Rosati}, {Rubin}, {Schlegel},
  {Spadafora}, {Stanford}, {Strovink}, {Suzuki}, {Takanashi}, {Tokita},
  {Yasuda}, \& {Supernova Cosmology Project}}]{2012ApJ...745...32B}
{Barbary}, K., {Aldering}, G., {Amanullah}, R., {et~al.} 2012, \apj, 745, 32

\bibitem[{{Bazin} {et~al.}(2009){Bazin}, {Palanque-Delabrouille}, {Rich},
  {Ruhlmann-Kleider}, {Aubourg}, {Le Guillou}, {Astier}, {Balland}, {Basa},
  {Carlberg}, {Conley}, {Fouchez}, {Guy}, {Hardin}, {Hook}, {Howell}, {Pain},
  {Perrett}, {Pritchet}, {Regnault}, {Sullivan}, {Antilogus}, {Arsenijevic},
  {Baumont}, {Fabbro}, {Le Du}, {Lidman}, {Mouchet}, {Mour{\~a}o}, \&
  {Walker}}]{2009A&A...499..653B}
{Bazin}, G., {Palanque-Delabrouille}, N., {Rich}, J., {et~al.} 2009, \aap, 499,
  653

\bibitem[{{Bell} {et~al.}(2003){Bell}, {McIntosh}, {Katz}, \&
  {Weinberg}}]{2003ApJS..149..289B}
{Bell}, E.~F., {McIntosh}, D.~H., {Katz}, N., \& {Weinberg}, M.~D. 2003, \apjs,
  149, 289

\bibitem[{{Bellm}(2014)}]{2014htu..conf...27B}
{Bellm}, E. 2014, in The Third Hot-wiring the Transient Universe Workshop, ed.
  P.~R. {Wozniak}, M.~J. {Graham}, A.~A. {Mahabal}, \& R.~{Seaman}, 27--33

\bibitem[{{Bertin}(2006)}]{2006ASPC..351..112B}
{Bertin}, E. 2006, in Astronomical Society of the Pacific Conference Series,
  Vol. 351, Astronomical Data Analysis Software and Systems XV, ed.
  C.~{Gabriel}, C.~{Arviset}, D.~{Ponz}, \& S.~{Enrique}, 112

\bibitem[{{Bertin} \& {Arnouts}(1996)}]{1996A&AS..117..393B}
{Bertin}, E. \& {Arnouts}, S. 1996, \aaps, 117, 393

\bibitem[{{Bertin} {et~al.}(2002){Bertin}, {Mellier}, {Radovich}, {Missonnier},
  {Didelon}, \& {Morin}}]{2002ASPC..281..228B}
{Bertin}, E., {Mellier}, Y., {Radovich}, M., {et~al.} 2002, in Astronomical
  Society of the Pacific Conference Series, Vol. 281, Astronomical Data
  Analysis Software and Systems XI, ed. D.~A. {Bohlender}, D.~{Durand}, \&
  T.~H. {Handley}, 228

\bibitem[{{Blondin} \& {Tonry}(2007)}]{2007ApJ...666.1024B}
{Blondin}, S. \& {Tonry}, J.~L. 2007, \apj, 666, 1024

\bibitem[{{Bolton} \& {Burles}(2003)}]{2003ApJ...592...17B}
{Bolton}, A.~S. \& {Burles}, S. 2003, \apj, 592, 17

\bibitem[{{Bolzonella} {et~al.}(2000){Bolzonella}, {Miralles}, \&
  {Pell{\'o}}}]{2000A&A...363..476B}
{Bolzonella}, M., {Miralles}, J.-M., \& {Pell{\'o}}, R. 2000, \aap, 363, 476

\bibitem[{{Botticella} {et~al.}(2008){Botticella}, {Riello}, {Cappellaro},
  {Benetti}, {Altavilla}, {Pastorello}, {Turatto}, {Greggio}, {Patat},
  {Valenti}, {Zampieri}, {Harutyunyan}, {Pignata}, \&
  {Taubenberger}}]{2008A&A...479...49B}
{Botticella}, M.~T., {Riello}, M., {Cappellaro}, E., {et~al.} 2008, \aap, 479,
  49

\bibitem[{{Botticella} {et~al.}(2012){Botticella}, {Smartt}, {Kennicutt},
  {Cappellaro}, {Sereno}, \& {Lee}}]{2012A&A...537A.132B}
{Botticella}, M.~T., {Smartt}, S.~J., {Kennicutt}, R.~C., {et~al.} 2012, \aap,
  537, A132

\bibitem[{{Broadhurst} {et~al.}(2005){Broadhurst}, {Ben{\'{\i}}tez}, {Coe},
  {Sharon}, {Zekser}, {White}, {Ford}, {Bouwens}, {Blakeslee}, {Clampin},
  {Cross}, {Franx}, {Frye}, {Hartig}, {Illingworth}, {Infante}, {Menanteau},
  {Meurer}, {Postman}, {Ardila}, {Bartko}, {Brown}, {Burrows}, {Cheng},
  {Feldman}, {Golimowski}, {Goto}, {Gronwall}, {Herranz}, {Holden}, {Homeier},
  {Krist}, {Lesser}, {Martel}, {Miley}, {Rosati}, {Sirianni}, {Sparks},
  {Steindling}, {Tran}, {Tsvetanov}, \& {Zheng}}]{2005ApJ...621...53B}
{Broadhurst}, T., {Ben{\'{\i}}tez}, N., {Coe}, D., {et~al.} 2005, \apj, 621, 53

\bibitem[{{Burns} {et~al.}(2011){Burns}, {Stritzinger}, {Phillips}, {Kattner},
  {Persson}, {Madore}, {Freedman}, {Boldt}, {Campillay}, {Contreras},
  {Folatelli}, {Gonzalez}, {Krzeminski}, {Morrell}, {Salgado}, \&
  {Suntzeff}}]{2011AJ....141...19B}
{Burns}, C.~R., {Stritzinger}, M., {Phillips}, M.~M., {et~al.} 2011, \aj, 141,
  19

\bibitem[{{Calzetti} {et~al.}(2000){Calzetti}, {Armus}, {Bohlin}, {Kinney},
  {Koornneef}, \& {Storchi-Bergmann}}]{2000ApJ...533..682C}
{Calzetti}, D., {Armus}, L., {Bohlin}, R.~C., {et~al.} 2000, \apj, 533, 682

\bibitem[{{Cappellaro} {et~al.}(2015){Cappellaro}, {Botticella}, {Pignata},
  {Grado}, {Greggio}, {Limatola}, {Vaccari}, {Baruffolo}, {Benetti}, {Bufano},
  {Capaccioli}, {Cascone}, {Covone}, {De Cicco}, {Falocco}, {Della Valle},
  {Jarvis}, {Marchetti}, {Napolitano}, {Paolillo}, {Pastorello}, {Radovich},
  {Schipani}, {Spiro}, {Tomasella}, \& {Turatto}}]{SUDARE}
{Cappellaro}, E., {Botticella}, M.~T., {Pignata}, G., {et~al.} 2015, \aap, 584, A62

\bibitem[{{Cappellaro} {et~al.}(1999){Cappellaro}, {Evans}, \&
  {Turatto}}]{Cappellaro}
{Cappellaro}, E., {Evans}, R., \& {Turatto}, M. 1999, \aap, 351, 459

\bibitem[{{Cardelli} {et~al.}(1989){Cardelli}, {Clayton}, \&
  {Mathis}}]{1989ApJ...345..245C}
{Cardelli}, J.~A., {Clayton}, G.~C., \& {Mathis}, J.~S. 1989, \apj, 345, 245

\bibitem[{{Casali} {et~al.}(2006){Casali}, {Pirard}, {Kissler-Patig},
  {Moorwood}, {Bedin}, {Biereichel}, {Delabre}, {Dorn}, {Finger}, {Gojak},
  {Huster}, {Jung}, {Koch}, {Lizon}, {Mehrgan}, {Pozna}, {Silber}, {Sokar}, \&
  {Stegmeier}}]{2006SPIE.6269E..0WC}
{Casali}, M., {Pirard}, J.-F., {Kissler-Patig}, M., {et~al.} 2006, in Society
  of Photo-Optical Instrumentation Engineers (SPIE) Conference Series, Vol.
  6269, Society of Photo-Optical Instrumentation Engineers (SPIE) Conference
  Series, 0

\bibitem[{{Chatzopoulos} {et~al.}(2011){Chatzopoulos}, {Wheeler}, {Vinko},
  {Quimby}, {Robinson}, {Miller}, {Foley}, {Perley}, {Yuan}, {Akerlof}, \&
  {Bloom}}]{2011ApJ...729..143C}
{Chatzopoulos}, E., {Wheeler}, J.~C., {Vinko}, J., {et~al.} 2011, \apj, 729,
  143

\bibitem[{{Coe} {et~al.}(2010){Coe}, {Ben{\'{\i}}tez}, {Broadhurst}, \&
  {Moustakas}}]{2010ApJ...723.1678C}
{Coe}, D., {Ben{\'{\i}}tez}, N., {Broadhurst}, T., \& {Moustakas}, L.~A. 2010,
  \apj, 723, 1678

\bibitem[{{Czoske}(2004)}]{2004ogci.conf..183C}
{Czoske}, O. 2004, in IAU Colloq. 195: Outskirts of Galaxy Clusters: Intense
  Life in the Suburbs, ed. A.~{Diaferio}, 183--187

\bibitem[{{Dahlen} {et~al.}(2008){Dahlen}, {Strolger}, \&
  {Riess}}]{2008ApJ...681..462D}
{Dahlen}, T., {Strolger}, L.-G., \& {Riess}, A.~G. 2008, \apj, 681, 462

\bibitem[{{Dahlen} {et~al.}(2012){Dahlen}, {Strolger}, {Riess}, {Mattila},
  {Kankare}, \& {Mobasher}}]{2012ApJ...757...70D}
{Dahlen}, T., {Strolger}, L.-G., {Riess}, A.~G., {et~al.} 2012, \apj, 757, 70

\bibitem[{{Dahlen} {et~al.}(2004){Dahlen}, {Strolger}, {Riess}, {Mobasher},
  {Chary}, {Conselice}, {Ferguson}, {Fruchter}, {Giavalisco}, {Livio}, {Madau},
  {Panagia}, \& {Tonry}}]{2004ApJ...613..189D}
{Dahlen}, T., {Strolger}, L.-G., {Riess}, A.~G., {et~al.} 2004, \apj, 613, 189

\bibitem[{{Di Carlo} {et~al.}(2002){Di Carlo}, {Massi}, {Valentini}, {Di
  Paola}, {D'Alessio}, {Brocato}, {Guidubaldi}, {Dolci}, {Pedichini},
  {Speziali}, {Li Causi}, {Caratti o Garatti}, {Cappellaro}, {Turatto},
  {Arkharov}, {Gnedin}, {Larionov}, {Benetti}, {Pastorello}, {Aretxaga},
  {Chavushyan}, {Vega}, {Danziger}, \& {Tornamb{\'e}}}]{2002ApJ...573..144D}
{Di Carlo}, E., {Massi}, F., {Valentini}, G., {et~al.} 2002, \apj, 573, 144

\bibitem[{{Diego} {et~al.}(2015){Diego}, {Broadhurst}, {Benitez}, {Umetsu},
  {Coe}, {Sendra}, {Sereno}, {Izzo}, \& {Covone}}]{2015MNRAS.446..683D}
{Diego}, J.~M., {Broadhurst}, T., {Benitez}, N., {et~al.} 2015, \mnras, 446,
  683

\bibitem[{{Dilday} {et~al.}(2010){Dilday}, {Smith}, {Bassett}, {Becker},
  {Bender}, {Castander}, {Cinabro}, {Filippenko}, {Frieman}, {Galbany},
  {Garnavich}, {Goobar}, {Hopp}, {Ihara}, {Jha}, {Kessler}, {Lampeitl},
  {Marriner}, {Miquel}, {Moll{\'a}}, {Nichol}, {Nordin}, {Riess}, {Sako},
  {Schneider}, {Sollerman}, {Wheeler}, {{\"O}stman}, {Bizyaev}, {Brewington},
  {Malanushenko}, {Malanushenko}, {Oravetz}, {Pan}, {Simmons}, \&
  {Snedden}}]{2010ApJ...713.1026D}
{Dilday}, B., {Smith}, M., {Bassett}, B., {et~al.} 2010, \apj, 713, 1026

\bibitem[{{Fabbro}(2001)}]{Sebastian}
{Fabbro}, S. 2001, PhD thesis, Universit\'{e} Denis Diderot, Paris, France

\bibitem[{{Fassia} {et~al.}(2000){Fassia}, {Meikle}, {Vacca}, {Kemp}, {Walton},
  {Pollacco}, {Smartt}, {Oscoz}, {Arag{\'o}n-Salamanca}, {Bennett}, {Hawarden},
  {Alonso}, {Alcalde}, {Pedrosa}, {Telting}, {Arevalo}, {Deeg}, {Garz{\'o}n},
  {G{\'o}mez-Rold{\'a}n}, {G{\'o}mez}, {Guti{\'e}rrez}, {L{\'o}pez}, {Rozas},
  {Serra-Ricart}, \& {Zapatero-Osorio}}]{2000MNRAS.318.1093F}
{Fassia}, A., {Meikle}, W.~P.~S., {Vacca}, W.~D., {et~al.} 2000, \mnras, 318,
  1093

\bibitem[{{Frye} {et~al.}(2012){Frye}, {Hurley}, {Bowen}, {Meurer}, {Sharon},
  {Straughn}, {Coe}, {Broadhurst}, \& {Guhathakurta}}]{2012ApJ...754...17F}
{Frye}, B.~L., {Hurley}, M., {Bowen}, D.~V., {et~al.} 2012, \apj, 754, 17

\bibitem[{{Gal-Yam} {et~al.}(2002){Gal-Yam}, {Maoz}, \&
  {Sharon}}]{2002MNRAS.332...37G}
{Gal-Yam}, A., {Maoz}, D., \& {Sharon}, K. 2002, \mnras, 332, 37

\bibitem[{{Gardner} {et~al.}(2006){Gardner}, {Mather}, {Clampin}, {Doyon},
  {Greenhouse}, {Hammel}, {Hutchings}, {Jakobsen}, {Lilly}, {Long}, {Lunine},
  {McCaughrean}, {Mountain}, {Nella}, {Rieke}, {Rieke}, {Rix}, {Smith},
  {Sonneborn}, {Stiavelli}, {Stockman}, {Windhorst}, \&
  {Wright}}]{2006SSRv..123..485G}
{Gardner}, J.~P., {Mather}, J.~C., {Clampin}, M., {et~al.} 2006, \ssr, 123, 485

\bibitem[{{Goobar} \& {Leibundgut}(2011)}]{2011ARNPS..61..251G}
{Goobar}, A. \& {Leibundgut}, B. 2011, Annual Review of Nuclear and Particle
  Science, 61, 251

\bibitem[{{Goobar} {et~al.}(2009){Goobar}, {Paech}, {Stanishev}, {Amanullah},
  {Dahl{\'e}n}, {J{\"o}nsson}, {Kneib}, {Lidman}, {Limousin}, {M{\"o}rtsell},
  {Nobili}, {Richard}, {Riehm}, \& {von Strauss}}]{Second}
{Goobar}, A., {Paech}, K., {Stanishev}, V., {et~al.} 2009, \aap, 507, 71

\bibitem[{{Graham} {et~al.}(2008){Graham}, {Pritchet}, {Sullivan}, {Gwyn},
  {Neill}, {Hsiao}, {Astier}, {Balam}, {Balland}, {Basa}, {Carlberg}, {Conley},
  {Fouchez}, {Guy}, {Hardin}, {Hook}, {Howell}, {Pain}, {Perrett}, {Regnault},
  {Baumont}, {LeDu}, {Lidman}, {Perlmutter}, {Ripoche}, {Suzuki}, {Walker}, \&
  {Zhang}}]{2008AJ....135.1343G}
{Graham}, M.~L., {Pritchet}, C.~J., {Sullivan}, M., {et~al.} 2008, \aj, 135,
  1343

\bibitem[{{Graur} {et~al.}(2015){Graur}, {Bianco}, \&
  {Modjaz}}]{2015MNRAS.450..905G}
{Graur}, O., {Bianco}, F.~B., \& {Modjaz}, M. 2015, \mnras, 450, 905

\bibitem[{{Graur} {et~al.}(2011){Graur}, {Poznanski}, {Maoz}, {Yasuda},
  {Totani}, {Fukugita}, {Filippenko}, {Foley}, {Silverman}, {Gal-Yam},
  {Horesh}, \& {Jannuzi}}]{Graur11}
{Graur}, O., {Poznanski}, D., {Maoz}, D., {et~al.} 2011, \mnras, 417, 916

\bibitem[{{Grillo} {et~al.}(2015){Grillo}, {Karman}, {Suyu}, {Rosati},
  {Balestra}, {Mercurio}, {Lombardi}, {Treu}, {Caminha}, {Halkola}, {Rodney},
  {Gavazzi}, \& {Caputi}}]{2015arXiv151104093G}
{Grillo}, C., {Karman}, W., {Suyu}, S.~H., {et~al.} 2016, \apj, 822, 78

\bibitem[{{Gunnarsson} \& {Goobar}(2003)}]{2003A&A...405..859G}
{Gunnarsson}, C. \& {Goobar}, A. 2003, \aap, 405, 859

\bibitem[{{Harrison} \& {Noonan}(1979)}]{1979ApJ...232...18H}
{Harrison}, E.~R. \& {Noonan}, T.~W. 1979, \apj, 232, 18

\bibitem[{{Hatano} {et~al.}(1998){Hatano}, {Branch}, \&
  {Deaton}}]{1998ApJ...502..177H}
{Hatano}, K., {Branch}, D., \& {Deaton}, J. 1998, \apj, 502, 177

\bibitem[{{Holz}(2001)}]{2001ApJ...556L..71H}
{Holz}, D.~E. 2001, \apjl, 556, L71

\bibitem[{{Horiuchi} {et~al.}(2011){Horiuchi}, {Beacom}, {Kochanek}, {Prieto},
  {Stanek}, \& {Thompson}}]{2011ApJ...738..154H}
{Horiuchi}, S., {Beacom}, J.~F., {Kochanek}, C.~S., {et~al.} 2011, \apj, 738,
  154

\bibitem[{{Hsiao} {et~al.}(2007){Hsiao}, {Conley}, {Howell}, {Sullivan},
  {Pritchet}, {Carlberg}, {Nugent}, \& {Phillips}}]{2007ApJ...663.1187H}
{Hsiao}, E.~Y., {Conley}, A., {Howell}, D.~A., {et~al.} 2007, \apj, 663, 1187

\bibitem[{{Jullo} {et~al.}(2007){Jullo}, {Kneib}, {Limousin},
  {El{\'{\i}}asd{\'o}ttir}, {Marshall}, \& {Verdugo}}]{2007NJPh....9..447J}
{Jullo}, E., {Kneib}, J.-P., {Limousin}, M., {et~al.} 2007, New Journal of
  Physics, 9, 447

\bibitem[{{Kelly} {et~al.}(2015){Kelly}, {Rodney}, {Treu},
  {Foley}, {Brammer}, {Schmidt}, {Zitrin}, {Sonnenfeld}, {Strolger}, {Graur},
  {Filippenko}, {Jha}, {Riess}, {Bradac}, {Weiner}, {Scolnic}, {Malkan}, {von
  der Linden}, {Trenti}, {Hjorth}, {Gavazzi}, {Fontana}, {Merten}, {McCully},
  {Jones}, {Postman}, {Dressler}, {Patel}, {Cenko}, {Graham}, \&
  {Tucker}}]{2015Sci...347.1123K}
{Kelly}, P.~L., {Rodney}, S.~A., {Treu}, T., {et~al.} 2015,
  Science, 347, 1123

\bibitem[{{Kelly} {et~al.}(2016){Kelly}, {Rodney}, {Treu},
  {Strolger}, {Foley}, {Jha}, {Selsing}, {Brammer}, {Bradac}, {Cenko},
  {Graham}, {Graur}, {Filippenko}, {Hjorth}, {Matheson}, {McCully}, {Molino},
  {Nonino}, {Riess}, {Schmidt}, {Tucker}, {von der Linden}, {Weiner}, \&
  {Zitrin}}]{2015arXiv151204654K}
{Kelly}, P.~L., {Rodney}, S.~A., {Treu}, T., {et~al.}, 2016, \apj, 819, L8

\bibitem[{{Kolatt} \& {Bartelmann}(1998)}]{1998MNRAS.296..763K}
{Kolatt}, T.~S. \& {Bartelmann}, M. 1998, \mnras, 296, 763

\bibitem[{{Kovner} \& {Paczynski}(1988)}]{1988ApJ...335L...9K}
{Kovner}, I. \& {Paczynski}, B. 1988, \apjl, 335, L9

\bibitem[{{Lagan{\'a}} {et~al.}(2008){Lagan{\'a}}, {Lima Neto},
  {Andrade-Santos}, \& {Cypriano}}]{2008A&A...485..633L}
{Lagan{\'a}}, T.~F., {Lima Neto}, G.~B., {Andrade-Santos}, F., \& {Cypriano},
  E.~S. 2008, \aap, 485, 633

\bibitem[{{Lemze} {et~al.}(2009){Lemze}, {Broadhurst}, {Rephaeli}, {Barkana},
  \& {Umetsu}}]{2009ApJ...701.1336L}
{Lemze}, D., {Broadhurst}, T., {Rephaeli}, Y., {Barkana}, R., \& {Umetsu}, K.
  2009, \apj, 701, 1336

\bibitem[{{Li} {et~al.}(2011){Li}, {Leaman}, {Chornock}, {Filippenko},
  {Poznanski}, {Ganeshalingam}, {Wang}, {Modjaz}, {Jha}, {Foley}, \&
  {Smith}}]{Li}
{Li}, W., {Leaman}, J., {Chornock}, R., {et~al.} 2011, \mnras, 412, 1441

\bibitem[{{Lien} \& {Fields}(2009)}]{2009JCAP...01..047L}
{Lien}, A. \& {Fields}, B.~D. 2009, \jcap, 1, 047

\bibitem[{{Limousin} {et~al.}(2007){Limousin}, {Richard}, {Jullo}, {Kneib},
  {Fort}, {Soucail}, {El{\'{\i}}asd{\'o}ttir}, {Natarajan}, {Ellis}, {Smail},
  {Czoske}, {Smith}, {Hudelot}, {Bardeau}, {Ebeling}, {Egami}, \&
  {Knudsen}}]{2007ApJ...668..643L}
{Limousin}, M., {Richard}, J., {Jullo}, E., {et~al.} 2007, \apj, 668, 643

\bibitem[{{LSST Science Collaboration} {et~al.}(2009){LSST Science
  Collaboration}, {Abell}, {Allison}, {Anderson}, {Andrew}, {Angel}, {Armus},
  {Arnett}, {Asztalos}, {Axelrod}, \& et~al.}]{2009arXiv0912.0201L}
{LSST Science Collaboration}, {Abell}, P.~A., {Allison}, J., {et~al.} 2009,
  ArXiv e-prints, [arXiv: 0912.0201]

\bibitem[{{Madau} \& {Dickinson}(2014)}]{2014ARA&A..52..415M}
{Madau}, P. \& {Dickinson}, M. 2014, \araa, 52, 415

\bibitem[{{Mannucci} {et~al.}(2007){Mannucci}, {Della Valle}, \&
  {Panagia}}]{2007MNRAS.377.1229M}
{Mannucci}, F., {Della Valle}, M., \& {Panagia}, N. 2007, \mnras, 377, 1229

\bibitem[{{Mannucci} {et~al.}(2008){Mannucci}, {Maoz}, {Sharon}, {Botticella},
  {Della Valle}, {Gal-Yam}, \& {Panagia}}]{2008MNRAS.383.1121M}
{Mannucci}, F., {Maoz}, D., {Sharon}, K., {et~al.} 2008, \mnras, 383, 1121

\bibitem[{{Maoz} {et~al.}(2010){Maoz}, {Sharon}, \&
  {Gal-Yam}}]{2010ApJ...722.1879M}
{Maoz}, D., {Sharon}, K., \& {Gal-Yam}, A. 2010, \apj, 722, 1879

\bibitem[{{Martini} {et~al.}(2007){Martini}, {Mulchaey}, \&
  {Kelson}}]{2007ApJ...664..761M}
{Martini}, P., {Mulchaey}, J.~S., \& {Kelson}, D.~D. 2007, \apj, 664, 761

\bibitem[{{Mattila} {et~al.}(2012){Mattila}, {Dahlen}, {Efstathiou}, {Kankare},
  {Melinder}, {Alonso-Herrero}, {P{\'e}rez-Torres}, {Ryder},
  {V{\"a}is{\"a}nen}, \& {{\"O}stlin}}]{Mattila12}
{Mattila}, S., {Dahlen}, T., {Efstathiou}, A., {et~al.} 2012, \apj, 756, 111

\bibitem[{{Melinder} {et~al.}(2012){Melinder}, {Dahlen}, {Menc{\'{\i}}a
  Trinchant}, {{\"O}stlin}, {Mattila}, {Sollerman}, {Fransson}, {Hayes},
  {Kankare}, \& {Nasoudi-Shoar}}]{Melinder}
{Melinder}, J., {Dahlen}, T., {Menc{\'{\i}}a Trinchant}, L., {et~al.} 2012,
  \aap, 545, A96

\bibitem[{{Narasimha} \& {Chitre}(1988)}]{1988ApJ...332...75N}
{Narasimha}, D. \& {Chitre}, S.~M. 1988, \apj, 332, 75

\bibitem[{{Neill} {et~al.}(2006){Neill}, {Sullivan}, {Balam}, {Pritchet},
  {Howell}, {Perrett}, {Astier}, {Aubourg}, {Basa}, {Carlberg}, {Conley},
  {Fabbro}, {Fouchez}, {Guy}, {Hook}, {Pain}, {Palanque-Delabrouille},
  {Regnault}, {Rich}, {Taillet}, {Aldering}, {Antilogus}, {Arsenijevic},
  {Balland}, {Baumont}, {Bronder}, {Ellis}, {Filiol}, {Gon{\c c}alves},
  {Hardin}, {Kowalski}, {Lidman}, {Lusset}, {Mouchet}, {Mourao}, {Perlmutter},
  {Ripoche}, {Schlegel}, \& {Tao}}]{2006AJ....132.1126N}
{Neill}, J.~D., {Sullivan}, M., {Balam}, D., {et~al.} 2006, \aj, 132, 1126

\bibitem[{{Nobili} {et~al.}(2005){Nobili}, {Amanullah}, {Garavini}, {Goobar},
  {Lidman}, {Stanishev}, {Aldering}, {Antilogus}, {Astier}, {Burns}, {Conley},
  {Deustua}, {Ellis}, {Fabbro}, {Fadeyev}, {Folatelli}, {Gibbons}, {Goldhaber},
  {Groom}, {Hook}, {Howell}, {Kim}, {Knop}, {Nugent}, {Pain}, {Perlmutter},
  {Quimby}, {Raux}, {Regnault}, {Ruiz-Lapuente}, {Sainton}, {Schahmaneche},
  {Smith}, {Spadafora}, {Thomas}, {Wang}, \& {Supernova Cosmology
  Project}}]{2005A&A...437..789N}
{Nobili}, S., {Amanullah}, R., {Garavini}, G., {et~al.} 2005, \aap, 437, 789

\bibitem[{{Nordin} {et~al.}(2014){Nordin}, {Rubin}, {Richard}, {Rykoff},
  {Aldering}, {Amanullah}, {Atek}, {Barbary}, {Deustua}, {Fakhouri},
  {Fruchter}, {Goobar}, {Hook}, {Hsiao}, {Huang}, {Kneib}, {Lidman}, {Meyers},
  {Perlmutter}, {Saunders}, {Spadafora}, {Suzuki}, \& {Supernova Cosmology
  Project}}]{2014MNRAS.440.2742N}
{Nordin}, J., {Rubin}, D., {Richard}, J., {et~al.} 2014, \mnras, 440, 2742

\bibitem[{{Oguri}(2007)}]{2007ApJ...660....1O}
{Oguri}, M. 2007, \apj, 660, 1

\bibitem[{{Oguri} \& {Kawano}(2003)}]{2003MNRAS.338L..25O}
{Oguri}, M. \& {Kawano}, Y. 2003, \mnras, 338, L25

\bibitem[{{Pan} \& {Loeb}(2013)}]{2013MNRAS.435L..33P}
{Pan}, T. \& {Loeb}, A. 2013, \mnras, 435, L33

\bibitem[{{Patel} {et~al.}(2014){Patel}, {McCully}, {Jha}, {Rodney}, {Jones},
  {Graur}, {Merten}, {Zitrin}, {Riess}, {Matheson}, {Sako}, {Holoien},
  {Postman}, {Coe}, {Bartelmann}, {Balestra}, {Ben{\'{\i}}tez}, {Bouwens},
  {Bradley}, {Broadhurst}, {Cenko}, {Donahue}, {Filippenko}, {Ford},
  {Garnavich}, {Grillo}, {Infante}, {Jouvel}, {Kelson}, {Koekemoer}, {Lahav},
  {Lemze}, {Maoz}, {Medezinski}, {Melchior}, {Meneghetti}, {Molino},
  {Moustakas}, {Moustakas}, {Nonino}, {Rosati}, {Seitz}, {Strolger}, {Umetsu},
  \& {Zheng}}]{2014ApJ...786....9P}
{Patel}, B., {McCully}, C., {Jha}, S.~W., {et~al.} 2014, \apj, 786, 9

\bibitem[{{Pirard} {et~al.}(2004){Pirard}, {Kissler-Patig}, {Moorwood},
  {Biereichel}, {Delabre}, {Dorn}, {Finger}, {Gojak}, {Huster}, {Jung}, {Koch},
  {Le Louarn}, {Lizon}, {Mehrgan}, {Pozna}, {Silber}, {Sokar}, \&
  {Stegmeier}}]{2004SPIE.5492.1763P}
{Pirard}, J.-F., {Kissler-Patig}, M., {Moorwood}, A., {et~al.} 2004, in Society
  of Photo-Optical Instrumentation Engineers (SPIE) Conference Series, Vol.
  5492, Ground-based Instrumentation for Astronomy, ed. A.~F.~M. {Moorwood} \&
  M.~{Iye}, 1763--1772

\bibitem[{{Porciani} \& {Madau}(2000)}]{2000ApJ...532..679P}
{Porciani}, C. \& {Madau}, P. 2000, \apj, 532, 679

\bibitem[{{Postman} {et~al.}(2012){Postman}, {Coe}, {Ben{\'{\i}}tez},
  {Bradley}, {Broadhurst}, {Donahue}, {Ford}, {Graur}, {Graves}, {Jouvel},
  {Koekemoer}, {Lemze}, {Medezinski}, {Molino}, {Moustakas}, {Ogaz}, {Riess},
  {Rodney}, {Rosati}, {Umetsu}, {Zheng}, {Zitrin}, {Bartelmann}, {Bouwens},
  {Czakon}, {Golwala}, {Host}, {Infante}, {Jha}, {Jimenez-Teja}, {Kelson},
  {Lahav}, {Lazkoz}, {Maoz}, {McCully}, {Melchior}, {Meneghetti}, {Merten},
  {Moustakas}, {Nonino}, {Patel}, {Reg{\"o}s}, {Sayers}, {Seitz}, \& {Van der
  Wel}}]{2012ApJS..199...25P}
{Postman}, M., {Coe}, D., {Ben{\'{\i}}tez}, N., {et~al.} 2012, \apjs, 199, 25

\bibitem[{{Refsdal}(1964)}]{1964MNRAS.128..307R}
{Refsdal}, S. 1964, \mnras, 128, 307

\bibitem[{{Richardson} {et~al.}(2002){Richardson}, {Branch}, {Casebeer},
  {Millard}, {Thomas}, \& {Baron}}]{2002AJ....123..745R}
{Richardson}, D., {Branch}, D., {Casebeer}, D., {et~al.} 2002, \aj, 123, 745

\bibitem[{{Richardson} {et~al.}(2014){Richardson}, {Jenkins}, {Wright}, \&
  {Maddox}}]{2014AJ....147..118R}
{Richardson}, D., {Jenkins}, III, R.~L., {Wright}, J., \& {Maddox}, L. 2014,
  \aj, 147, 118

\bibitem[{{Riehm} {et~al.}(2011){Riehm}, {M{\"o}rtsell}, {Goobar}, {Amanullah},
  {Dahl{\'e}n}, {J{\"o}nsson}, {Limousin}, {Paech}, \& {Richard}}]{Third}
{Riehm}, T., {M{\"o}rtsell}, E., {Goobar}, A., {et~al.} 2011, \aap, 536, A94

\bibitem[{{Riello} \& {Patat}(2005)}]{2005MNRAS.362..671R}
{Riello}, M. \& {Patat}, F. 2005, \mnras, 362, 671

\bibitem[{{Rodney} {et~al.}(2015{\natexlab{a}}){Rodney}, {Patel}, {Scolnic},
  {Foley}, {Molino}, {Brammer}, {Jauzac}, {Bradac}, {Coe}, {Broadhurst},
  {Diego}, {Graur}, {Hjorth}, {Hoag}, {Jha}, {Johnson}, {Kelly}, {Lam},
  {McCully}, {Medezinski}, {Meneghetti}, {Merten}, {Richard}, {Riess},
  {Sharon}, {Strolger}, {Treu}, {Wang}, {Williams}, \&
  {Zitrin}}]{2015arXiv150506211R}
{Rodney}, S.~A., {Patel}, B., {Scolnic}, D., {et~al.} 2015{\natexlab{a}}, ApJ, 811, 70

\bibitem[{{Rodney} {et~al.}(2015{\natexlab{b}}){Rodney}, {Patel}, {Scolnic},
  {Foley}, {Molino}, {Brammer}, {Jauzac}, {Brada{\v c}}, {Broadhurst}, {Coe},
  {Diego}, {Graur}, {Hjorth}, {Hoag}, {Jha}, {Johnson}, {Kelly}, {Lam},
  {McCully}, {Medezinski}, {Meneghetti}, {Merten}, {Richard}, {Riess},
  {Sharon}, {Strolger}, {Treu}, {Wang}, {Williams}, \&
  {Zitrin}}]{2015ApJ...811...70R}
{Rodney}, S.~A., {Patel}, B., {Scolnic}, D., {et~al.} 2015{\natexlab{b}}, \apj,
  811, 70

\bibitem[{{Rodney} {et~al.}(2014){Rodney}, {Riess}, {Strolger}, {Dahlen},
  {Graur}, {Casertano}, {Dickinson}, {Ferguson}, {Garnavich}, {Hayden}, {Jha},
  {Jones}, {Kirshner}, {Koekemoer}, {McCully}, {Mobasher}, {Patel}, {Weiner},
  {Cenko}, {Clubb}, {Cooper}, {Filippenko}, {Frederiksen}, {Hjorth},
  {Leibundgut}, {Matheson}, {Nayyeri}, {Penner}, {Trump}, {Silverman}, {U},
  {Azalee Bostroem}, {Challis}, {Rajan}, {Wolff}, {Faber}, {Grogin}, \&
  {Kocevski}}]{2014AJ....148...13R}
{Rodney}, S.~A., {Riess}, A.~G., {Strolger}, L.-G., {et~al.} 2014, \aj, 148, 13

\bibitem[{{Salpeter}(1955)}]{1955ApJ...121..161S}
{Salpeter}, E.~E. 1955, \apj, 121, 161

\bibitem[{{Scalzo} {et~al.}(2010){Scalzo}, {Aldering}, {Antilogus}, {Aragon},
  {Bailey}, {Baltay}, {Bongard}, {Buton}, {Childress}, {Chotard}, {Copin},
  {Fakhouri}, {Gal-Yam}, {Gangler}, {Hoyer}, {Kasliwal}, {Loken}, {Nugent},
  {Pain}, {P{\'e}contal}, {Pereira}, {Perlmutter}, {Rabinowitz}, {Rau},
  {Rigaudier}, {Runge}, {Smadja}, {Tao}, {Thomas}, {Weaver}, \&
  {Wu}}]{2010ApJ...713.1073S}
{Scalzo}, R.~A., {Aldering}, G., {Antilogus}, P., {et~al.} 2010, \apj, 713,
  1073

\bibitem[{{Scannapieco} \& {Bildsten}(2005)}]{2005ApJ...629L..85S}
{Scannapieco}, E. \& {Bildsten}, L. 2005, \apjl, 629, L85

\bibitem[{{Sharon} {et~al.}(2010){Sharon}, {Gal-Yam}, {Maoz}, {Filippenko},
  {Foley}, {Silverman}, {Ebeling}, {Ma}, {Ofek}, {Kneib}, {Donahue}, {Ellis},
  {Freedman}, {Kirshner}, {Mulchaey}, {Sarajedini}, \&
  {Voit}}]{2010ApJ...718..876S}
{Sharon}, K., {Gal-Yam}, A., {Maoz}, D., {et~al.} 2010, \apj, 718, 876

\bibitem[{{Sharon} {et~al.}(2007){Sharon}, {Gal-Yam}, {Maoz}, {Filippenko}, \&
  {Guhathakurta}}]{2007ApJ...660.1165S}
{Sharon}, K., {Gal-Yam}, A., {Maoz}, D., {Filippenko}, A.~V., \&
  {Guhathakurta}, P. 2007, \apj, 660, 1165

\bibitem[{{Smith} {et~al.}(2009){Smith}, {Silverman}, {Chornock}, {Filippenko},
  {Wang}, {Li}, {Ganeshalingam}, {Foley}, {Rex}, \&
  {Steele}}]{2009ApJ...695.1334S}
{Smith}, N., {Silverman}, J.~M., {Chornock}, R., {et~al.} 2009, \apj, 695, 1334

\bibitem[{{Spergel} {et~al.}(2013){Spergel}, {Gehrels}, {Breckinridge},
  {Donahue}, {Dressler}, {Gaudi}, {Greene}, {Guyon}, {Hirata}, {Kalirai},
  {Kasdin}, {Moos}, {Perlmutter}, {Postman}, {Rauscher}, {Rhodes}, {Wang},
  {Weinberg}, {Centrella}, {Traub}, {Baltay}, {Colbert}, {Bennett},
  {Kiessling}, {Macintosh}, {Merten}, {Mortonson}, {Penny}, {Rozo},
  {Savransky}, {Stapelfeldt}, {Zu}, {Baker}, {Cheng}, {Content}, {Dooley},
  {Foote}, {Goullioud}, {Grady}, {Jackson}, {Kruk}, {Levine}, {Melton},
  {Peddie}, {Ruffa}, \& {Shaklan}}]{2013arXiv1305.5422S}
{Spergel}, D., {Gehrels}, N., {Breckinridge}, J., {et~al.} 2013, ArXiv e-prints

\bibitem[{{Stanishev} {et~al.}(2009){Stanishev}, {Goobar}, {Paech},
  {Amanullah}, {Dahl{\'e}n}, {J{\"o}nsson}, {Kneib}, {Lidman}, {Limousin},
  {M{\"o}rtsell}, {Nobili}, {Richard}, {Riehm}, \& {von Strauss}}]{First}
{Stanishev}, V., {Goobar}, A., {Paech}, K., {et~al.} 2009, \aap, 507, 61

\bibitem[{{Strolger} {et~al.}(2015){Strolger}, {Dahlen}, {Rodney}, {Graur},
  {Riess}, {McCully}, {Ravindranath}, {Mobasher}, \& {Shahady}}]{Strolger15}
{Strolger}, L.-G., {Dahlen}, T., {Rodney}, S.~A., {et~al.} 2015, \apj,  813, 93

\bibitem[{{Sullivan} {et~al.}(2000){Sullivan}, {Ellis}, {Nugent}, {Smail}, \&
  {Madau}}]{2000MNRAS.319..549S}
{Sullivan}, M., {Ellis}, R., {Nugent}, P., {Smail}, I., \& {Madau}, P. 2000,
  \mnras, 319, 549

\bibitem[{{Taddia} {et~al.}(2013){Taddia}, {Stritzinger}, {Sollerman},
  {Phillips}, {Anderson}, {Boldt}, {Campillay}, {Castell{\'o}n}, {Contreras},
  {Folatelli}, {Hamuy}, {Heinrich-Josties}, {Krzeminski}, {Morrell}, {Burns},
  {Freedman}, {Madore}, {Persson}, \& {Suntzeff}}]{2013A&A...555A..10T}
{Taddia}, F., {Stritzinger}, M.~D., {Sollerman}, J., {et~al.} 2013, \aap, 555,
  A10

\bibitem[{{Taylor} {et~al.}(2014){Taylor}, {Cinabro}, {Dilday}, {Galbany},
  {Gupta}, {Kessler}, {Marriner}, {Nichol}, {Richmond}, {Schneider}, \&
  {Sollerman}}]{2014ApJ...792..135T}
{Taylor}, M., {Cinabro}, D., {Dilday}, B., {et~al.} 2014, \apj, 792, 135

\bibitem[{{Trenti} \& {Stiavelli}(2008)}]{2008ApJ...676..767T}
{Trenti}, M. \& {Stiavelli}, M. 2008, \apj, 676, 767

\bibitem[{{Treu} {et~al.}(2015){Treu}, {Schmidt}, {Brammer}, {Vulcani}, {Wang},
  {Brada{\v c}}, {Dijkstra}, {Dressler}, {Fontana}, {Gavazzi}, {Henry}, {Hoag},
  {Huang}, {Jones}, {Kelly}, {Malkan}, {Mason}, {Pentericci}, {Poggianti},
  {Stiavelli}, {Trenti}, \& {von der Linden}}]{2015ApJ...812..114T}
{Treu}, T., {Schmidt}, K.~B., {Brammer}, G.~B., {et~al.} 2015, \apj, 812, 114

\bibitem[{{Umetsu} {et~al.}(2015){Umetsu}, {Sereno}, {Medezinski}, {Nonino},
  {Mroczkowski}, {Diego}, {Ettori}, {Okabe}, {Broadhurst}, \&
  {Lemze}}]{2015ApJ...806..207U}
{Umetsu}, K., {Sereno}, M., {Medezinski}, E., {et~al.} 2015, \apj, 806, 207

\bibitem[{{Zitrin} {et~al.}(2014){Zitrin}, {Redlich}, \&
  {Broadhurst}}]{2014ApJ...789...51Z}
{Zitrin}, A., {Redlich}, M., \& {Broadhurst}, T. 2014, \apj, 789, 51

\bibitem[{{Zwicky}(1937)}]{1937ApJ....86..217Z}
{Zwicky}, F. 1937, \apj, 86, 217

\bibitem[{{Zwicky}(1938)}]{1938ApJ....88..529Z}
{Zwicky}, F. 1938, \apj, 88, 529

\end{thebibliography}
\end{document}